\documentclass{aa}

\usepackage{graphicx}
\usepackage{txfonts}

\graphicspath{{figures/}}

\usepackage{siunitx}

\newcommand{\dd}{\mathrm{d}}
\newcommand\dint{\mathrm{d}}
\newcommand\dx{\dint x}
\newcommand\ds{\dint s}

\newcommand{\Spectractor}{\texttt{Spectractor }}
\newcommand{\DCCD}{D_{\mathrm{CCD}}}
\newcommand{\Neff}{N_{\mathrm{eff}}}
\newcommand{\GCCD}{G_{\mathrm{CCD}}}

\begin{document}

\title{Slitless spectrophotometry with forward modelling: Principles and application to measuring atmospheric transmission}
\author{J.~Neveu\inst{1,2}\and V.~Brémaud\inst{2}\and P.~Antilogus\inst{1}\and F.~Barret\inst{3}\and S.~Bongard\inst{1}\and Y.~Copin\inst{4}\and S.~Dagoret-Campagne\inst{2}\and C.~Juramy\inst{1}\and L.~Le~Guillou\inst{1}\and M.~Moniez\inst{2}\and E.~Sepulveda\inst{1}\and The LSST Dark Energy Science Collaboration}
\institute{Sorbonne Universit\'e, CNRS, Universit\'e de Paris, LPNHE, 75252 Paris Cedex 05, France
\and Universit\'e Paris-Saclay, CNRS, IJCLab, 91405, Orsay, France
\and MODAL'X, UPL, Univ. Paris Nanterre, CNRS, F92000 Nanterre France
\and Univ Lyon, Univ Claude Bernard Lyon 1, CNRS/IN2P3, IP2I Lyon, UMR 5822, F-69622, Villeurbanne, France
}

\authorrunning{J. Neveu et al.}
\titlerunning{Slitless spectrophotometry with forward modelling}


\abstract
{In the next decade, many optical surveys will aim to answer the question of the nature of dark energy by measuring its equation-of-state parameter at the per mill level. This requires trusting the photometric calibration of the survey with a precision never reached so far on many sources of systematic uncertainties. The measurement of the on-site atmospheric transmission for each exposure, or for each season or for the full survey on average, can help reach the per mill precision for the magnitudes.}
{This work aims at proving the ability to use slitless spectroscopy for standard-star spectrophotometry and its use to monitor on-site atmospheric transmission as needed, for example, by the Vera C. Rubin Observatory Legacy Survey of Space and Time supernova cosmology program. We fully deal with the case of a disperser in the filter wheel, which is the configuration chosen in the Rubin Auxiliary Telescope.} 
{The theoretical basis of slitless spectrophotometry is at the heart of our forward-model approach to extract spectroscopic information from slitless data. We developed a publicly available software called \texttt{Spectractor}\thanks{https://github.com/LSSTDESC/Spectractor}, which implements each ingredient of the model and finally performs a fit of a spectrogram model directly on image data to obtain the spectrum.}
{We show through simulations that our model allows us to understand the structure of spectrophotometric exposures. We also demonstrate its use on real data by solving specific issues and illustrating that our procedure allows the improvement of the model describing the data. Finally, we discuss how this approach can be used to directly extract atmospheric transmission parameters from the data and thus provide the base for on-site atmosphere monitoring. We show the efficiency of the procedure in simulations and test it on the limited available data set.} %
{}
{}

\keywords{Astronomical instrumentation, methods and techniques - Atmospheric effects - Instrumentation: spectrographs -  Techniques: imaging spectroscopy - Cosmology: observations}

\maketitle

\section{Introduction}\label{sec:intro}

Cosmology measures and interprets the evolution of the whole Universe. To probe
its dynamics and understand the nature of dark energy, observers need to
compute distances at different epochs from the light they receive in
telescopes. The evolution of cosmological distances with time indicates how dark
energy, dark matter, and matter interact and how they can be modelled.

Optical surveys use magnitude and colour comparisons to build a relative
distance scale. For instance, type Ia supernovae (SNe~Ia) revealed the presence
of a dark energy component because they appeared fainter in the early Universe
than was thought \citep{Riess1998,Perlmutter1999,Betoule2014,Scolnic2018}. More precisely, because SNIa colours shift with the expansion of the
Universe, high-redshift supernovae were fainter in red bands than
what can be inferred from low-redshift supernovae observed in blue bands. This
case underlines that colours need to be accurately calibrated in an optical
survey to display the dynamics of the Universe (see e.g. \citealt{Betoule2013}). Every chromatic effect that alters the
astral light distorts our dynamic perception of the expansion of the Universe, such as
the galactic dust, the instrumental response, or the local atmospheric
conditions.

In this paper, we present a forward-modelling method to analyse and extract data
gathered with a dispersing element (grating or hologram) in the filter wheel of
a telescope. We label our approach {forward modelling} because we
implement a numerical simulation of the data-taking procedure that includes as much a priori knowledge as available, and then estimate model parameters
  from likelihood maximisation. This method is fundamentally different from the traditional flux-weighted sum orthogonal to the dispersion axis \citep{Horne1986,Robertson1986} or the algebraic method that uses multiple images \citep{Ryan2018} because it relies on physical modelling to directly describe the footprint of the spectrum on the imaging sensor. Deconvolution techniques and point spread function (PSF) modelling have been explored for optical fibre spectrographs \citep{2010PASP..122..248B,2019MNRAS.484.2403L}, but in our forward model, we proceed to build a physical model for the extraction of the spectra, and in particular, the atmospheric transmission from spectra. Our approach was inspired by the forward modelling developed in \cite{Outini2019}, and we applied it for punctual sources with the ultimate goal of measuring atmospheric transmission.

The scientific context of our work is the study of the atmospheric transmission
variability via the repeated observation of stable (standard) stars. 
\cite{Burke_2010} opened the path to controlling the optical survey photometry with dedicated 
measurements of atmospheric components by observing standard stars. Then,
\cite{Burke_2017} reached a 5 per mill (i.e. per thousand)
relative photometric calibration between filters that covered the full optical and
the near-infrared (near-IR) range and accounted for linear temporal variations of the atmosphere
transmission over the nights. Because of the scale of the new SNe~Ia surveys, the
number of observed objects now reaches a point where even such an exquisite
calibration becomes the dominant source of systematics. One of the
challenges of modern SNe~Ia cosmology thus is to be able to
accurately measure and estimate the chromatic variations of the atmospheric
transmission at the per-thousand level that allow us to probe for systematic variations, either nightly,
seasonal, or directional.

While spectrophotometry at the required precision has been hinted at in
\cite{Buton_2013,Rubin2022}, it relies on
the dedicated use of a specifically designed integral field spectrograph. Our
current approach instead focuses on exploring the spectrophotometric
possibilities offered by a much simpler design: We consider a slitless
spectrograph, where a disperser (either a grating or a hologram) is inserted in
the converging beam of a telescope in a regular filter wheel. This
implementation is used in the Rubin Auxiliary Telescope (AuxTel) \citep{2020SPIE11452E..0UI}, as well as on the Star Direct Illumination Calibration Experiment (StarDICE) telescope. StarDICE is an experiment aiming at
transferring to stars the unit of optical power (watt) defined at the National Institute of Standards and Technology (NIST) with a reference cryogenic radiometer, the Primary Optical Watt Radiometer (POWR) \citep{Houston_2006,hazenberg:tel-02950846,StarDICE_bench}.

This paper describes the preparatory work for these projects and presents the
analysis procedure we developed and tested on a few nights of data gathered at the Cerro Tololo Inter-American Observatory (CTIO). It describes some implementation choices and demonstrates that the forward
modelling approach allows us to incrementally build a detailed understanding of the
data that in the end can permit the direct extraction of the parameters used to
describe the atmospheric transmission variability.

The first section of this paper describes the theory of slitless spectrophotometry, the basic implementation
in \texttt{Spectractor}, and the data and simulation sets we used to assess the quality of the algorithm. Section~\ref{sec:spectractor}
details the different ingredients of the \Spectractor software, the assumptions and the implementation choices. In particular, we detail the regularised deconvolution technique at the heart of the process to obtain a prior for the forward model and to qualify the code on simulations. The application of \Spectractor to extract spectra from on-sky CTIO data is described in section~\ref{sec:data},
while section~\ref{sec:atm} focuses on the measurement of the atmospheric transmission. Discussions and summaries conclude the paper in section~\ref{sec:conclusion}.

\section{Forward modelling of a slitless spectrogram}\label{sec:slitless}

There are many different possible configurations for gathering spectroscopic data
without the use of a slit. 
The slitless spectrograph configuration that we considered (a grating or hologram in a converging beam) can be implemented in different
ways that could require special care in the forward-modelling analysis. For
example, the field of view can be small and may contain only one star, or it can be
crowded, with many source-dispersed images (so-called spectrograms hereafter) super-imposed on each other. Different detectors might span the field of view, with responses that
need to be mapped and gaps that need to be accounted for. We did not try to abstractly solve all different situations. We therefore defer all the technical issues that we did not encounter to further work and
concentrate on those we did and solved.

Within these restrictions, we consider in the following only the case of point
sources and do not discuss, except for a passing mention in the next section,
the case of extended sources such as galaxies or resolved planetary
  nebulae, nor the deblending of these extended sources with point sources.

\subsection{Description and geometry of a slitless spectrograph}

\begin{figure}[!ht]
\begin{center}
\includegraphics[width=0.8\columnwidth]{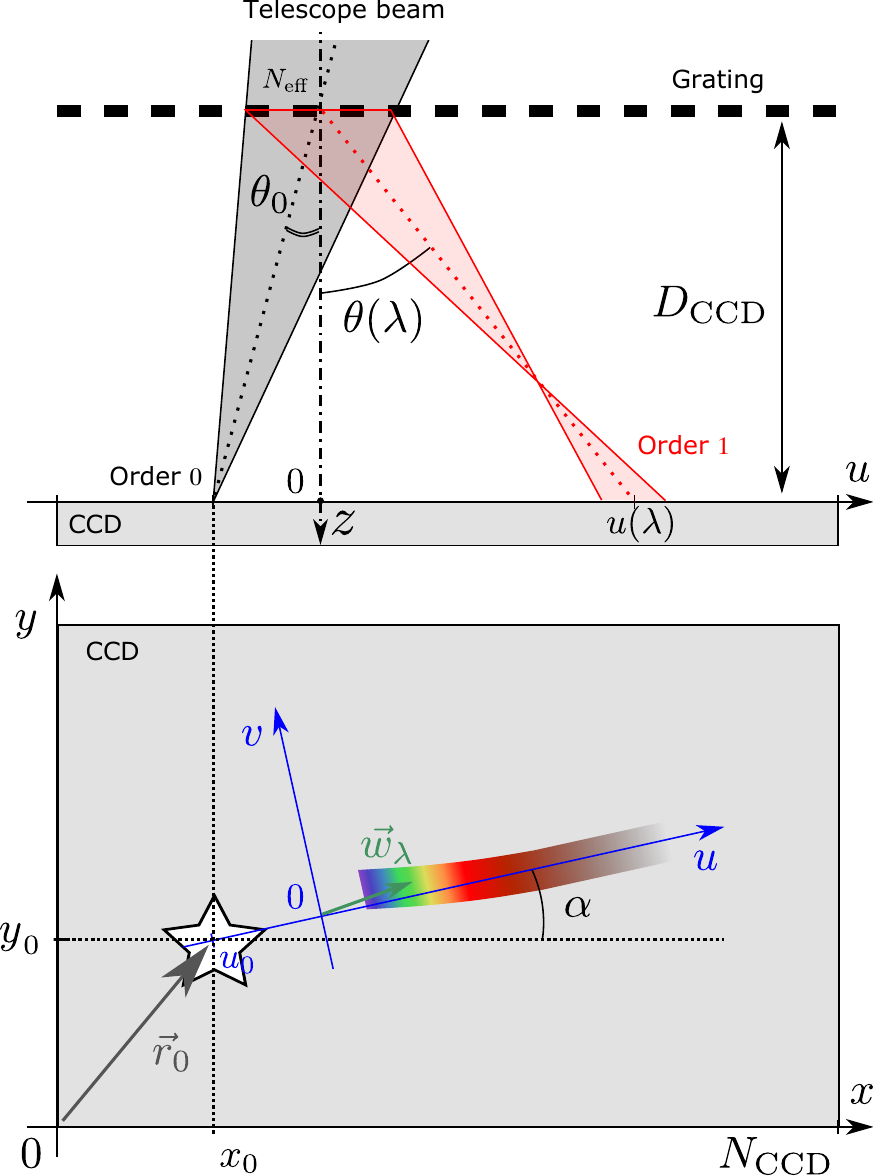}
\end{center}
\caption[] 
{Geometry of a simple slitless spectrograph in the plane orthogonal to the CCD (top) and parallel to the CCD (bottom). A convergent beam with incident angle $\theta_0$ is focused on a CCD at position $\vec r_0=(x_0,y_0)$, but also passed through a grating with $\Neff$ grooves per millimetre at a distance $\DCCD$ above the CCD. The beam is deflected at an angle $\theta(\lambda)$ along the mean dispersion axis $u$, which forms an angle $\alpha$ with respect to the CCD $x$ axis, but is focused somewhere above the sensor. Atmospheric refraction adds a supplementary dispersion along and transversally to the mean dispersion axis.}
\label{fig:dispersion}
\end{figure}

The slitless spectrograph we consider can be seen as a grating with $N$ grooves
per millimetre, placed in a telescope beam at a distance $\DCCD$ from 
a charge-coupled device (CCD) sensor. In the following, the positions in the sensor plane are parametrised with
the coordinates $\vec r = (x, y)$, and the $z$-axis points orthogonally toward
the CCD. The disperser can be more or less complex and may be used in a convergent or in
a parallel beam, and its resolution can vary, but in the end, it will
spread the source light in different diffraction
orders superimposed on the CCD, with a sky background that is also
diffracted. Depending on the choice of $N$, on the size $N_{\mathrm{CCD}}$ (in pixels) of
the sensor, and on $\DCCD$, different diffraction orders will finally be
recorded by the sensor.

The special case of the zeroth order is worth mentioning. While its presence on the
image is not mandatory, knowing its centroid position $\vec r_0$ can
significantly facilitate the setting of the zero of the wavelength calibration.

The positive and negative diffraction orders are placed on each side of the zeroth
order on a line forming an angle $\alpha$ with the $x$-axis. We parametrised the
position along this dispersion axis with the coordinate $u$ and transversally
with the coordinate $v$. The zeroth order stands at coordinates $(u_0,0)$ in the
$(u,v)$ coordinate system. If the instrument wavelength coverage spans more
than one octave in wavelength, the different diffraction orders are superimposed
on each other.

The source spectrogram is the total 2D CCD image formed by the cummulation of the dispersed light. All the notations are illustrated in Figure~\ref{fig:dispersion}.

For a periodic grating placed inside a convergent beam instead of a parallel beam, this optical system response is astigmatic, that is, the image of a point source
such as a star is not point-like on the sensor. Usually, the image 
is elliptical, and the redder the wavelength, the wider the ellipse (see e.g. \citealt{holo}).
However, the centroid of these ellipses is still given by the classical grating formula (see
e.g. \citealt{Murty:62,Hall:66,schroeder2000astronomical}),
\begin{align}
    & \sin \theta_p(\lambda) - \sin \theta_0 = p N_{\rm eff} \lambda,\label{eq:grating} \\
    &  \tan \theta(\lambda) = \frac{u(\lambda)}{\DCCD}, \tan \theta_0 = \frac{u_0}{\DCCD}.
\end{align}
The angles are those of the projection in the plane perpendicular to the grating
lines (see Figure~\ref{fig:dispersion}); $\theta_0$ is the angle of the
projected telescope beam axis with respect to the normal to the grating surface,
$p$ is the diffraction order, $\theta_p(\lambda)$ is the corresponding projected
diffracted angle, and $\Neff$ is the effective spatial frequency of the
grating lines at the position of the central ray\footnote{For some dispersers
  this number of lines per millimetre can depend on the beam position on the
  grating.} of the light beam (hereafter called chief ray).
  
The observer can choose the focus of the telescope. One common
choice is to set the focus using the zeroth order, but if this is done, the spectrogram can be affected by defocusing effects at increasingly redder wavelengths (for periodic gratings). To minimise the defocusing effect
and increase the spectrograph resolution, the focus
for a particular wavelength can be optimised, but it is then more difficult to set the zero of
the wavelength calibration with a defocused zeroth-order image.  In the following, we
assume that the focus has been made on the zeroth order unless otherwise
specified.

\subsection{Theoretical model of a spectrogram}

\begin{figure}
  \begin{center}
    \includegraphics[width=1\columnwidth]{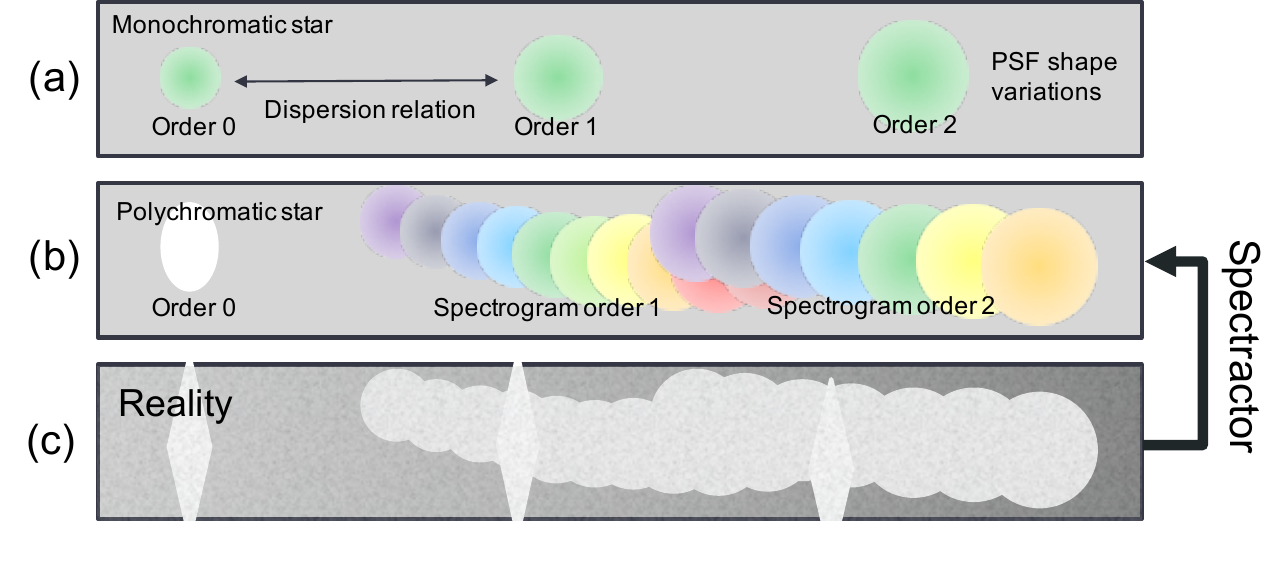}
  \end{center}
  \caption[] {General illustration of the spectrograph model and of the purpose
    of the spectrum extraction. (a) Illustration of the image formation for a
    monochromatic star observed through a slitless spectrograph. (b) Same for a
    polychromatic star. (c) Real acquired image: The detector generally is a
    black-and-white sensor, with noise, a structured and chromatic background,
    and field star zeroth- and first-order spectrograms. The goal of
    \Spectractor is to extract the colour information from
    this image to obtain the stellar spectrum.}
\label{fig:spectractor_illustration}
\end{figure}

A theoretically perfect image of a light source can be modelled by its spatio-spectral flux
density $C(\vec r, \lambda)$. For a point source, we can separate the spectral
and spatial distributions as
\begin{equation}
  S_*(\vec r,\lambda) = 
    S_{*}(\lambda) \times \delta\left(\vec r - \vec r_0\right),
\end{equation}
with $S_*(\lambda)$ spectral
energy density (SED) of the astrophysical object, and $\delta$ the Dirac
distribution. The observed spectral and spatial distribution is then
\begin{equation}
  C_p(\vec r,\lambda) = 
    T_{\text{inst}, p}(\lambda) \, T_{\text{atm}}(\lambda | \vec P_a) \,
    S_*(\vec r,\lambda),
\end{equation}
where $ T_{\rm atm}(\lambda| \vec P_a)$ is the atmospheric
transmission, depending on a set of atmospheric parameters $\vec P_a$, and
$T_{\rm inst, p}(\lambda)$ is the instrumental transmission (including the CCD
quantum efficiency) for the diffraction order $p$.

In the case of a monochromatic point source, we have $C_0(\vec r,\lambda) =
A\,\delta(\lambda- \lambda_0)\times \delta\left(\vec r - \vec r_0\right)$, with $A$ the received source flux at $\lambda_0$. The PSF describes the
optical response of the telescope and of the atmospheric turbulence on a
sensitive surface such as a CCD (see
Figure~\ref{fig:spectractor_illustration}~(a)). It depends a priori on the
wavelength $\lambda$ and can be modelled by a function $\phi_0(\vec r,
\lambda)$ whose spatial integral is normalised to one. Therefore, by definition
of the PSF, the image of a monochromatic point source centred on $\vec r_0$ on
the CCD can be described as a convolution product,
\begin{align}
I_0(\vec r) & = \int \dd \lambda \iint \dd^2 \vec r'\ C_0(\vec r', \lambda)\, \phi_{0}(\vec r - \vec r', \lambda) \notag \\
 & = A\,\phi_{0}(\vec r - \vec r_0, \lambda_0).
\end{align}

With a slitless spectrograph, the mechanism is the same, but the incoming light
is dispersed in several diffraction orders $p$, and light from all diffraction orders of the sky background is superimposed. The position of the point-source
image depends on the wavelength and on the order $p$. Moreover, the shape of the PSF itself can
depend on the order $p$ and on the wavelength $\lambda$. Here and everywhere else, the dispersed-imaging PSF integrates both the seeing and the instrumental PSF. We introduce the
dispersion relation $\vec{\Delta}_p(\lambda)$ as the 2D vectorial quantity that
describes the position of the PSF centroids $\left\lbrace x_{c,p}(\lambda)-x_0,
y_{c,p}(\lambda)-y_0 \right\rbrace$ on the CCD with respect to the zeroth-order position $(x_0,y_0)$, for a diffraction order $p$ and a
wavelength $\lambda$. This quantity can be computed by applying
the classical grating formula \ref{eq:grating}. With a point-like monochromatic
source, the image recorded on the CCD is modelled as
\begin{align}
I(\vec r) & =  \sum_p \int \dd \lambda \iint \dd^2 \vec r'\ C_p(\vec r', \lambda)\, \phi_{p}(\vec r - \vec \Delta_{p}(\lambda) - \vec r', \lambda) \notag \\
 & = \sum_p A_p\,\phi_{p}(\vec r - \vec{\Delta}_p(\lambda_0), \lambda_0),
\end{align}
with
$\vec{\Delta}_0(\lambda_0) = \vec{r}_0$ and $A_p$ the flux density at
wavelength $\lambda_0$ for the order $p$. On the image, we expect a series of
spots, one per order $p$, of different intensities, with different sizes, but
containing the same spectral information $S_*(\lambda_0)$ about the source.

The observed sources are naturally polychromatic. For point-like sources, the
theoretical description above holds at each wavelength, and the image can be
described as
\begin{equation}\label{eq:spectrogram_model}
    I(\vec{r}) = \sum_{p} \int \mathrm{d}\lambda\ S_p(\lambda)\phi_p\left(\vec{r} - \vec{\Delta}_p(\lambda), \lambda  \right),
\end{equation}
with
\begin{align}
  \vec{\Delta}_p(\lambda) & = \left\{
                            x_{c,p}(\lambda) -x_0,
                            y_{c,p}(\lambda) - y_0
                            \right\}, \\
  S_p(\lambda) & = T_{\text{inst}, p}(\lambda) \, T_{\text{atm}}(\lambda | \vec P_a) \, S_{*}(\lambda). \label{eq:Sp}
\end{align}
Therefore, the spectrogram of a polychromatic source can be viewed as a
stack of monochromatic images with different centroids, or as a dispersed image with a very chromatic PSF (see
Figure~\ref{fig:spectractor_illustration}~(b)). This description of the image
and of the spectrogram formation is the base of our forward model for slitless spectrograph data.

To obtain the  $S_p(\lambda)$ spectra, a process to un-stack
the monochromatic images spread across the image by the disperser is needed
(Figure~\ref{fig:spectractor_illustration}~(c))). This in turn requires that ingredients such as the PSF $\phi_p(\vec r, \lambda)$ and the dispersion
relation $\vec{\Delta}_p(\lambda)$ are sufficiently well known, either a priori, through a specific data analysis, or directly fitted on data. The hardest point usually is the determination of the PSF kernel as a function of wavelength. As illustrated in Figures~\ref{fig:dispersion} and~\ref{fig:spectractor_illustration}, in the general case, the PSF is blurred and defocused. A simple Moffat profile can be used to model the PSF as long as the defocus is small; we used this approximation throughout this paper and leave the general defocus case for future work. Nevertheless, we studied and discuss two cases: The use of a standard grating, and the use of an holographic disperser that limits the defocus.  

In this paper, we describe our implementation of this process in the form of
the \texttt{python3} \Spectractor code \cite{spectractorascl} (see Section~\ref{sec:spectractor}). We also show
how we tested it on simulations and data and how it could be
used in order to ingest lacking modelling information from specific data analysis.


\begin{figure*}[!t]
\begin{center}
\includegraphics[width=0.8\textwidth]{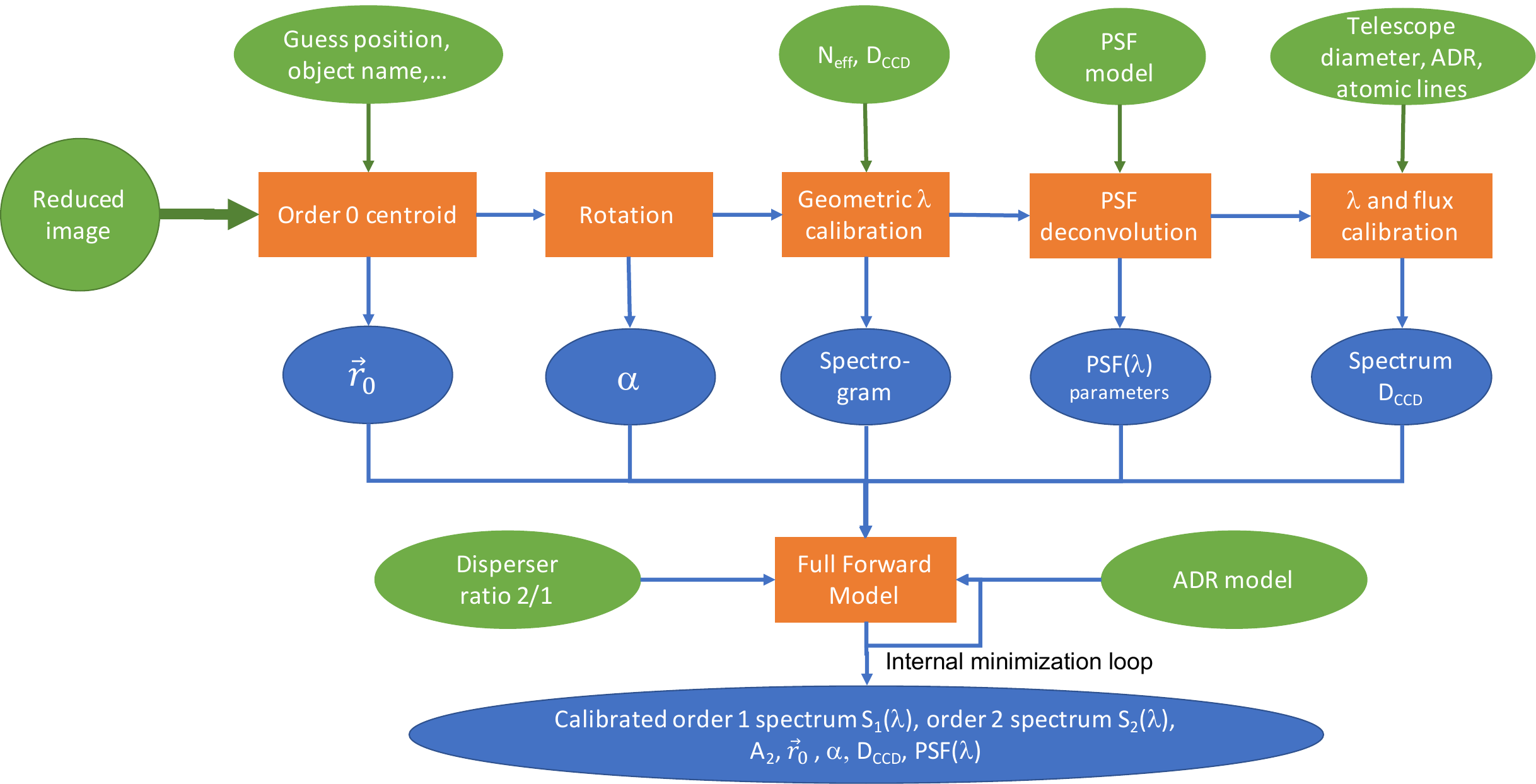}
\end{center}
\caption[] 
{Overview of the \Spectractor  pipeline. The green ellipses represent external inputs needed for the spectrum extraction, and blue ellipses stand for the \Spectractor products. The bottom stage represents the full forward-modelling method for extracting the spectrum from the raw spectrogram.}
\label{fig:spectractor_graph}
\end{figure*}

\subsection{\Spectractor}

We call
\texttt{Spectractor}\footnote{\url{https://github.com/LSSTDESC/Spectractor}}$^{,}$\footnote{\url{https://spectractor.readthedocs.io}}
the computer suite we wrote to analyse the future AuxTel images and the
images obtained on the Cerro Tololo Inter-American Observatory (CTIO) $0.9\,$m
telescope. It was trained on CTIO data, but with the purpose of being easily configurable for slitless spectrophotometry with other telescopes. The main steps, inputs, and outputs of the extraction part are illustrated in Figure~\ref{fig:spectractor_graph} and are described in detail in Section~\ref{sec:spectractor}. To facilitate reading of the following, we summarise the main \Spectractor steps below.

\paragraph{Zeroth-order centroid}
The main inputs of \Spectractor are a pre-processed image (overscan subtracted,
debiased, and spatial flat fielded) obtained with a slitless spectrograph, and a
configuration file setting the main geometrical and spectrographic properties of
the instrument ($\DCCD$, $N$, telescope diameter, pixel size, the PSF model,
etc.) At the time of writing this paper, the PSF models we implemented were either a
Gaussian profile, a Moffat profile, or a Moffat minus a Gaussian profile. Additional and
more detailed PSF models are planned to describe the AuxTel data more accurately
as they are analysed. The centroid of the zeroth
order is contained in the image. We thus implemented a searching procedure to
inform us on the origin of the location of the spectrum and set the zero of the
wavelength scale.

\paragraph{Rotation}
Because of the geometry of the spectrograph and the dispersion properties of
the grating, the spectrogram was cropped from the image and was de-rotated so that
the dispersion axis was along the horizontal axis of the cropped image. The rotation angle is fitted later in the full model without explicitly rotating
the image.

\paragraph{Geometric wavelength calibration}
On this image, a first fit of a 1D sliced PSF model transverse to the
dispersion axis is performed. The PSF shape parameters are represented by a
polynomial function \footnote{A polynomial function of the fourth order is
  sufficient to absorb the main chromatic variations of the PSF shape.}  as a
function of the distance to the zeroth order.

\paragraph{PSF deconvolution}
The procedure is continued by a deconvolution that uses a 2D PSF model and the 1D result as a
prior to regularise it.

\paragraph{Wavelength calibration}
A first wavelength calibration is performed using the detection of the principal
absorption or emission lines in the extracted spectrum (astrophysical or
telluric lines).

\paragraph{Flux calibration} 
The spectrum flux in ADU is converted into $\SI{}{erg/s/cm^2/nm}$ units using the telescope collecting area, CCD gain, and exposure time.

\paragraph{Full forward model}
Finally, given all these prior ingredients, a full forward model is initiated
on the pre-processed\footnote{For future developments, one
  could model directly the unpreprocessed
  exposure, for instance introducing the chromatic flat fields directly in
  the forward model.}
but not rotated exposure using a 2D PSF model, and a model for the atmospheric
differential refraction (ADR)\footnote{ADR is also called Differential Chromatic Refraction (DCR).}. The PSF shape parameters as well as the
wavelength calibration are refitted in the process. The main output is a
calibrated spectrum in wavelength and amplitude, but \Spectractor also returns
a host of useful fitted parameters such as the PSF shape chromaticity or the
$\DCCD$ distance to perform extraction quality analyses, and in the end,
to improve the forward modelling or its initialisation.

All steps but the last are implemented in order to provide the
required ingredients to the full forward models, and they are thus completely
contingent to the availability of understanding the instrument. This
is again be addressed below when we discuss how the optical
transmission of the instrument can be obtained.

\subsection{Data examples}

While we use the natural ability of the forward-model implementation to easily
provide simulations to validate the code, we also present the use of
\Spectractor on real data.

\subsubsection{CTIO data}

In order to test the approach of slitless spectroscopy, in particular, using a
holographic disperser, we benefited of a run of 17 nights in May-June 2017 at the
Cerro Tololo Inter-American Observatory (CTIO) $0.9\,$m Cassegrain telescope
($f/13.7$, scale at focal plane $\SI{60}{\micro\meter}/\arcsec$). This telescope is equipped
with a cooled Tek2K CCD device of $2048\times 2046$
pixels, read by four amplifiers\footnote{\url{https://noirlab.edu/science/programs/ctio/instruments/Tek2K}}. Two filter wheels are installed. The first wheel
in the light path was used to insert broad-band filters, and the second wheel
holds different dispersers.

While many gratings were tried, we focused on two of them here: a
Thorlabs blazed grating (300 lines/mm) ref. GT50-03, and an amplitude holographic
optical element (around 350 lines/mm) especially designed for this
telescope. This hologram is fully described and analysed based on the CTIO
data in~\cite{holo}. Its main advantage is that the defocusing described in
Figure~\ref{fig:dispersion} is very limited, which allowed us to model its
chromatic PSF with simpler mathematical models.

By using these dispersers in the upstream filter wheel, we readily transformed the
CTIO 0.9 telescope into a spectrophotometric instrument with a resolution of about
150 -- 200 \citep{holo}.

\begin{figure}[!h]
  \begin{center}
    \includegraphics[width=\columnwidth]{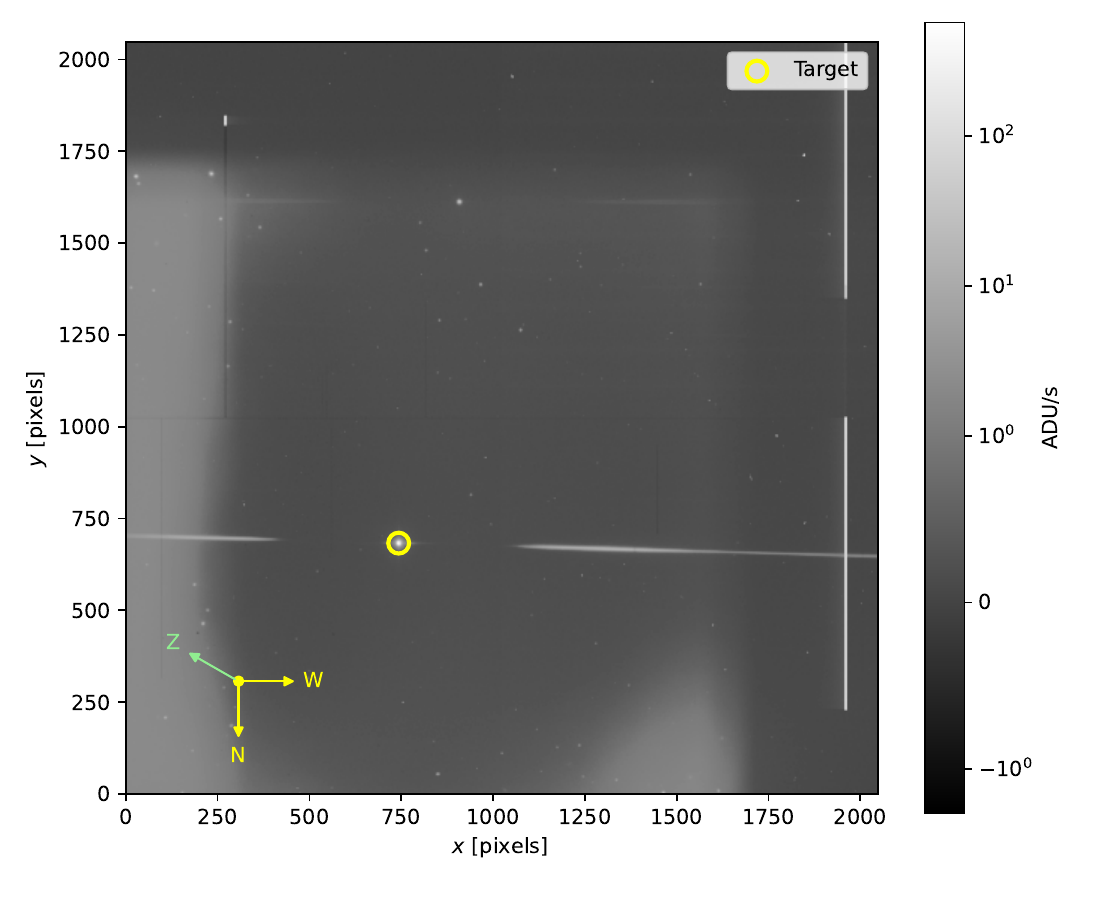}
  \end{center}
  \caption{Processed exposure of CALSPEC star HD111980 observed at CTIO with an amplitude-hologram grating of 350 lines/mm on 2017 May 30. The greyscale
    gives the exposure intensity in ADU/s. The yellow circle indicates the zeroth-order position of HD111980.}
  \label{fig:reduc_201705_134}
\end{figure}

Figure~\ref{fig:reduc_201705_134} shows an example of the data we obtained: The
dispersion axes are nearly horizontal along the $x$-axis of the CCD, and for
an optimal focusing of the amplitude hologram, the target star was placed around pixel coordinates
$(750,700)$. The spectrum covers two amplifiers. Field stars are present, and
the sky background is also dispersed (brighter in the middle).

Dome flats were taken with two different filters, blue ($\lambda < \SI{550}{\nano\meter}$) and red ($\lambda > \SI{715}{\nano\meter}$). We saw no difference in the extraction of the spectra using one or the other. In order to be more precise about the extraction of absorption bands in water in the red part of the spectrum, we chose to use only the red flat to flatten out the pixel-to-pixel fluctuations in our exposures. Combined bias was taken at the beginning of each night for the bias subtraction. We made sure that the meta data contained informations about the properties of the CCD and the on-site meteorological station.

We mainly analysed the performances of the holographic element we brought during
this run. Fortunately, we had one very good night on 2017 May 30 with very
stable conditions in terms of temperature and seeing, which we used to
estimate the atmospheric transmission. During that night, we essentially monitored the
CALSPEC\footnote{\url{https://www.stsci.edu/hst/instrumentation/reference-data-for-calibration-and-tools/astronomical-catalogs/calspec}} \citep{Bohlin_2014,Bohlin_2020} star HD111980 with an amplitude hologram. The main characteristics of
these data are
summarised in Table~\ref{tab:reduc134}.

\begin{table}[!h]
\caption{Principal characteristics of the exposure used as an example in
  section~\ref{sec:spectractor}.}\label{tab:reduc134}
\resizebox{\hsize}{!}{\begin{tabular}{cc}
\hline\hline
Properties & Values  \\
\hline
Observatory & CTIO 0.9\,m \\
Disperser & Amplitude hologram $\approx 350\,$lines/mm\\
Dispersion axis angle $\alpha$ & $\approx \SI{-1.6}{\degree}$ \\
$\DCCD$ & $\approx \SI{56}{\milli\meter}$ \\
Pixel size & \SI{24}{\micro\meter} \\
Pixel spectral scale & $\approx\SI{1}{\nano\meter}$ \\
Pixel angular scale & $0\farcs401$ \\
Target & CALSPEC HD111980 \\
Exposure & \SI{120}{\second} \\
Background level & $\approx 25\,$ADU\\
Seeing & $0\farcs65$ \\
Airmass & 1.13 \\
Outside pressure & \SI{784}{\hecto\pascal} \\
Outside temperature & \SI{8.6}{\celsius} \\
\hline
\end{tabular}}
\end{table}

\subsubsection{Simulations}

To test the \Spectractor pipeline, we used the full forward model for CTIO spectrograms (see
section~\ref{sec:ffm}) to simulate observations of CALSPEC stars (in particular,
HD111980). 

The simulation used in Section~\ref{sec:spectractor} shares the same known
characteristics as the real-data image presented in Table~\ref{tab:reduc134}, but with variations in the unknown parameters such as the PSF
model, the amount of second-order diffraction contamination, and the atmospheric
parameters.

For the simulations, a 2D Moffat circular
PSF kernel $\phi(x,y, \lambda)$ was chosen. 
To model the widening of the PSF due to defocusing or chromatic seeing effects, the shape parameters $\gamma$ and $\alpha$
evolved with wavelength as a $n_{\mathrm{PSF}}$ order polynomial function,
\begin{align}
\phi(x, y |\vec r_c, \vec P) & = A\left[1+\left(\frac{x-x_c}{\gamma(z(x))}\right)^2+\left(\frac{y-y_c}{\gamma(z(x))}\right)^2\right]^{-\alpha(z(x))} \\
A & = \frac{\alpha-1}{\pi\gamma^2}\quad\text{with}\quad \alpha > 1, \\
\gamma(z) & = \sum_{i=0}^{n_{\mathrm{PSF}}} \gamma_i L_i(z), \quad \alpha(z)  = \sum_{i=0}^{n_{\mathrm{PSF}}} \alpha_i L_i(z).
\end{align}
The integral of this PSF kernel is exactly $A$,
and its centre is at $\vec r_c = (x_c,y_c)$. The PSF shape parameters
$(\gamma(x), \alpha(x))$ are themselves sets of polynomial
coefficients $\gamma_i$ and $\alpha_i$, respectively. The $L_i(x)$ functions are
the order $i$ Legendre polynomials. We call $x_{\mathrm{min}}$ and $x_{\mathrm{max}}$ the left and right pixel
positions of the spectrogram edges on the $x$-axis. The parameter $z\in[-1,1]$ was rescaled
proportionally on the desired pixel range
$[x_{\mathrm{min}}, x_{\mathrm{max}}]$, set to approximately encompass the wavelength
range $[\SI{350}{\nano\meter}, \SI{1100}{\nano\meter}]$ with the formula
\begin{equation}
z(x) = \frac{x-(x_{\mathrm{max}}+x_{\mathrm{min}})/2}{ (x_{\mathrm{max}}-x_{\mathrm{min}})/2}.
 \end{equation} 
 Parametrisation with Legendre
polynomials has the advantage to give an equal weight to all polynomial
coefficients during $\chi^2$ minimisation, regardless of the degree of the polynomial
functions. The chosen $n_{\mathrm{PSF}}$, $\gamma_i$, and $\alpha_i$ values are quoted for each simulation. The simulation suite is fully available in the \Spectractor code.

\section{Forward-model extraction of a spectrum}\label{sec:spectractor}

In this section, we follow and describe in detail the steps of the \Spectractor
pipeline in order to obtain the first-order spectrum $S_1(\lambda)$ of a point
source, calibrated in flux and wavelength. These steps cover the orange boxes of figure \ref{fig:spectractor_graph}. The process starts from a preprocessed image containing the 2D image that is formed by a star observed through a slitless spectrograph, which we call a spectrogram.

\subsection{Uncertainty evaluation}\label{sec:uncertainties}

When it is not given, we must start to build the
uncertainty map of the exposure when the exposure is pre-processed. The uncertainties on the pixel values are
estimated using the CCD gain $\GCCD(x,y)$ (in electrons/ADU) and its read-out noise
$\sigma_{\mathrm{ro}}$ (in electrons). Other sources of noise, such as those coming from the flat-fielding or the dark current, are subdominant. The correlations between pixels are also negligible. The exposure
unit is considered to be in ADU at that point. The uncertainty $\sigma(x,y)$ on
the pixel value $I(x,y)$ is then
\begin{equation}
\sigma(x,y) = \frac{1}{\GCCD(x,y)}\sqrt{\sigma_{\mathrm{ro}}^2 + \GCCD(x,y)I(x,y)}
\end{equation}
because we assume that the number of photoelectrons in each pixel follows a Poisson
distribution. The uncertainty map can be inverted to obtain a weight map, on which
we can superimpose a mask to remove bad pixels. Assuming no correlations between
pixels, we then assemble the weight matrix $\mathbf{W}$ as the diagonal matrix of
the inverted pixel variances. 

The computed noise variance uses the pixel value itself, which incorporates the noise fluctuation. Using weights incorporating the fluctuations can introduce a bias on the recovered quantities because in the case of positive fluctuation, the weight is increased, while in the case of negative fluctuation, the weight is decreased. The bias is smaller with spectrograms with a high signal-to-noise ratio (S/N). For this reason, we only considered the case with a high S/N in simulations and data, and we avoid the bias due to this uncertainty evaluation (even at the spectrum edges, where flux is low). The case with a low S/N is postponed for future developments.

\subsection{Zeroth-order centroid}\label{sec:order0}

\begin{figure}[!h]
\begin{center}
\includegraphics[width=\columnwidth]{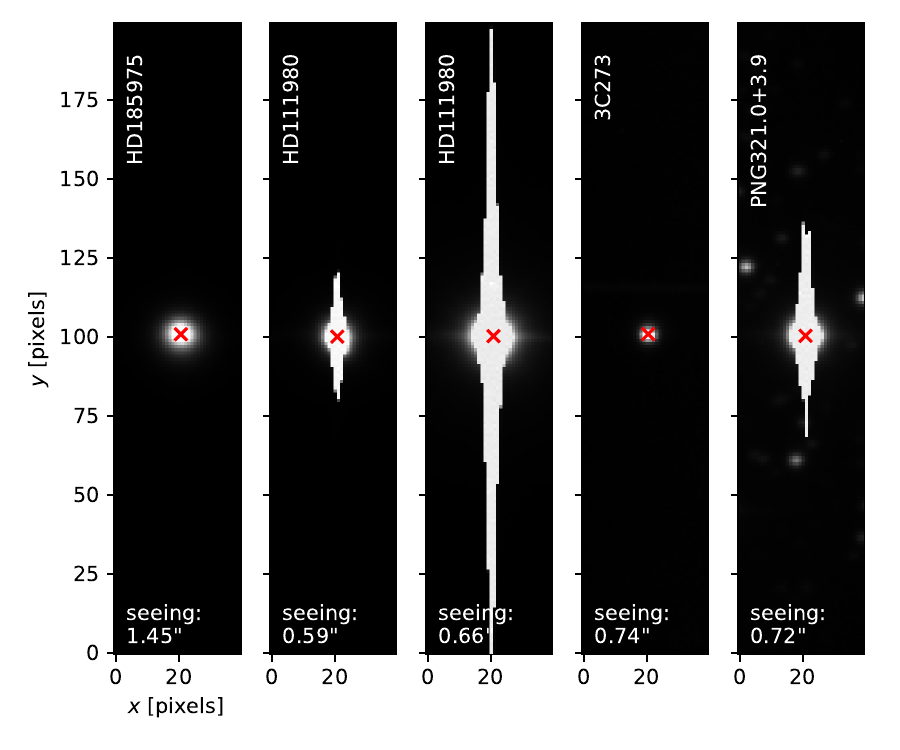}
\end{center}
\caption[] 
{Zeroth-order images from different celestial objects, observed in different seeing conditions with several dispersers. The red cross shows the fitted centroid using the \Spectractor method from section~\ref{sec:order0}. All images have saturated pixels.}
\label{fig:order0_hall_of_shame}
\end{figure}

Because the zeroth order is included in the observed images, we
used its centroid to set the zero of the wavelength scale. Therefore, an error in the
determination of the zeroth-order centroid $(x_0, y_0)$ causes a systematic
shift of the wavelength calibration.

A subset of different situations encountered in the CTIO data is presented in
Figure~\ref{fig:order0_hall_of_shame}. To obtain a high S/N in the spectrogram, the exposure time was set at such a value
that the zeroth order is saturated, causing bleeding spikes. If the exposure has astrometric coordinates in the World
Coordinate System (WCS), then we can obtain the precise position of the star on
the CCD.  However, in most images we obtain from CTIO, the WCS associated with the
images is incorrect because the mount calibration is not correct.

\begin{figure*}[!h]
\begin{center}
\includegraphics[width=\textwidth]{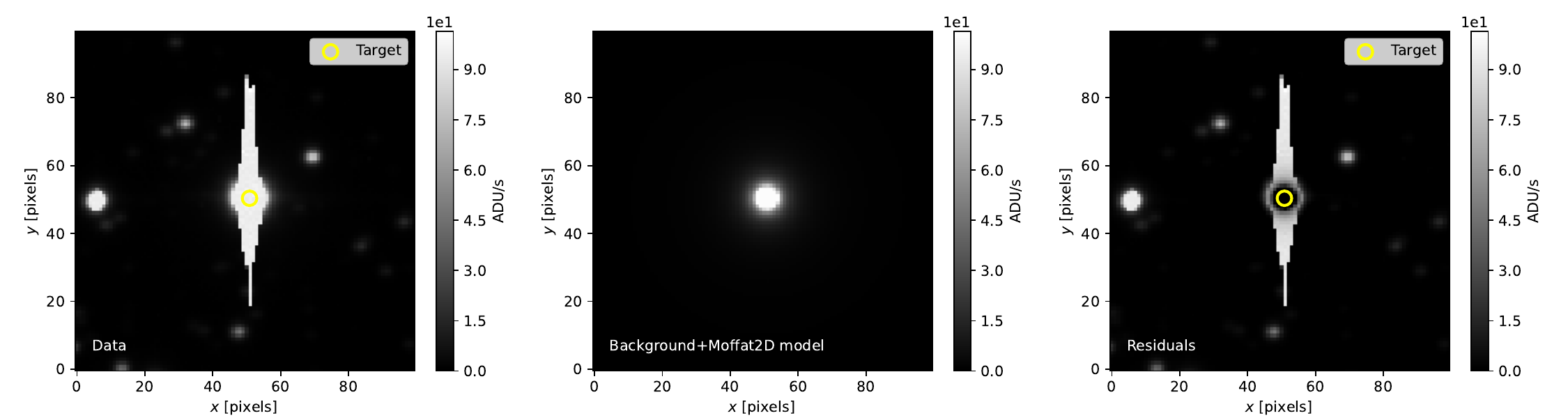}
\end{center}
\caption[] 
{Illustration of the fitting process to find the zeroth-order centroid for the planetary nebula PNG321.0+3.9. Left: zeroth-order data image of the target. Middle: Best-fitting 2D circular Moffat model. Right: Residuals.}
\label{fig:order0_centroid_fit}
\end{figure*}

While not directly part of the extraction
  procedure, we present the algorithm we implemented to find the zeroth-order
centroid in these difficult cases as a useful example of how to improve the
starting point from a preliminary analysis of the data.

First, the image is cropped around the supposed position of the zeroth-order
image, as close as needed so that the targeted star is the brightest
object. Then the cropped image is projected onto the $x$- and $y$-axis: The
maximum of the two projections sets a new approximation of the zeroth-order position. From there, saturated pixels are detected, and a null weight is
associated with them. A 2D second-order polynomial background with a $3\sigma$
outlier removal is fitted and subtracted from the cropped image. Finally, a 2D
circular Moffat profile is fitted on the weighted pixels: Only the crown of
non-saturated pixels counts and locks the fit. This last step is then repeated a
second time on a new cropped image for which the width and height are divided by two, centred
on the last fitted centroid. This step is illustrated in
Figure~\ref{fig:order0_centroid_fit}. We tested this process on many images,
most of which were pathological, and we visually confirmed that the accuracy of this
algorithm is finer than the pixel size on CTIO images.

\begin{figure}[!h]
\begin{center}
\includegraphics[width=\columnwidth]{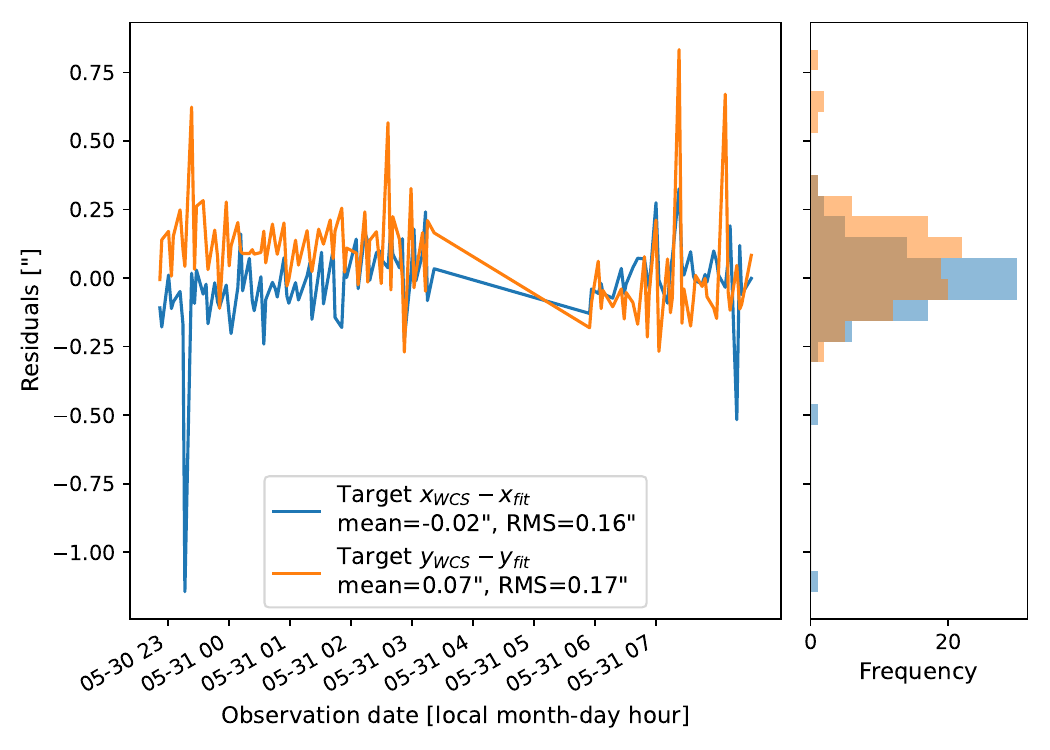}
\end{center}
\caption[] 
{Difference between the fitted centroids of the target stars and the centroid recovered using the WCS estimate locked on the Gaia catalogue (blue and orange lines) on the $x$- and $y$-axis at CTIO during the night of 2017 May 30, as a function of the exposure date. The histogram of the differences is presented on the right.}
\label{fig:astrometry_residuals}
\end{figure}

Another issue we solved is that when a disperser is added to the
telescope beam path, the WCS associated with the image can be shifted or
distorted. In the case of a crowded field, for faint objects, or for a very pathological
zeroth order, we developed a method for estimating the WCS using the field stars, the
library
\texttt{astrometry.net}\footnote{\url{http://astrometry.net/doc/readme.html}}
\citep{astrometrynet}, and the Gaia DR2 catalogue. The process is described in the
appendix~\ref{sec:astrometry}, but it was not used by default for the CALSPEC
stars we observed. However, when the centroids obtained with both
methods are compared, their difference has a scatter of 0\farcs15 (around one-half of a pixel) on CTIO
data, which confirms that both methods are accurate at the 0\farcs15
level at least (Figure~\ref{fig:astrometry_residuals}).

This accuracy is converted into a prior on the zeroth-order shift $\delta u_0$ used
during the wavelength-calibration process to account for a mistake in the evaluation of the zeroth-order centroid.

\subsection{Rotation}

Unless special care has been taken to that end when mounting the disperser in
the filter wheel, the spectrogram image can be tilted (possibly intentionally) with respect to the $x$-axis, with an angle, which can be poorly known depending on the mounting of the
disperser into the telescope beam.

This dispersion direction can either be known {a priori} or fitted in the full
forward-model step. We needed a supplementary step to estimate a
good starting point for this angle because our case is the latter case. 

In addition, we found it extremely useful that the
spectrogram was roughly aligned with the $x$-axis of the exposure in our procedure, with the
wavelength increasing with $x$. To this end, the exposure must be flipped and rotated
accordingly before the process of extracting more information could be continue.  That was useful
both for diagnosing the data quality and for refining the starting point of the full
forward model.

The spectrogram of sources that are sufficiently continuous in wavelength, such as the
thermal emission component of stars, displays filament shapes on the
2D image that can be detected using a Hessian analysis inspired by the one
developed in \citet{PlanckFilament}. The advantage of this technique is that it
comes with an analytical expression of the angle of the detected shape with
respect to the horizontal or vertical axis of the CCD grid.

\begin{figure}[!h]
\begin{center}
\includegraphics[width=\columnwidth]{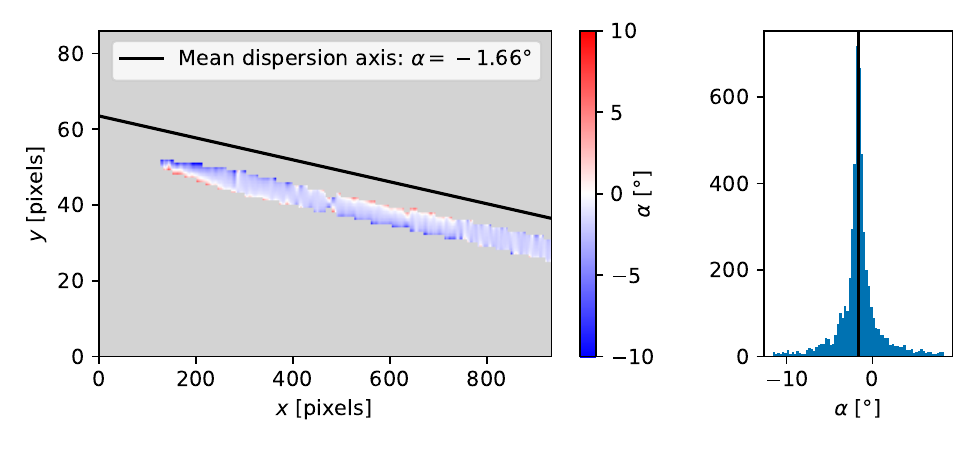}
\end{center}
\caption[] 
{Disperser rotation angle estimation. Left: Angle $\alpha$ on the spectrogram image for the 5\% pixels with the highest $\lambda_-$ values found in the Hessian matrix, and the dispersion axis as the median of the angles (black line, shifted upward for clarity). The masked pixel with low $\lambda_-$ values are indicated (light grey). Right: Histogram of the selected angles from the left panel and its median (vertical black line).}
\label{fig:rotation_hessian}
\end{figure}

The Hessian matrix $H(x,y)$ of the image is computed for each pixel value $I(x,y)$ as
\begin{equation}
\displaystyle    {H(x,y) = \begin{pmatrix}
    H_{xx} & H_{xy} \\
    H_{xy} & H_{yy}
    \end{pmatrix} = \begin{pmatrix}
    \cfrac{\partial^2 I}{\partial x^2} & \cfrac{\partial^2 I}{\partial x \partial y} \\
    \cfrac{\partial^2 I}{\partial x \partial y} &  \cfrac{\partial^2 I}{\partial y^2} 
    \end{pmatrix} }.
\end{equation}
The two eigenvalues of the Hessian matrix $H$ are calculated as
\begin{equation}
    \lambda_{\pm}(x,y) = \frac{1}{2} \left(H_{xx} + H_{yy} \pm h  \right),
\end{equation}
with $h = \sqrt{(H_{xx}-H_{yy})^2 + 4H_{xy}^2}$. The eigenvalue $\lambda_-$ is
associated with the eigenvector directed along the spectrum dispersion axis, and
$\lambda_+$ corresponds to the eigenvector with the largest change in intensity
value, that is, transverse to the dispersion axis. The orientation angle of these
eigenvectors with respect to the $x$-axis can be computed analytically. For
instance, we have for $\lambda_-$ 
\begin{align}
    \alpha(x,y) & = \arctan \left( \frac{H_{yy} - H_{xx} - h}{2 H_{xy}}  \right) \notag \\ & = \frac{1}{2} \arctan \left( \frac{2 H_{xy}}{H_{xx} - H_{yy}}  \right), 
\end{align}
with the trigonometric formula $\tan 2\alpha = 2 \tan\alpha / (1- \tan^2
\alpha)$. After selecting the 5\% pixels with the highest $\lambda_-$ value
above a reasonable threshold, the median $\alpha$ of the remaining $\alpha(x,y)$
values gives the orientation of the spectrum with respect to the $x$-axis. A
linear fit can also be performed across the selected pixels, and the slope gives
an angle very close to the one estimated with the median of the angle
values. This process is illustrated Figure~\ref{fig:rotation_hessian}.

Because of the atmospheric differential refraction (ADR), the
spectrogram can be sheared transversally to the dispersion axis. The ADR shear
is about 2$\arcsec$ across the visible spectrum, which is 5 pixels at
CTIO, while the spectrogram is $\approx 700$ pixels long, and it is thus neglected
at this step of this analysis. On the other hand, it is fully accounted for
in the full forward model.

\subsection{Background estimation}

\begin{figure}[!h]
\begin{center}
\includegraphics[width=\columnwidth]{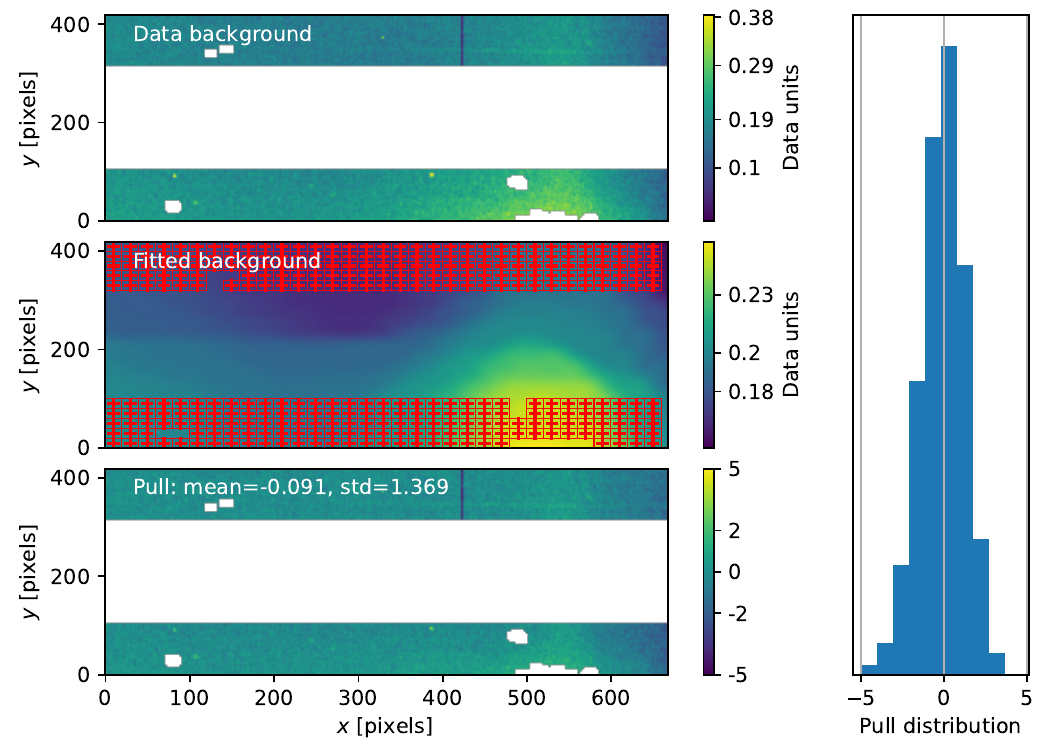}
\end{center}
\caption[] 
{Background estimation on a CTIO image (in arbitrary units). {Top:} Laterals bands to the spectrogram (masked behind the rectangular grey region) where the background is estimated. The grey patches indicate masked sources. {Middle:} Fitted background using the \texttt{SExtractor} method with the evaluation boxes in red (here final size is $20\times 20$ pixels). {Bottom:} Residuals normalised to their uncertainties. {Right:} distribution of the normalised residuals in units of $\sigma$.}
\label{fig:background_extraction}
\end{figure}

The background of the image must be carefully subtracted to avoid bias in the
estimation of the spectrum flux and its PSF. However, due to optical vignetting,
it can be non-flat, and it is dispersed. It can also contain additional field stars
and their corresponding spectrograms.

To estimate the spectrogram background, we first selected two lateral bands above and
below the spectrogram region that had the same length and width
$N_y^{(\mathrm{bgd})}$ (see Figure~\ref{fig:background_extraction}). First, we masked the sources detected above a $3\sigma$
threshold. Then we divided the two lateral regions in a few square boxes of size
$(N_y^{(\mathrm{bgd})}/2, N_y^{(\mathrm{bgd})}/2)$. The box dimensions must be
larger than the typical size of a field star PSF or of a spectrogram width, but
small enough to account for the spatial variations in the background.

To obtain a first estimate of the background, we used a Python wrapper of the
\texttt{SExtractor}\footnote{\url{http://www.astromatic.net/software/sextractor}
} algorithm for background extraction, which is roughly based on the bilinear
interpolation of the median value inside the boxes, after a sigma-clipping
rejection of the outlier pixels (more details in
\citealt{1996A&AS..117..393B}). This process is illustrated in
Figure~\ref{fig:background_extraction}.

In a second step, we analysed the distribution of the background residuals
normalised with their uncertainties in the two regions, with the sources being
masked. When the histogram of the residuals normalised by their errors (the pull distribution) departs from a distribution of zero mean and
standard deviation equal to 1, we refined the background estimate by dividing the box size by 2, and the process was
continued iteratively until the mean was below 1 and the standard deviation was
below 2, or until the box size decreases below a threshold of 5 pixels.

At the end of this process, the estimated background was interpolated between the
two lateral bands to obtain the background below the spectrogram, and it was finally
subtracted. We call $B(\vec r)$ the background map. The background RMS was
also evaluated by \texttt{SExtractor} and was added quadratically as a background uncertainty to
the error budget of the spectrogram.

At this stage of the pipeline, with $\alpha$ and $\vec r_0$ given and using a
first geometrical wavelength calibration to approximately define the left and right
margins of the spectrogram, we cropped the exposure to extract a background-subtracted spectrogram. This is extremely useful for diagnosis purposes and can also be directly studied with an atmospheric forward model (see Section~\ref{sec:atm}).

\subsection{First spectrum extraction}\label{sec:spectrum_deconvolution}

The next step of the pipeline we implemented is devoted to
extracting the first-order spectrum $S_1(\lambda)$ and estimating the
wavelength-dependent PSF. The extraction of the spectral information $S_1(\lambda)$ from the spectrogram
image is a delicate process. The intuitive and traditional way to extract slitless spectra at this stage of the pipeline is to sum over the cross-dispersion direction, possibly with weights, to form what we call a cross-dispersion spectrum. \cite{Horne1986} and \cite{Robertson1986} presented an optimal method to achieve this. However, this method leads to distorted spectra in case of a wavelength-dependent PSF (neighbouring wavelengths contaminate each other on the sensor), which is not an issue for spectroscopy, but is problematic for spectrophotometry. In the following, we first describe a method for deconvolving the spectrum for the PSF (Section~\ref{sec:spectrum_deconvolution}). The products of that process are then very useful to inform the full forward model, finalising the unbiased extraction of the spectrum (Section~\ref{sec:ffm}).

\subsubsection{First-order model} of the spectrogram

To start, we only consider the spectrogram of the first
diffraction order, potentially with the superposition of a second diffraction order, as
provided by the previous step of the pipeline. The cropped spectrogram has the shape
$(N_x, N_y)$ pixels.

Inspired by equation~\eqref{eq:spectrogram_model}, we model the spectrogram as a
discrete stack of $N_x$ 2D PSF realisations of amplitude $A_i$, separated by one
pixel along the $x$-axis,
\begin{equation}\label{eq:spectrogram_model_I1}
\vec I_1(\vec{r} | \vec A, \vec r_c, \vec{P}) = \sum_{i=0}^{N_x} A_i\, \phi(\vec{r} | \vec r_{c,i}, \vec{P}_i),
\end{equation}
with $\vec{r}=(x,y)$ the vector of the pixel coordinates, $\vec{A}$ the amplitude
parameter vector along the dispersion axis of the spectrogram, and $\phi(\vec{r}
| \vec r_{c,i}, \vec{P}_i)$ the 2D PSF kernel whose integral is
normalised to one. This kernel depends non-linearly on the shape parameter vector
$\vec{P}_i$ and on the centroid position vector $\vec r_{c,i}=(x_{c,i},
y_{c,i})$, where only the $y_{c,i}$ coordinate is considered unknown. The
$x_{c,i}$ coordinate is set directly to the pixel index $i$. This choice of
implementation can be discussed and changed, but it was found to be practical
because the PSF is then well sampled by the pixel grid. In theory, we could
choose another sampling for $x_{c,i}$, however, to increase the speed of the
spectrum extraction or to enhance the spectral resolution.

In some way, the array of vectors $\vec r_{c}$ is a sampled precursor of the
dispersion relation $\vec \Delta_1(\lambda)$, and the vector $A$ is the flux
density $S_1(\lambda)$ integrated within the pixels.  When we index all the $N_{x} N_{y}$
spectrogram pixels as a long vector
$\vec{Z} = \left(\zeta_{1}, \cdots, \zeta_{N_{x} N_{y}}\right)$, the
equation~\ref{eq:spectrogram_model_I1} takes a matricial form, 
\begin{equation}
  \vec I_1(\vec{Z} | \vec A,\vec r_{c}, \vec{P}) =
  \mathbf{M}(\vec{Z} | \vec r_{c}, \vec{P} )\, \vec{A}
  \label{eq:1},
\end{equation}
with
\begin{multline}
  \mathbf{M}(\vec{Z} | \vec r_{c}, \vec{P} )  = \\
  \left(\begin{array}{ccc}
 \phi(\zeta_1 | \vec{r}_{c,1}, \vec{P}_1) &  \cdots
 & \phi(\zeta_1 | \vec{r}_{c,N_x}, \vec{P}_{N_x}) \\
 \vdots &  \ddots & \vdots\\
 \phi(\zeta_{N_x N_y} | \vec{r}_{c,1}, \vec{P}_1) &  \cdots
 & \phi(\zeta_{N_x N_y} | \vec{r}_{c,N_x}, \vec{P}_{N_x})  \\
\end{array}\right).
\end{multline}
The  $(N_x N_y, N_x)$ matrix $\mathbf{M}$ is called the design matrix.

This model of the spectrogram is designed to deconvolve the spectrum
$S_1(\lambda)$ from the PSF. In principle, we can choose to sample it with PSF
kernels separated by arbitrary distances. However, if the PSF is correctly
sampled by the pixel grid, it is difficult to extract the spectrum with a
spatial resolution below the typical PSF width, and therefore, below the pixel
size. 

At this stage, we wish to extract the spectrum from a spectrogram that is potentially contaminated by higher diffraction orders, as yielded
by the pipeline steps discussed above, or that is distorted by atmospheric differential refraction. 
The final extraction using a full forward model that takes these physical effects entirely into account is the last step of the
\Spectractor pipeline. This is presented in section~\ref{sec:ffm}.

\subsubsection{Preparation for deconvolution}\label{sec:fit_1D}

To initialise the deconvolution with parameters close to the final best model, a
first PSF fit is performed transversally to the dispersion axis with 1D PSF
kernels on the rotated spectrogram image,
\begin{equation}\label{eq:spectrogram_model_I1D}
\vec I_1^{(1D)}(\vec{r} | \vec A^{(1D)}, \vec r_c, \vec{P}) = \sum_{i=0}^{N_x} A_i^{(1D)}\, \phi^{(1D)}(\vec{r} | \vec r_{c,i}, \vec{P}_i).
\end{equation}

This procedure is done in two steps. The first step fits the 1D parameters
independently for each pixel column by applying a $5\sigma$ clipping to reject
field stars and other CCD defects. In the second step, a polynomial evolution of
the 1D PSF parameter vector along the dispersion axis is proposed. For
instance, a polynomial evolution of the $y_{c,i}(x_{c,i})$ positions can model
the transverse ADR, and a polynomial evolution with the width of the PSF can model a
defocusing effect. The polynomial coefficients are fitted using a
Gauss-Newton minimisation of a $\chi^2$
(equation~\ref{eq:chromaticspsf1d_chi2}) for the non-linear parameters (see
appendix~\ref{sec:newton}), alternating with a linear resolution for the
amplitude parameters as follows.

Assuming that we gather all the spectrogram pixel values in a long vector $\vec D$ (as $\vec{Z}$), we can model it as
\begin{equation}
\vec{D} = \mathbf{M}(\vec{Z} | \vec r_{c}, \vec{P} )  \vec A + \vec{\epsilon},
\end{equation}
with $\vec{\epsilon}$ the random noise vector. The $\chi^2$ function to minimise is
\begin{equation}\label{eq:chromaticspsf1d_chi2}
\chi^2(\vec{A} | \vec P)= \left(\vec{D} - \mathbf{M} \, \vec{A}\right)^T \mathbf{W}
    \left(\vec{D} -  \mathbf{M} \, \vec{A} \right),
\end{equation}
with $\mathbf{W}$ the weight matrix of dimension $(N_xN_y, N_xN_y)$, which is the inverse of the data covariance matrix (see
section~\ref{sec:uncertainties}). In most cases, this matrix is diagonal because the
pixels are all considered to be independent. The minimum of equation
\ref{eq:chromaticspsf1d_chi2} is reached for the set of amplitude parameters
$\hat{\vec{A}}$ given by
\begin{equation}
\hat{\vec{A}}^{(1D)} =  (\mathbf{M}^T \mathbf{W} \mathbf{M})^{-1} \mathbf{M}^T \mathbf{W} \vec{D}.
\end{equation}
The covariance matrix associated with the $\hat{\vec{A}}$ coefficients is
$\mathbf{C}^{(1D)} = (\mathbf{M}^T \mathbf{W} \mathbf{M})^{-1}$. At the end of this process, we obtain
a first guess of the $\vec r_c$ and $\vec P$ parameters and a first guess
of the amplitudes $\hat{\vec{A}}^{(1D)}$ with their uncertainties $\vec
\sigma_{A^{(1D)}}$, which form what we call a transverse cross-spectrum.

\begin{figure*}[!ht]
  \textbf{\hspace{1cm}After 1D cross-dispersion extraction\hspace{0.2\textwidth}After 2D PSF deconvolution}
  \begin{center}
    \includegraphics[width=0.47\textwidth]{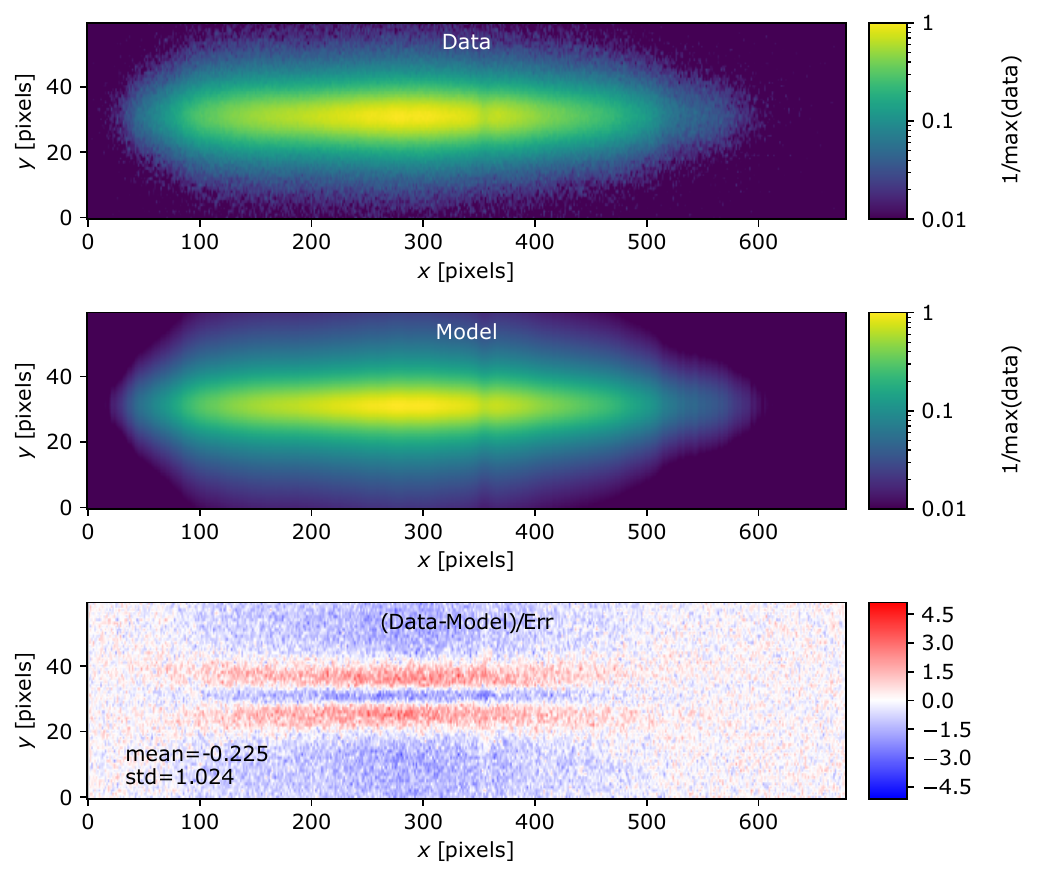}\hfill
    \includegraphics[width=0.47\textwidth]{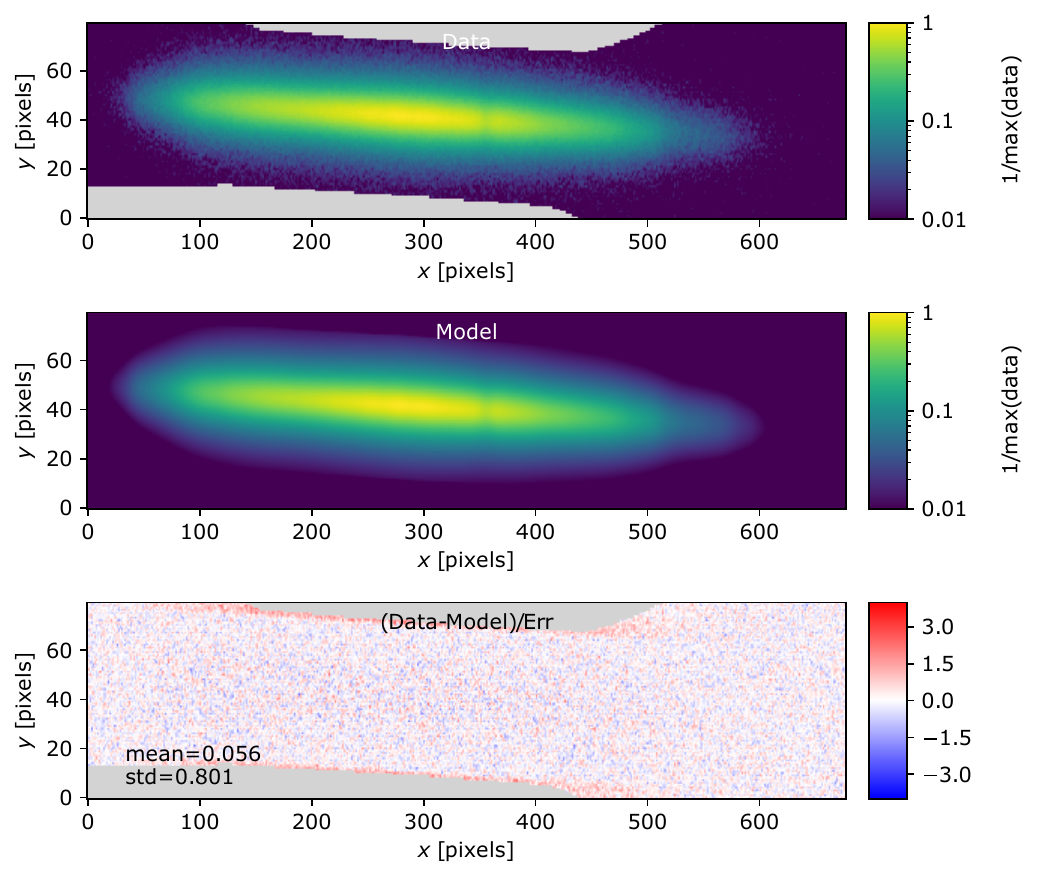}
  \end{center}
  \caption{Results from the deconvolution of a simulated spectrogram of the CALSPEC star
    HD111980 with a Moffat PSF kernel, without second-order diffraction contamination, $(\gamma_0, \gamma_1, \gamma_2)=(10,2,5)$ and
    $(\alpha_0,\alpha_1, \alpha_2)=(2,0,0)$. {Left:} Simulated
    spectrogram (top), best-,fitting spectrogram model (middle) and residuals in
    units of $\sigma$ (bottom) after the rotation process and the 1D transverse
    fit.
    {Right:} Same simulated spectrogram with its original rotation
    (top), best-fitting spectrogram model (middle), and residuals in units of
    $\sigma$ (bottom) after the deconvolution process. All colour maps are
    normalised by the maximum of the simulated spectrogram. The grey areas designate masked pixels.}
\label{fig:deconvolution_residuals}
\end{figure*}

The result of this extraction is illustrated on simulated data in
Figure~\ref{fig:deconvolution_residuals} left. For this simulation, we chose to ignore the second-order diffraction, and we used a wide PSF kernel with strong chromatic
variations to enhance the visibility of the residuals for the 1D extraction:
$(\gamma_0, \gamma_1, \gamma_2)=(10,2,5)$ and $(\alpha_0,\alpha_1,
\alpha_2)=(2,0,0)$.  At first glance, the model in the top panel appears to be accurate,
but the residuals show that this 1D model failed to capture the
true evolution of the PSF shape with wavelength. As expected, the residuals are
less dramatic with a thinner PSF, but remain measurable. This shows that a 2D PSF extraction method has to be used.

At this stage, we obtained a spectrum close to a cross-dispersion spectrum as in \cite{Horne1986} and \cite{Robertson1986}, but informed with a fitted PSF model with a smooth polynomial wavelength evolution.

\subsubsection{PSF deconvolution}\label{sec:spectrum_deconvolution_psf2d}

Having provided ourselves with this first 1D estimate of the spectrum, we can
resort to a more accurate 2D PSF modelling.  When using 2D PSF kernels, the
latter linear regression method enters the category of the deconvolution
problems or inverse problems. Because the spectral amplitude information is mixed
and diluted at a scale below the typical size of the PSF, the computation of the
$\hat{\vec{A}}$ vector sampled at the pixel scale inverting
equation~\ref{eq:1} using 2D PSF kernels
may lead to results that are far from the reality, while giving an apparently good
fit to data (low $\chi^2$). When this is done, a common symptom is the alternation of
positive and negative values in $\hat{\vec{A}}$, or at least with large
variations, demonstrating that the problem is ill-posed. As we know by the
physics that well-sampled spectra are rather continuous and differentiable
functions, we enforce a regularisation method to smooth the resulting
$\hat{\vec{A}}$ vector. 


A first fit to the data, without any prior on $\vec A$, using a 1D transverse
PSF model fitted independently to each column of data along the dispersion axis
thus yields a vector $\hat{\vec{A}}^{(1D)}$ that contains most of the spectral
flux, especially in the smooth parts of the spectrum, but lacks precision in
the rapidly varying parts. 
The most visible effect of the PSF is to smooth the absorption
lines, and more generally, to deform the spectral information where the spectral
energy density evolves rapidly (e.g. in the blue part of the
spectrum), while conserving the total flux. It nonetheless provides useful information that we
use as a prior $\vec A_0$ on $\vec A$ when performing a fit using a 2D PSF
kernel with a Tikhonov regularisation.

The Tikhonov regularisation method proposes to add a regularisation quantity to the $\chi^2$ and minimise a new cost function,
\begin{align}
\mathcal{E}(\vec{A} | \vec r_c, \vec P)& = \left(\vec{D} - \mathbf{M}\, \vec{A}\right)^T \mathbf{W}
    \left(\vec{D} - \mathbf{M}\,\vec{A} \right) \notag \\ &\ \ \  +  r (\vec A - \vec A_{0} )^T \mathbf{Q} (\vec A - {\vec A}_{0}) \notag \\ 
    & = \chi^2(\vec{A} | \vec r_c, \vec P) + r \chi^2_{\rm pen}(\vec{A} | \vec A_{0}),\label{eq:chi2_deconv} \\
    \vec A_0 = \hat{\vec{A}}^{(1D)}
\end{align}
where $\mathbf{Q}$ is the weight matrix.  The last term favours $\hat{\vec A}$ to
be close to prior vector ${\vec A}_{0}$, with a positive regularisation
hyper-parameter $r$.

Minimising this $\mathcal{E}(\vec{A} | \vec r_c, \vec P)$ function is still a linear regression for the $\vec A$ parameters, whose optimal value now is
\begin{equation}
  \hat{\vec{A}} =  (\mathbf{M}^T \mathbf{W} \mathbf{M} +  r\mathbf{Q})^{-1} (\mathbf{M}^T \mathbf{W} \vec{D}  +r\mathbf{Q}{\vec A}_{0} ).
  \end{equation}
The covariance matrix associated with $\hat{\vec  A}$ directly is
\begin{equation}\label{eq:covariance_matrix}
\mathbf{C} = (\mathbf{M}^T \mathbf{W} \mathbf{M} + r\mathbf{Q})^{-1}.
\end{equation}

We tested different $\mathbf{Q}$ matrices on CTIO simulations, and the matrix that
gives the most satisfying results when $\hat{\vec A}$ is compared to the true
amplitudes uses the Laplacian operator $\mathbf{L}$,
\begin{align}
\mathbf{L} & = \begin{pmatrix}
-1 & 1 & 0 & 0 & \cdots & 0 & 0\\
1 & -2 & 1 & 0 &\cdots & 0 & 0\\
0 & 1 & -2 & 1 &\cdots & 0 & 0\\
\vdots & \ddots & \ddots & \ddots & \vdots& \vdots \\
0 & 0 & 0 &0 & \cdots & -2 & 1\\ 
0 & 0 & 0 & 0 &\cdots & 1 & -1\\ 
\end{pmatrix}\\
  \mathbf{Q} & =  \mathbf{L}^T \mathbf{U}^T \mathbf{U} \mathbf{L},
\end{align}
with $ \mathbf{U}^T \mathbf{U}$ the matrix proposed in
equation~\ref{eq:Qsimple}, 
\begin{equation}\label{eq:Qsimple}
\mathbf{U}^T \mathbf{U} = \begin{pmatrix}
1/\sigma^2_{A_{1D}^{(1)}} & 0 & \cdots & 0\\
0 & 1/\sigma^2_{A_{1D}^{(2)}} & \cdots & 0\\
\vdots & \vdots & \ddots & \vdots \\
0 & 0 & \cdots & 1/\sigma^2_{A_{1D}^{(N_x)}}\\ 
\end{pmatrix}.
\end{equation}


The total variation regularisation is known to be able to retrieve very sharp
features (e.g. steps or edges) when deconvolving an image. The Laplacian
regularisation cannot do the same, but discontinuities are not expected in the
physical spectra we observe. Moreover, the norm-2 regularisation offers
analytical solution, while a norm-1 regularisation needs to conduct an iterative
minimisation process (more details in appendix~\ref{sec:laplacian_reg}).


For the 1D transverse estimate, the deconvolution fit is performed using a
Gauss-Newton minimisation of $\mathcal{E}(\vec{A} | \vec P)$ for the
non-linear $\vec P$ parameters (see appendix~\ref{sec:newton}) and a
linear resolution for the $\vec A$ parameters (see
section~\ref{sec:fit_1D}). The Gauss-Newton minimisation is repeated three times
with a clipping rejection of bad pixels to remove field stars or CCD
defects that can pull the final parameters in undesired directions. At the
end of this process, we obtain the measured $\hat{\vec r}_c$ and $\hat{\vec
  P}$ parameters, and a measurement of the amplitudes $\hat{\vec A}(r)$
with their covariance matrix $\mathbf{C}(r)$. The vector $\hat{\vec A}(r)$ forms the searched
spectrum, but it might be contaminated by the second diffraction order at this stage. To accelerate the
computation, regions farther away than twice the PSF FWHM from the centroids are
masked and set to zero with null weights (the grey regions in
Figure~\ref{fig:deconvolution_residuals}). For this process, a default value of
the $r$ regularisation parameter was chosen, without fully exploring how it
could be optimised.

The PSF deconvolution problem was solved in another way in \cite{Ryan2018} for slitless spectroscopy. Instead of using a PSF model, the mathematical model was inverted using multiple exposures of the same spectrum, ideally taken at different orientations, dispersion directions, or dithered positions. In this way, the number of data leads to an invertible problem, but a damping hyper-parameter equivalent to $r$ still needs to be introduced in equation~\ref{eq:chi2_deconv}. \cite{Ryan2018} tuned this hyper-parameter manually and determined the breaking point of a specific L-curve (see \citealt{Hansen1992}). This method is mathematically similar to the one we used (described in Section~\ref{sec:regularisation}) because it leads to an equilibrium between information from the prior and data.

\subsubsection{Optimisation of the regularisation}\label{sec:regularisation}

The correct amount of information for which the data can be searched might be thought hard to determine. It may seem hopeless to find information at a spatial scale below the typical width of the convolution kernel with a unique spectrogram. Therefore, we searched for the way to determine the optimal hyper-parameter $r$.

The regularisation parameter $r$ is optimised via the study of the resolution operator
\begin{equation}
\mathbf{R} =  \mathbf{I} - r \mathbf{C} \mathbf{Q},
\end{equation}
with $\mathbf{I}$ the identity matrix. A fruitful interpretation of the
  $\mathbf{R}$ operator is given in \cite{hansen2010discrete} with
\begin{align}
\mathrm{Tr}\,\mathbf{I} & = \mathrm{Tr}\,\mathbf{R} + \mathrm{Tr}\left(r \mathbf{C} \mathbf{Q}\right) \Leftrightarrow \notag \\
\left[\text{\# parameters}\right] & = \left[\text{\# parameters resolved by data}  \right]\notag \\ &\ \ \  +  \left[\text{\# parameters resolved by prior}  \right],
\end{align}
where the trace of the resolution operator gives the effective number of
degrees of freedom that can be extracted from the data for a given amount of
prior information. We set
\begin{equation}
N_{\mathrm{dof}} = \mathrm{Tr}\,\mathbf{R}.
\end{equation}
The optimal $r$ parameter is found by minimising the following $G(r)$ function, which behaves like a reduced $\chi^2$,
\begin{equation}
G(r) = \frac{\chi^2(\hat{\vec{A}} | \hat{\vec r}_c,\hat{\vec P})}{(N_x N_y - N_{\mathrm{dof}})^2}.
 \end{equation}
This method, known as general cross-validation (GCV), is extensively presented
for example in \cite{Golub1979,wahba1990spline,hansen2010discrete}. It is demonstrated
that the minimum of $G(r)$ corresponds to the minimum of the distance $|
\mathbf{M} \hat{\vec{A}} - \mathbf{M} \vec{A}_{\text{truth}}|^2$, where
$\vec{A}_{\text{truth}}$ is the true amplitude vector hidden in the data.


We tested different ways to implement the optimisation of the regularisation
process in \Spectractor. The most efficient result (in terms
of rapidity and bias of the final result) is obtained by setting a default reasonable
regularisation parameter for the fitting procedure of the amplitude $\hat{\vec
  A}$ and PSF $\hat{\vec r}_c,\ \hat{\vec P}$ parameters, and finally to
find the minimum of the $G(r)$ function. We observed in the simulations that
regardless of the $r$ hyper-parameter that is chosen at the beginning, the process
reconstructs an unbiased spectrum at the end of the $\mathcal{E}(\vec{A} |
\vec r_c,\vec P)$ minimisation. The level of regularisation of the solution
can thus be set {a posteriori} by finding the optimal $r$ after minimising the $G(r)$
function.

The result of a 2D deconvolution process is presented in the right panel of
Figure ~\ref{fig:deconvolution_residuals}  for a simulation with a
wide PSF kernel, but without second diffraction order to stress the benefit of
the deconvolution. The residual map between the best-fitting model and the
simulated spectrogram shows that the 1D transverse fit cannot extract
the spectrum from the spectrogram image correctly (Figure~\ref{fig:deconvolution_residuals} left), while the 2D deconvolution ends with a
quasi-unstructured residual map that respects the expected Gaussian distribution (Figure~\ref{fig:deconvolution_residuals} right). Because our model is informed by a PSF model, we are able to extract spectra with a single exposure, which involves rather small matrices compared with \cite{Ryan2018}. This allows us to tune the $r$ hyper-parameter automatically in a few seconds with a standard laptop.

If the spectrogram is not fully contained in the sensor area, the spectrum exhibits a discontinuity that causes the norm-2 regularisation to fail (the second derivative from the Laplacian operator is undefined). For a given instrumental PSF, it is therefore better to use a more dispersive grating to feed the deconvolution with more data and increase the wavelength resolution, but the spectrogram must fit inside the sensor area for regularisation techniques to be use.

\subsubsection{The spectrophotometric uncertainty principle}

\begin{figure}[!ht]
\begin{center}
\includegraphics[width=\columnwidth]{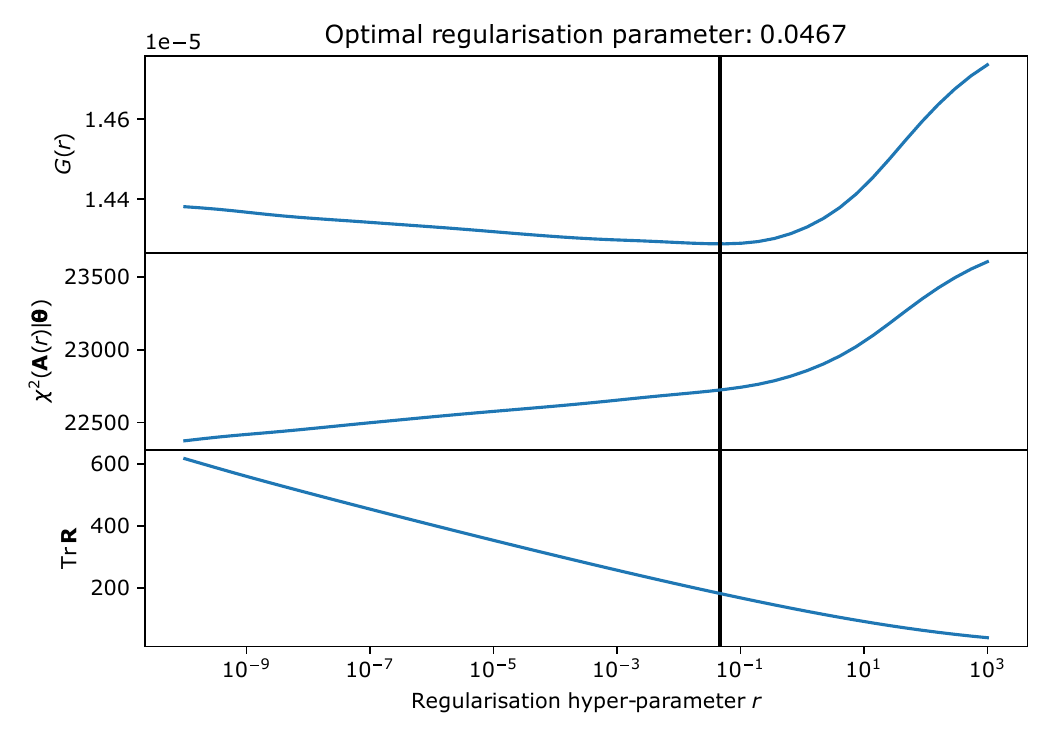}
\end{center}
\caption[] {Optimisation of the $r$ hyper-parameter for a simple simulation
  with $n_{\mathrm{PSF}}=0, \gamma_0=5, \alpha_0=3$ (and the same characteristics
  as in Table~\ref{tab:reduc134}). {Top:} $G(r)$ function (blue) and
  the minimum position $r = 0.0467$ (vertical black line). {Middle:}
  $\chi^2$ evolution with $r$. {Bottom:} Trace of the resolution matrix
  $\text{Tr}\,\mathbf{R}$. The intersection with the black line gives the
  effective number of parameters fitted by data, here, 180.}
\label{fig:deconvolution_regularisation}
\end{figure}

The regularity of the deconvolved solution depends on the hyper-parameter
$r$. The optimal $r$ parameter is chosen as the minimum of the $G(r)$ function
represented in the top panel of Figure~\ref{fig:deconvolution_regularisation}
for a simple simulation with $n_{\mathrm{PSF}}=0, \gamma_0=5, \alpha_0=3$ (and
the same characteristics as in Table~\ref{tab:reduc134}),
\begin{align}
\phi(x, y| \vec r_c, \vec P) = \frac{\alpha_0-1}{\pi\gamma_0^2}\left[1+\left(\frac{x-x_c}{\gamma_0}\right)^2+\left(\frac{y-y_c}{\gamma_0}\right)^2\right]^{-\alpha_0}.
\end{align}
The second panel displays
the $\chi^2(\hat{\vec{A}}(r) | \hat{\vec r}_c,\hat{\vec P})$ function,
which shows that the optimum $\vec A(r)$ solution does not minimise the
agreement with the data (minimum $\chi^2$) , but makes a compromise with a regularisation
scheme (modelled by the $\chi^2_{\mathrm{pen}}(\vec A | \vec A_0)$ penalty
term). The lower panel illustrates that the effective number of amplitude
parameters fitted by data with the optimum regularisation hyper-parameter is
approximately 180. For this simulation, about 680 amplitude
parameters were fitted in a spectrogram built with a constant PSF FWHM of about
5.5 pixels. Intuitively, we can conjecture that an optimum
relation must exist between the typical width of the PSF kernel and the amount of
information that can be searched for in data,
\begin{align}
\left[\text{PSF width}\right]& \times \left[\text{\# effective degrees of freedom} \right] \notag \\ &\qquad \approx \left[\text{\# parameters}\right].
\end{align}

If too many parameters are searched for, the problem
becomes ill-posed. If too few parameters are searched for, it should be compensated for with a wide PSF
kernel. We thus postulate that we should have a spectrophotometric uncertainty
principle of the type
\begin{equation}
\sigma_{\mathrm{PSF}} \times \frac{N_\mathrm{dof}}{N_x} \gtrsim h,
\end{equation}
where the optimum is reached at equality, and $h$ has the same units as
$\sigma_{\mathrm{PSF}}$ the width of the PSF kernel. This formula gives the minimum number of the degrees of freedom required to describe data given a PSF width.

\begin{figure}[!h]
\begin{center}
\includegraphics[width=0.85\columnwidth]{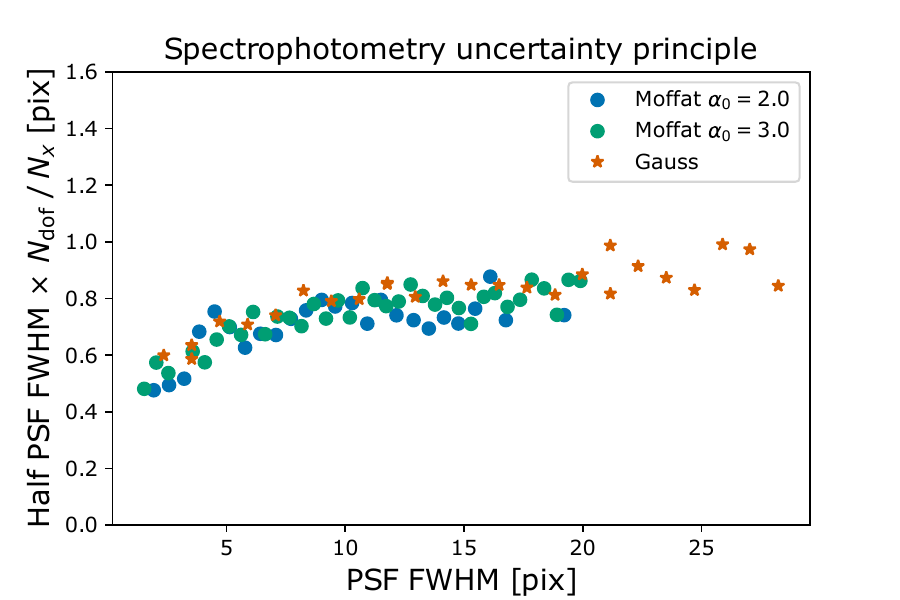}
\end{center}
\caption[] 
{Product of the PSF FWHM with $N_{\mathrm{dof}} / N_x$ as a function of the PSF FWHM for three different PSF models: a Gaussian kernel (orange stars), a Moffat kernel with $\alpha_0=2$ (blue points), and  a Moffat kernel with $\alpha_0=3$ (green points).}
\label{fig:uncertainty_principle}
\end{figure}

We tested the deconvolution and regularisation process on a large number of
simulated spectrograms with a constant width ($n_{\mathrm{PSF}}=0$), but without
second diffraction order. For the sake of simplicity, we tested this without any loss of
generality. We tried a Gaussian PSF kernel and a Moffat kernel with two
different exponents ($\alpha_0=2$ and $3$) and various $\gamma_0$ values. The
results are summarised in Figure~\ref{fig:uncertainty_principle}.  For the three models, $\sigma_{\mathrm{PSF}}$ was chosen to be half of the PSF
FWHM. The figure shows that for any PSF kernel, the measure of the number of
degrees of freedom $N_\mathrm{dof}/N_x$ scales as the inverse of the PSF
width. They show a
definite trend for the product $\sigma_{\mathrm{PSF}} \times N_\mathrm{dof}
/N_x$ that appears to be asymptotically constant and equal to the number $h$
close to 0.8 pixel when the PSF size is significantly greater than a few pixels. This $h$ value varies with the S/N of the spectrogram, but for a situation, it locks the relation between $\sigma_{\mathrm{PSF}}$ and  $N_\mathrm{dof}$.
This flatness of the relation shows that our procedure agrees when the extraction of information at the scale of the PSF kernel is considered optimal.

It is also noteworthy that this equation could be exploited to accelerate the computation of the PSF cubes: Instead of computing a PSF kernel for each pixel column $i$, we could compute it for each $N_\mathrm{dof}/N_x$ pixel. Because the computation times were not an issue in this paper, we left this investigation for a future project in which computation speed counts.

\subsection{Wavelength calibration}\label{sec:wavelength_cal}

\begin{figure}[!h]
\begin{center}
\includegraphics[width=0.9\columnwidth]{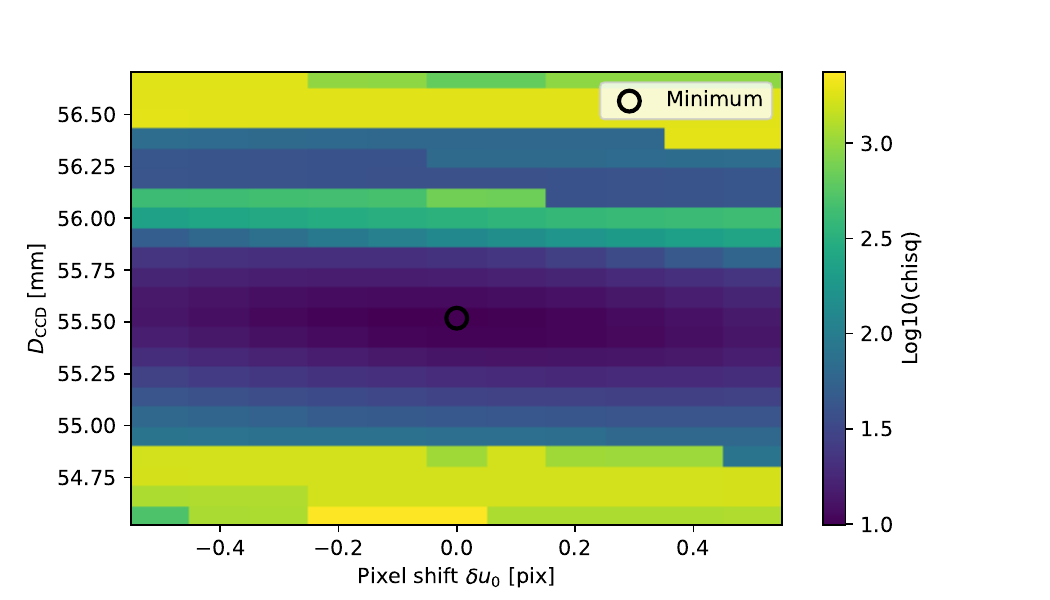}
\includegraphics[width=\columnwidth]{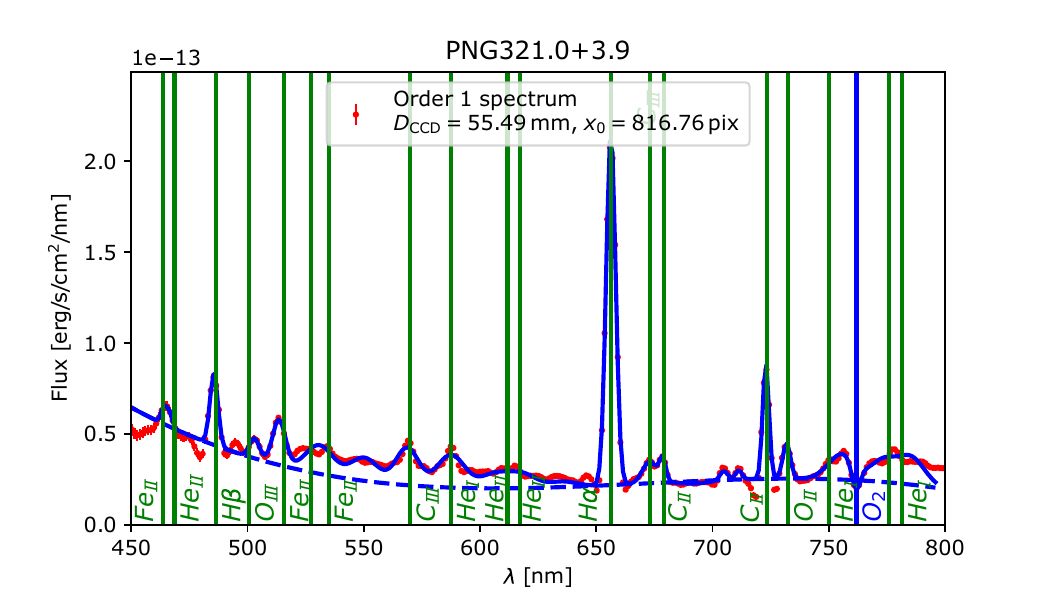}
\end{center}
\caption[] 
{Wavelength-calibration process on a planetary nebula spectrum. {Top:} Global $\chi^2(\delta u_0^{(\mathrm{fit})},D_\mathrm{CCD})$ function and its minimum (black circle) for the planetary nebula PNG321.0+3.9 observed at CTIO. The sharp steps at the top of the plot arise when certain lines are detected. {Bottom:} Calibrated spectrum of PNG321.0+3.9. The vertical lines indicate emission or absorption lines that are detected, positioned at their tabulated values. The dashed blue lines show the fitted background (whose degree depends on its length), and the plain blue lines show the Gaussian profiles fitted on data.}
\label{fig:D2CCD_x0_fit}
\end{figure}

Despite the astigmatism of the system, to
  first approximation, the slitless spectrograph obeys the usual grating
formula (Eq.~\ref{eq:grating}; see
e.g. \citealt{Murty:62,Hall:66,schroeder2000astronomical}).  Using the notations
of Figure~\ref{fig:dispersion}, the grating formula can be inverted to find the
relation between the $u$ coordinate along the dispersion axis and $\lambda$.

First, we assume that the true zeroth-order position is at $u_0$ along the
dispersion axis, but that a misfit of its centroid (see
Section~\ref{sec:order0}) can shift the position by a quantity $\delta
u_0^{(\mathrm{fit})}$.

The ADR also slightly spreads the {zeroth-order}
image along the local constant azimuth line in a deterministic way depending on
the airmass and the spectrum of the source. It also depends on the
parallactic and camera angles, the atmosphere temperature, pressure, and humidity. It is incorporated in the wavelength calibration process as a wavelength-dependent shift $\delta
u^{(\mathrm{ADR})}(\lambda)$ of the PSF centroid position along the dispersion axis with respect to a reference wavelength $\lambda_{\mathrm{ref}}$: $\delta
u^{(\mathrm{ADR})}(\lambda_{\mathrm{ref}}) = 0$.

We model this effect using the NIST metrology
toolbox\footnote{\url{https://emtoolbox.nist.gov/Wavelength/Documentation.asp}}
recommendation of using a modified version of the Edlén
equation~\citep{1966Metro...2...71E} by Birch and Downs
\citep{1993Metro..30..155B,Birch_1994} (see appendix~\ref{sec:adr}).

The distance $d(\lambda)$ between
an abscissa of the spectrogram and the zeroth order then reads
\begin{align}
& \delta u(\lambda) = \delta u_0^{(\mathrm{fit})} + \delta u^{(\mathrm{ADR})}(\lambda),\\
& u_0 = \DCCD \tan \theta_0, \\
& d(\lambda) = u(\lambda)-u_0-\delta u(\lambda),
\end{align}
so that
\begin{multline}\label{eq:dispersion}
d(\lambda | \DCCD, \delta u_0^{(\mathrm{fit})}) = \DCCD\left[\tan\left(\arcsin(p \Neff \lambda+\sin \theta_0 )\right)\right.\\  \left. - \tan \theta_0\right] -  \delta u^{(\mathrm{ADR})}(\lambda) - \delta u_0^{(\mathrm{fit})}.
\end{multline}
The bijection between the position on the CCD and the wavelength is thus
parametrised with two unknown parameters, $\DCCD$ and $\delta
u_0^{(\mathrm{fit})}$, that need to be fitted.

As a starting point, we compute a first wavelength array $\lambda_0$ from the array of distances $d$ to the {zeroth order} along the dispersion axis, assuming $\delta
u_0=0$ and given a prior value of $\DCCD$. To obtain a wavelength array given $\DCCD$ and $\delta u_0$, equation~\ref{eq:dispersion} is inverted as
\begin{equation}
\lambda = \frac{1}{p \Neff}\left[\sin\arctan\left(\frac{d + \delta u^{(\mathrm{ADR})}(\lambda) + \delta u_0^{(\mathrm{fit})}}{\DCCD}\right) - \sin \theta_0   \right].
\end{equation}
To remove the ambiguity with ADR, which also depends on wavelength, we iterate this computation five times starting from $\lambda_0$ and updating $\lambda$. We verified that it is enough to converge toward a stable wavelength solution. In these steps, we associated a wavelength array $\lambda$ with the amplitude array $\vec A$. From this calibration, $\lambda_{\mathrm{ref}}$ is computed as the mean wavelength weighted by the spectrum $\vec A$ itself, in order to associate the maximum amplitude of the {zeroth order} with its brighter wavelength. With the fit of $\delta u_0^{(\mathrm{fit})}$ and this setting of $\lambda_{\mathrm{ref}}$, we ensure that we are not sensitive to the slight dispersion of the {zeroth order} itself.

The parameters $\DCCD$ and $\delta u_0^{(\mathrm{fit})}$ are fit on data using the most prominent absorption (or emission) lines on the observed stellar spectrum (typically, the hydrogen lines $H\alpha$ and $H\beta$, and the dioxygen lines at 762.1\,nm and 686.7\,nm). They are locally fitted with a polynomial background plus a Gaussian profile of unknown height, centroid, and width. A partial $\chi^2$ quantity is computed for each spectroscopic line and added into a global $\chi^2$.

A penalty defined as the squared distance between the fitted Gaussian centroids
and the tabulated values for the detected lines, weighted by the squared {S/N}, is then added to the $\chi^2$.

Finally, the full $\chi^2$ and its penalty are normalised to the number of detected lines. This normalisation of the global $\chi^2$ is necessary to avoid solutions that
favour a lower number of detected lines, while the penalty gives more weights to
the well-detected lines and anchors them on their tabulated values. The two
parameters $\delta u_0^{(\mathrm{fit})}$ and $\DCCD$ are varied to minimise the
penalised global $\chi^2$ and find the best solution for the wavelength
calibration.

\begin{figure*}[!h]
  \begin{center}
  \textbf{After the full forward-model fit}
    \includegraphics[width=0.8\textwidth]{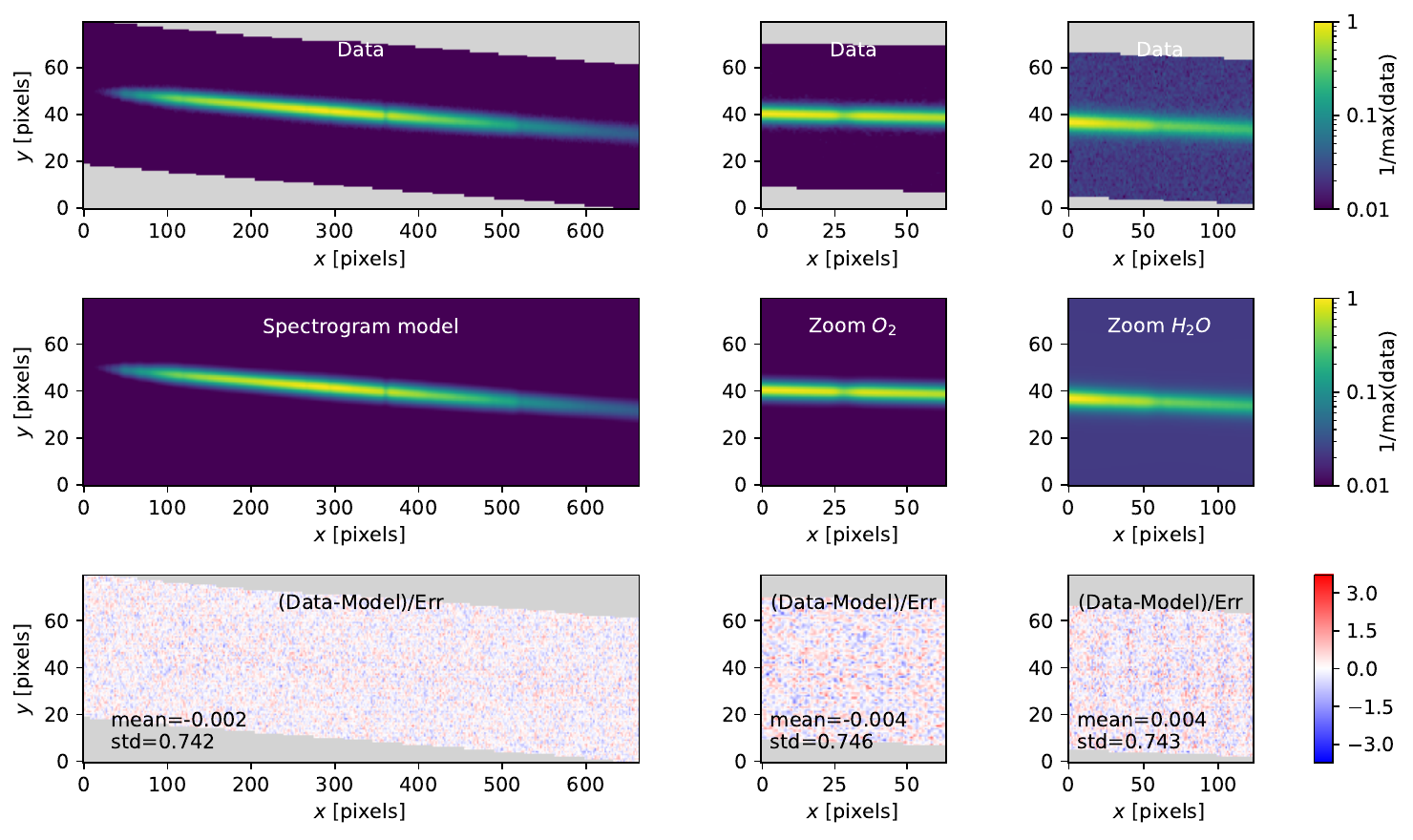}\hfill
  \end{center}
  \caption{Same as Figure~\ref{fig:deconvolution_residuals}, but with the addition of a second-order diffraction spectrogram and after the full forward-model fitting procedure.}
  \label{fig:ffm_sim}
\end{figure*}

The result of this process is illustrated in Figure~\ref{fig:D2CCD_x0_fit}. At the top, the global $\chi^2$ is represented for the wavelength calibration of the planetary nebula PNG321.0+3.9 observed at CTIO. The sharp steps at high or low $\DCCD$ values reflect situations in which some emission lines are detected or are not detected, which emphasizes the need to normalise the global $\chi^2$ by the number of detected lines. The smoothness around the minimum is due to the penalty term on $\delta u_0^{(\mathrm{fit})}$. In the calibrated spectrum, we can observe many emission lines that have been detected by the algorithm, and a good alignment between the tabulated values (represented by the vertical lines), the extrema of the Gaussian profiles, and those of the data curve. {A summary of the detected lines with high S/N is presented in Table~\ref{tab:lines}, reporting their fitted wavelength and their equivalent width (EQW).}

\begin{table}[!h]
\caption{Emission lines detected in the spectrum of planetary nebula PNG321.0+3.9 with {a S/N} above 10. }\label{tab:lines}
\resizebox{\hsize}{!}{\begin{tabular}{cccccccc}
\hline\hline
Line & Tabulated & Detected & Shift & FWHM & Amplitude & SNR & EQW \\
& nm & nm & nm & & erg/s/cm$^2$/nm & & nm \\
\hline
$He_{I}$ & 388.8 & 383.8 & -5.0 & 14.1 & 1.1e-13 & 46.0 & -28.3 \\
$H\beta$ & 486.3 & 485.6 & -0.7 & 4.0 & 3.9e-14 & 17.1 & -3.0 \\
$Fe_{II}$ & 515.8 & 513.6 & -2.2 & 6.5 & 2.5e-14 & 10.9 & -9.6 \\
$H\alpha$ & 656.3 & 656.3 & -0.0 & 4.8 & 1.9e-13 & 84.6 & -44.5 \\
$C_{II}$ & 723.5 & 722.9 & -0.6 & 3.2 & 6.4e-14 & 28.0 & -7.1 \\
$He_{I}$ & 861.7 & 865.6 & 3.9 & 8.7 & 5e-15 & 10.2 & -2.5 \\
$C_{III}$ & 970.5 & 968.8 & -1.7 & 6.4 & 1.1e-14 & 22.9 & -1.5 \\
$He_{I}$ & 1023.5 & 1024.3 & 0.8 & 10.4 & 5.6e-15 & 13.2 & -4.7 \\
\hline
\end{tabular}}
\tablefoot{The third column gives the centroid of the fitted Gaussian profile, and the fourth column lists the distance to the tabulated value. The EQW is reported in the last column.}
\end{table}

\subsection{Flux calibration}

With the fitted wavelength solution, the spectrum amplitude $\hat{\vec{A}}$ can be converted from
ADU units into flux densities in $\SI{}{erg/s/cm^2/nm}$, assuming that the telescope collecting area $\mathcal{S}_T$, the exposure
time $\tau$, and the CCD gain $\GCCD$ (in $e^-/$ADU) are known,
\begin{equation}\label{eq:flux_cal}
S_1(\lambda) = \hat{\vec{A}} \times \frac{hc\, \GCCD}{\mathcal{S}_T \tau \lambda \delta \lambda},
\end{equation}
where $\delta \lambda$ is the local variation in $\lambda$ within one pixel.

The end product of this pipeline is thus a
background-subtracted spectrum, calibrated in wavelength and flux, that
is the product of three quantities: the object SED, the instrumental
transmission, and the atmospheric transmission, which might be contaminated by a
second diffraction order because this latter has not yet been taken into account. 

\begin{figure*}[!h]
\begin{center}
\includegraphics[width=0.8\textwidth]{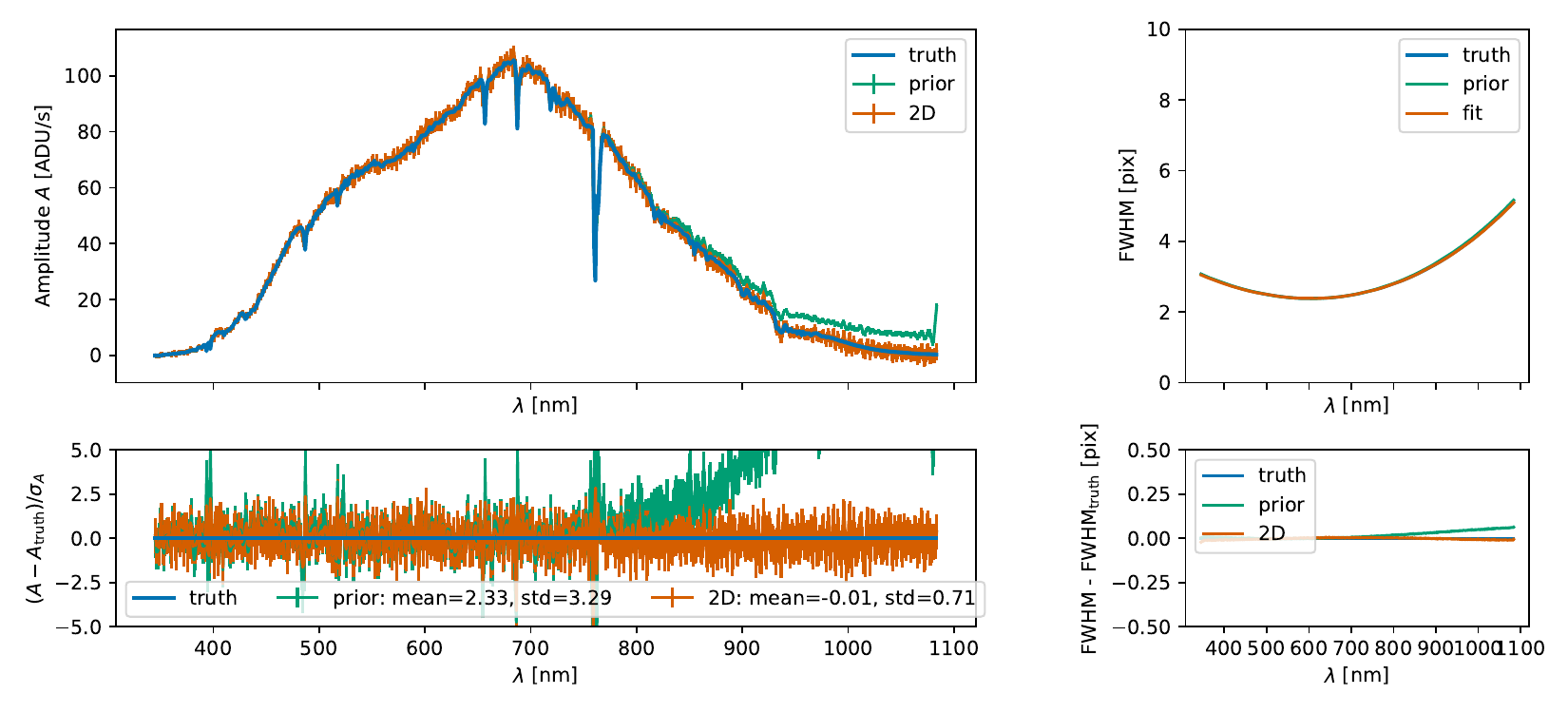}
\end{center}
\caption[] 
{Results from the full forward-model fit of a simulated  spectrogram of the CALSPEC star HD111980 with a Moffat PSF kernel. A second diffraction order was simulated. {Top left:} Spectrum output from the full forward-model process (orange) compared with the true spectrum injected in the simulation (blue) and the intermediate 1D transverse fit used as the prior vector $\vec A_0$ (green). {Bottom left:} Residuals between the true spectrum and the 1D and 2D fits normalised by their estimated uncertainties. {Top right:} FWHM of the true PSF (blue, right below the green curve) compared with the fitted PSF during the full forward-model procedure (green) or the 1D transverse fit (orange). {Bottom right:} Difference between the true PSF FWHM injected in the spectrogram simulation and the 1D and 2D fits. All blue curves are covered by the orange curves.}
\label{fig:ffm_truth}
\end{figure*}

\subsection{{Full forward model of the spectrogram}}\label{sec:ffm}

At the end of the previous steps, we also have first {estimates} of the two instrumental model functions $\phi_1(\vec r, \lambda)$ and $\vec \Delta_p(\lambda)$, and of geometric parameters such as the zeroth-order position $\vec r_0$ and the dispersion angle $\alpha$. With all these ingredients, we can implement a full forward model of the data that also takes the atmospheric differential refraction (ADR) and the superposition with the second diffraction order (last stage of Figure~\ref{fig:spectractor_graph}) into account.

In practice, we enrich the forward model described in the steps above with the
knowledge of the ADR physics (see Sections~\ref{sec:wavelength_cal}
and~\ref{sec:adr}) and with the knowledge of the second-order to first-order
transmission ratio $r_{2/1}(\lambda)$ of the spectrograph disperser. The ADR
model replaces a polynomial approach to predict $\vec \Delta_p(\lambda)$. In
other words, it is used to predict the trace of the spectrogram on the sensor
without free parameters, as long as airmass, outside pressure, outside
temperature, and humidity are given. With this, two free parameters remain for the spectrogram trace on the sensor to be fully constrained: the dispersion axis angle
$\alpha$ and $\delta y^{(\mathrm{fit})}$, which compensates for a misfit of the zeroth-order centroid along the $y$-axis. {The ratio $r_{2/1}(\lambda)$ can} be
measured on an optical test bench (see Section~\ref{sec:thorlabs}) or on on-sky
data. In the full forward model, we use a new design matrix $\vec{\tilde{M}}$
that is defined as
\begin{equation}
\vec{\tilde{M}} = \vec{M}(\vec{Z}|  \vec r_c, \vec P ) + A_2\vec R_{2/1}\vec{M}(\vec{Z}|  \vec r_c^{p=2}, \vec P^{p=2} ),
\end{equation}
where $A_2$ is a {safety normalisation} parameter,
$\vec R_{2/1}$ is the transmission ratio vector computed for  a given wavelength calibration, $\vec r_c^{p=2}$ are the centroid positions
of the second diffraction order PSF kernels, and $\vec P^{p=2}$ are their shape
parameters. The vector $\vec r_c^{p=2}$ is computed using the grating
formula~\ref{eq:grating} and the ADR model.  $\vec P^{p=2}$ can be fitted
independently of $\vec P^{p=1}$, but we chose to assume that the PSF shape
depends more on the spectrograph defocusing towards the infrared than on the
atmospheric chromatic seeing. The PSF shape parameters for the second
diffraction order are thus considered the same as for the first
diffraction order at the same distance of the {zeroth order}. We chose to set the
$\vec P^{p=2}$ vector accordingly\footnote{Another choice could have been to assume that the spectrograph does not suffer
from defocus, and thus arguing that the PSF shape parameters for the second
diffraction order are the same than that of the first diffraction order at the
same wavelength whatever the distance to the {zeroth order}. For CTIO images, our first choice leads to better fits to data.}. Therefore, the full forward model now includes both first- and second-order spectrogram models as
\begin{equation}
  \vec I_1(\vec{Z} | \vec A,\vec r_{c}, \vec{P}) + \vec I_2(\vec{Z} | \vec A,\vec r_{c}^{p=2}, \vec{P}^{p=2}) =
  \vec{\tilde{M}}\, \vec{A}.
  \label{eq:12}
\end{equation}


We implemented a two-step iterative method that alternates the wavelength
calibration described in Section~\ref{sec:wavelength_cal} and the full forward
model described just above (by combining a linear fit for the $\vec
A$ spectrum amplitudes and a Gauss-Newton descent for the non-linear
parameters $\vec P$) with the same $r$ hyper-parameter that
was fitted (in Section~\ref{sec:regularisation}). The $\vec A$ parameter vector determined with PSF 2D deconvolution previously (Section~\ref{sec:spectrum_deconvolution_psf2d}) is seeded in the forward-model fit as a new prior $\vec A_0$. A $20\sigma$ clipping is performed to reject the field stars and their concomitant
spectrograms as well as other sensor defects.
This procedure ensures that all the forward-model parameters are fitted again together on
data, using the more complete model including $\vec A$, $\vec P$, $\DCCD$, $\delta u_0^{(\mathrm{fit})}$, and $A_2$. Their values replaced all those that were fitted previously. The residual map obtained in this way is flatter than
before, with an even smaller final $\chi^2$, and consequently, with a better accuracy of
all the fitted parameters.

The two main products of this step are a first-order diffraction spectrum  $S_1(\lambda)$ separated
from the second-order spectrum, and the second-order spectrum
$S_2(\lambda)$ with
\begin{equation}
S_2(\lambda) = r_{2/1}(\lambda) S_1(\lambda),
\end{equation}
in $\SI{}{erg/s/cm^2/nm}$ (following Eq. ~\eqref{eq:flux_cal}). 

At this point, we consider the second diffraction order not
as a nuisance, but as a useful signal. With a strong bending due to ADR (e.g. with the dispersion axis orthogonal to the zenith direction), it can be detached from the first diffraction order on purpose to maximise the statistical power of the exposure. 

In summary, a full forward model takes advantage of the higher diffraction orders as a
redundant piece of data to fit all parameters, especially in the bluer part, where
absorption lines are twice wider in pixels than in the first-order spectrum. This is a key
advantage of the forward approach compared to the direct approach.

\subsection{Validation on simulations}\label{sec:ffm_validation}

\begin{table}[!h]
\caption{Parameters of a spectrogram simulation and \Spectractor estimations, recovered with a full forward-model approach.}\label{tab:ffm_sim134}
\begin{tabular}{ccc}
\hline\hline
Parameter & Simulation value & Recovered value \\
\hline
$x_0$ [pix] & 743.89 & $743.76 $  \\
$y_0$ [pix] & 682.92 & $682.93 $ \\
$\delta y^{(\mathrm{fit})}$ [pix] & 0 & $-0.199\pm 0.003$ \\
$\alpha$ [$^\circ$] & $-1.5653$ & $-1.5654\pm 0.0004$ \\
$\left\langle B(\vec r) \right\rangle$ [ADU/s] & 0.1930 & 0.1930 \\
$\DCCD$ [mm] & 56.322 & $56.329$\\
$A_2$ & 1 & $0.98 \pm 0.02$\\
$\gamma_0$ & 3 & $3.00 \pm 0.01$ \\
$\gamma_1$ & 1 & $0.99 \pm 0.02$ \\
$\gamma_2$ & 1 & $1.02 \pm 0.02$ \\
$\alpha_0$ & 3 & $3.01 \pm 0.01$ \\
$\alpha_1$ & 0 & $-0.02 \pm 0.01$ \\
$\alpha_2$ & 0 & $0.03 \pm 0.02$ \\
$N_{\mathrm{dof}}$ & 669 & 298\\
Reduced $\chi^2$ & -- & 0.53\\
\hline
\end{tabular}
\tablefoot{Parameters with uncertainties are fitted during the Gauss-Newton descent of the full forward model, while others are just estimated on data. The simulation was done with a spectrum made from 669 amplitude parameters, and the regularisation process recovered 303 of them. The $\delta y^{(\mathrm{fit})}$ centroid correction is included in the $y_0$ quoted value.}
\end{table}

\begin{figure}[!h]
\begin{center}
\includegraphics[width=0.85\columnwidth]{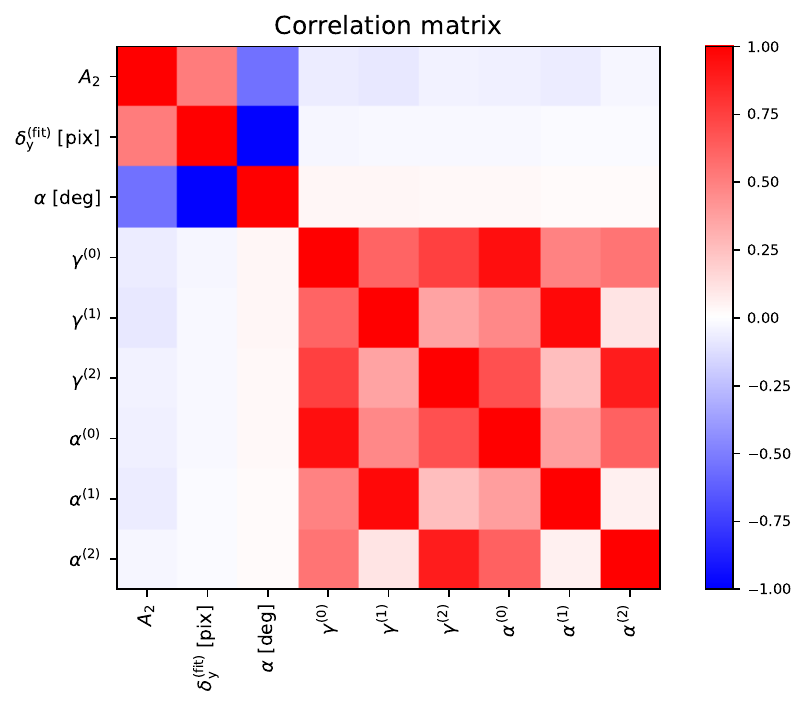}
\end{center}
\caption[] {Correlation matrix of the full forward-model fitting of a spectrogram simulation at the
  end of the Gauss-Newton descent.}
\label{fig:sim_134_cov}
\end{figure}

\begin{figure}[!ht]
  \begin{center}
    \includegraphics[width=\columnwidth]{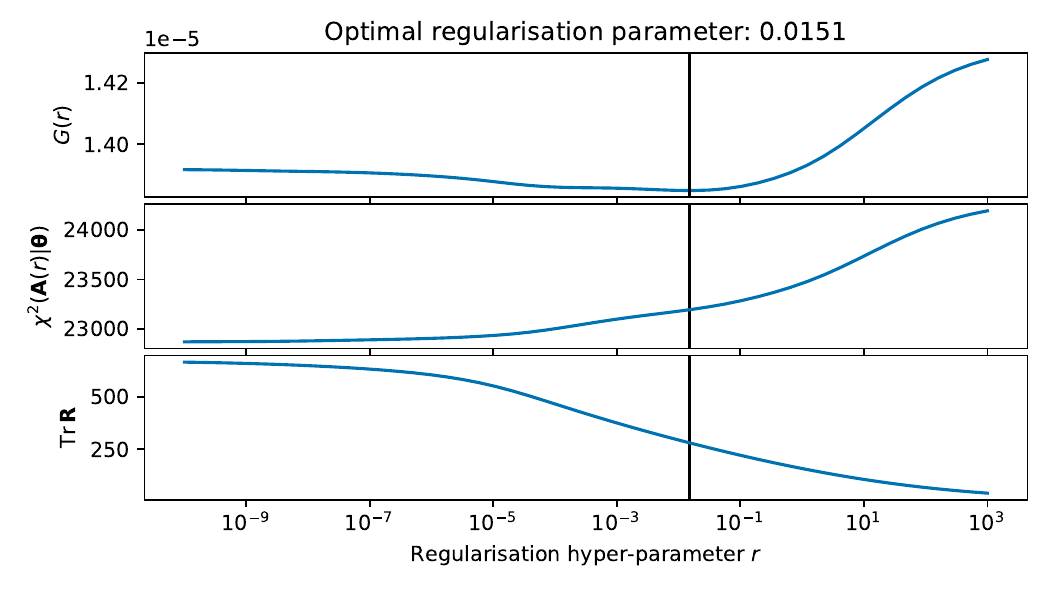}
  \end{center}
  \caption{Same as Figure~\ref{fig:deconvolution_regularisation}, but for the simulation
    illustrated in Figure~\ref{fig:ffm_truth}.}
\label{fig:deconvolution_regularisation_sim134}
\end{figure}

To test the full forward model, we simulated a spectrogram with a second
diffraction order and a sharp Moffat PSF kernel to increase the spectral
resolution (see the PSF parameter values in Table~\ref{tab:ffm_sim134}), whose
shape evolves as a second-order polynomial
function. Figure~\ref{fig:ffm_sim} compares the simulated data with the
fitted spectrogram model and focuses on some atmospheric absorption lines: The
residuals follow the expected Gaussian distribution again, even in the fast-varying regions
of the spectrum.

In Figure~\ref{fig:ffm_truth}, we show that the process recovered the
true spectrum injected in the simulation within the estimated uncertainties
(diagonal elements from the $\mathbf{C}$ matrix from
equation~\ref{eq:covariance_matrix}).
The agreement is excellent at all wavelengths, even around the fast-varying
absorption lines. The right panel of the figure also shows the FWHM of the true PSF. The reconstructed PSF displays the same
wavelength-dependent PSF FWHM as the simulated one.
It is also remarkable that while the cross-spectrum issued from the transverse 1D fit described in Section~\ref{sec:fit_1D} failed to recover the true spectrum and the true PSF profile because of the presence of the second diffraction order (orange curves), it still proved to be an important seed for the regularisation process because only its regularity is used because of the Laplacian operator $\mathbf{L}$.

The recovered parameters are compared with the simulation values in Table~\ref{tab:ffm_sim134}. They are fitted together with their uncertainties in the full forward-model minimisation, which provides their full covariance matrix (Figure~\ref{fig:sim_134_cov}), while the other parameters such as the star centroid are just estimated on data. 

The regularisation quantities are given in
Figure~\ref{fig:deconvolution_regularisation_sim134}. For a PSF FWHM between 2 to 4 pixels, we again obtain
$N_{\mathrm{dof}}\approx 300$ out of $\approx 700$ parameters. This confirms the rule of thumb given by the spectrophotometric uncertainty principle. For spectrograms such as those presented in the
previous figures, the end-to-end pipeline takes 2 minutes on a standard laptop.

\begin{figure}[!ht]
  \begin{center}
    \includegraphics[width=\columnwidth]{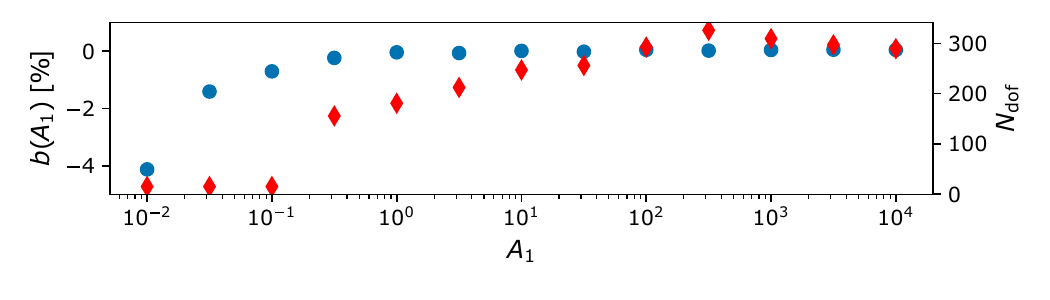}
    \includegraphics[width=\columnwidth]{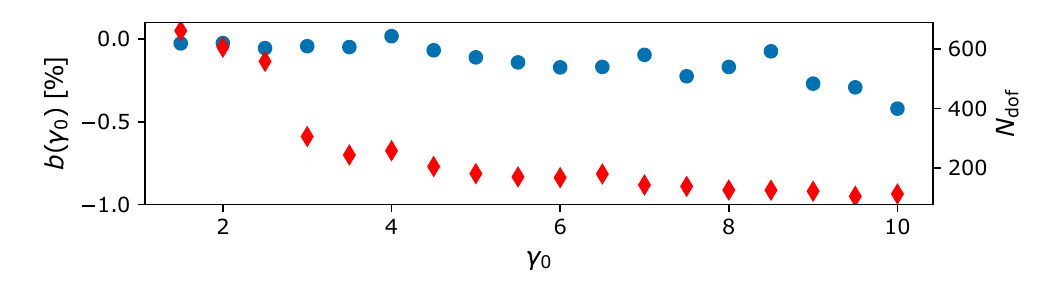}
    \includegraphics[width=\columnwidth]{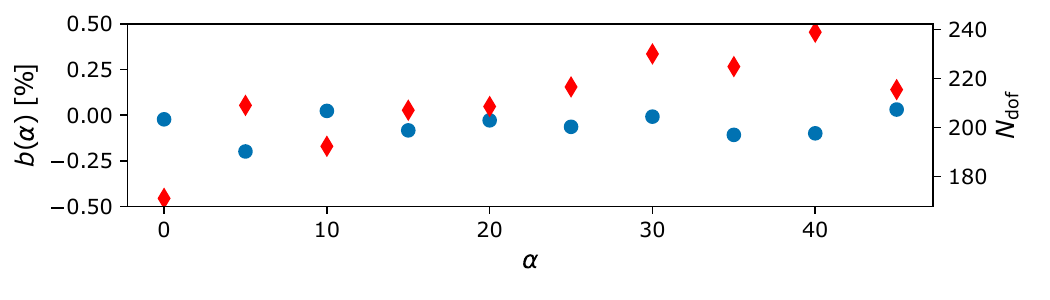}
  \end{center}
  \caption{Extraction bias $b$ (blue points) as a function of the {S/N} materialised by the amplitude $A_1$ factor (top), of the PSF width $\gamma_0$ (middle), and of the rotation angle $\alpha$ (bottom). For completeness, the effective number of degrees of freedom $N_{\mathrm{dof}}$ is represented with red diamonds for each simulation.}
\label{fig:extraction_bias}
\end{figure}

The \Spectractor implementation was tested on many simulations and
recovered the simulation parameters within the estimated uncertainties (68\% confidence interval) for sets of parameters that were not too extreme (smooth wavelength dependence on the PSF, or the PSF kernel sampled over a few pixels). We also evaluated the extraction bias $b$ between the true spectrum given in the simulation $S_1^{\rm truth}(\lambda)$ and the extracted spectrum $S_1(\lambda)$ as
\begin{equation}
b = \frac{ \int \dd \lambda \left(S_1(\lambda)-  S_1^{\rm truth}(\lambda) \right) }{\int \dd \lambda S_1^{\rm truth}(\lambda)}. 
\end{equation}
The extraction bias was evaluated with many simulations in different cases in terms of {S/N}, resolution, and  geometry. For the first case, the variation in {S/N} was simulated by multiplying the simulated spectrum by an arbitrary grey factor $A_1$, keeping the image background at the same level. The {S/N} of the simulation presented in Figure~\ref{fig:ffm_truth} and Table~\ref{tab:ffm_sim134} corresponds to $A_1=1$. We found no significant bias in the $A_1>1$ regime (Figure~\ref{fig:extraction_bias}), but a small bias appears for spectrograms with a lower {S/N} at the percent level. {This is because a Gaussian model was used in the evaluation of the uncertainty map (see~\ref{sec:uncertainties}) while a Poisson distribution is more accurate at low S/N.} The variation in spectrogram resolution was simulated by changing the PSF width $\gamma_0$, and a small negative bias appeared for low-resolution spectra (large $\gamma_0$). These spectra exhibit wider and shallower absorption lines than the true spectrum, which leads to these negative $b$ values, but only at the sub-percent level. However, except for the absorption lines, the overall spectrum shape from blue to red is recovered perfectly. Finally, geometry variations were also simulated using different dispersion axis angles $\alpha$, but we found no bias. In conclusion, the most important condition for extracting unbiased spectra from slitless spectrophotometry is a sufficient {S/N}, closely followed by a sufficiently fine spectral resolution. The adequacy of these parameters is easy to test a priori with a forward-model simulation.

For completeness, we represent the reconstructed number of degrees of freedom for all these simulations in Figure~\ref{fig:extraction_bias}. As expected, for spectrograms with a very low {S/N}, this number is close to zero{. In that case, the extracted spectra are close to the $\vec A_0 = \hat{\vec{A}}^{(1D)}$ prior spectra. When} the signal increases, then $N_{\mathrm{dof}}$ strongly increases until it saturates because of pixelisation. Conversely, this number decreases when the PSF width increases because the spectrogram has a lower spectral resolution.

\begin{figure*}[!ht]
\begin{center}
\textbf{Thorlabs blazed periodic grating\hspace{0.33\textwidth} Hologram\hspace{0.12\textwidth}~}\\
\includegraphics[width=0.47\textwidth]{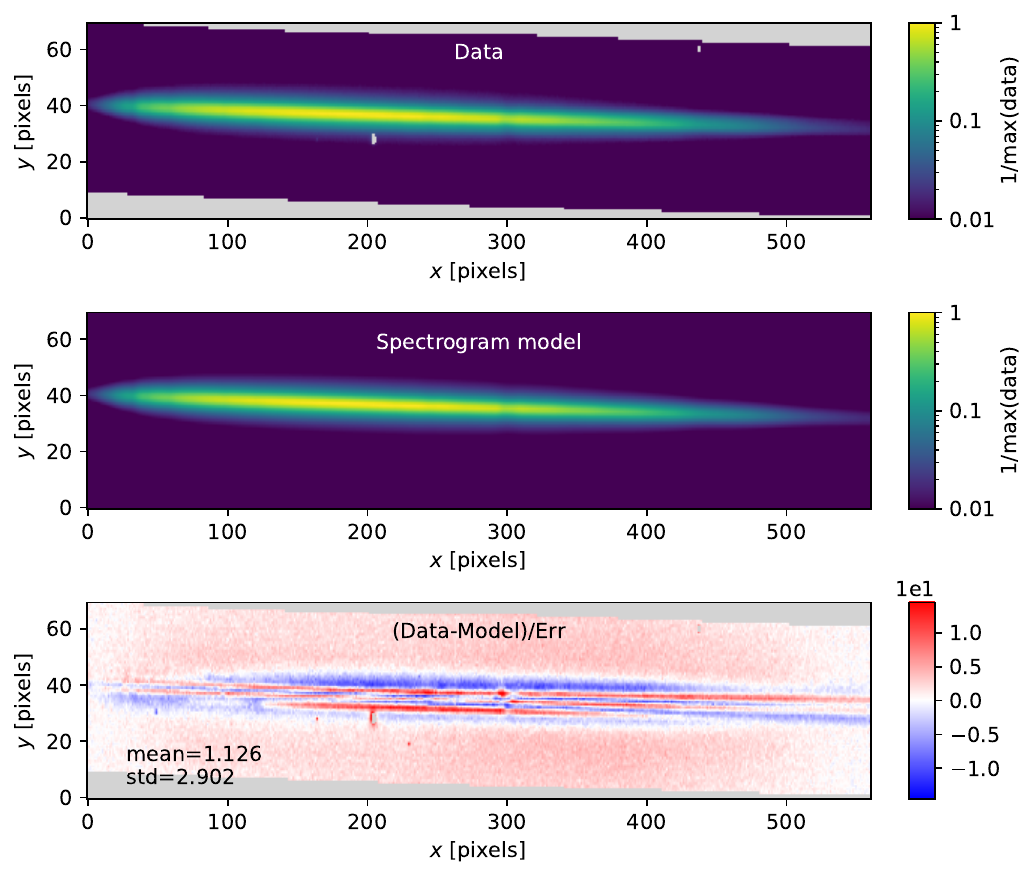}\hfill
\includegraphics[width=0.47\textwidth]{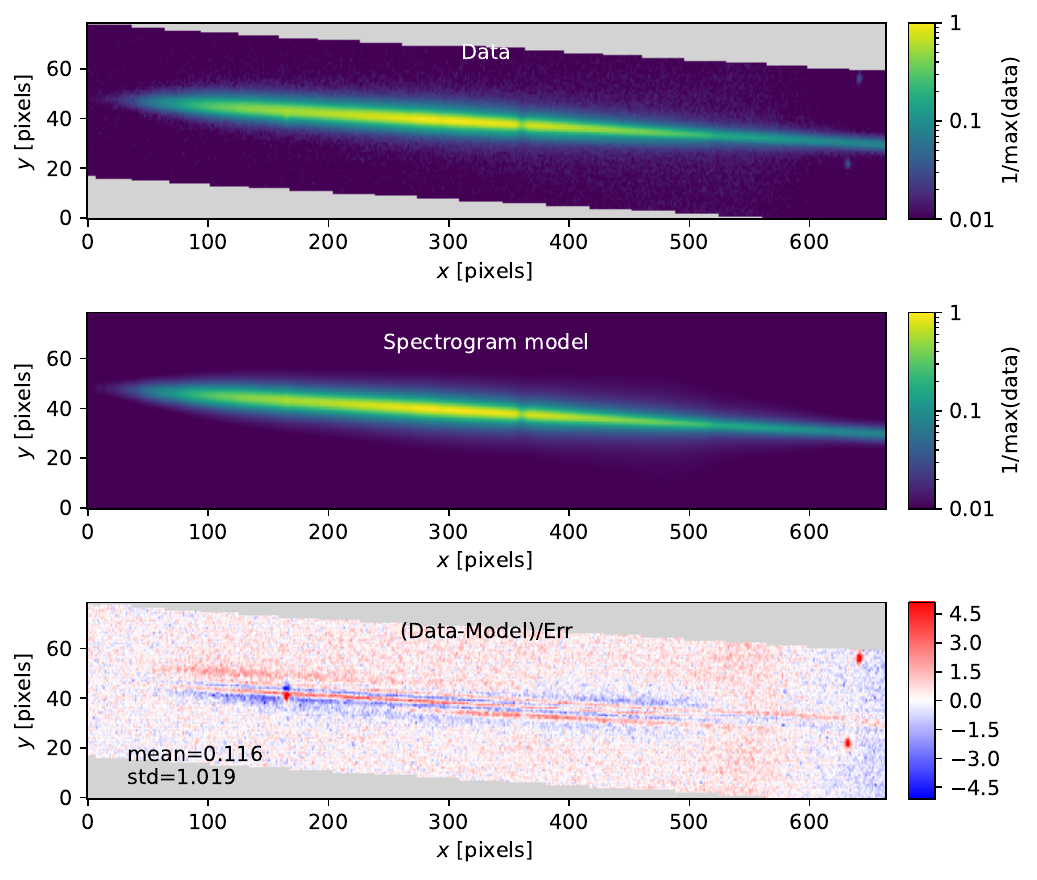}
\end{center}
\caption[] 
{Same as in Figure~\ref{fig:ffm_sim}, but for CTIO data: CALSPEC star HD111980 observed on 2017 May 30 with a blazed Thorlabs grating 300 lines/mm (left), for which the PSF is out of focus and deviates from a Moffat model, and an amplitude hologram of 350 lines/mm (right), better focused on the CCD.}
\label{fig:ffm_data}
\end{figure*}

\begin{figure*}[!ht]
\begin{center}
\includegraphics[width=0.5\textwidth]{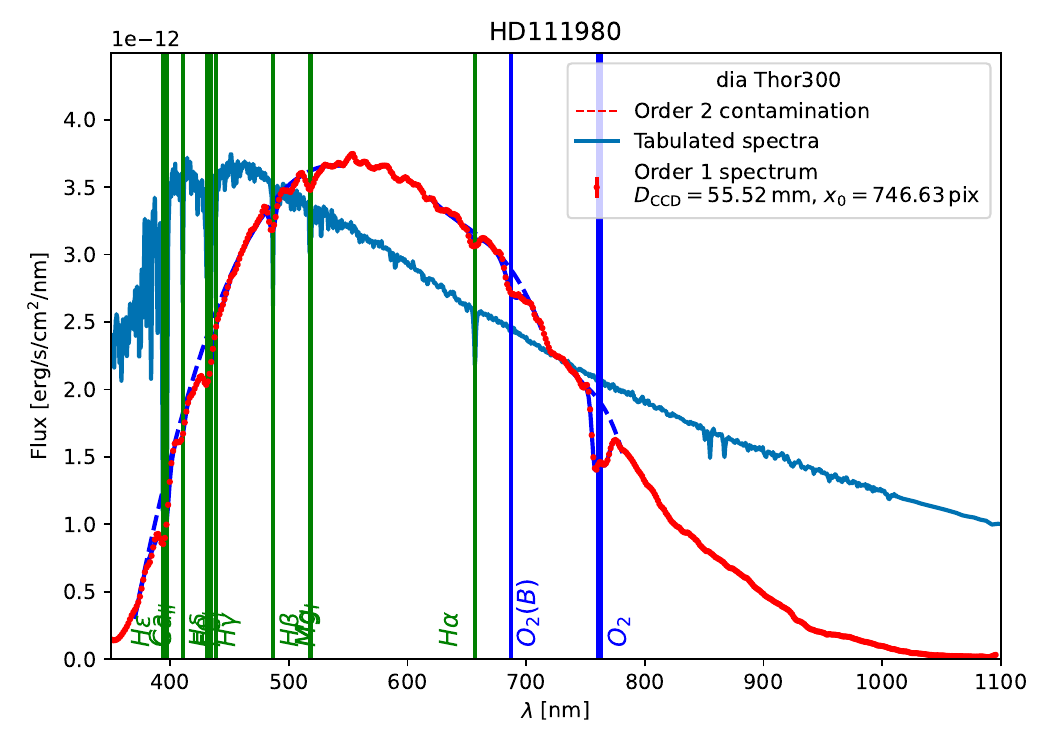}\hfill
\includegraphics[width=0.5\textwidth]{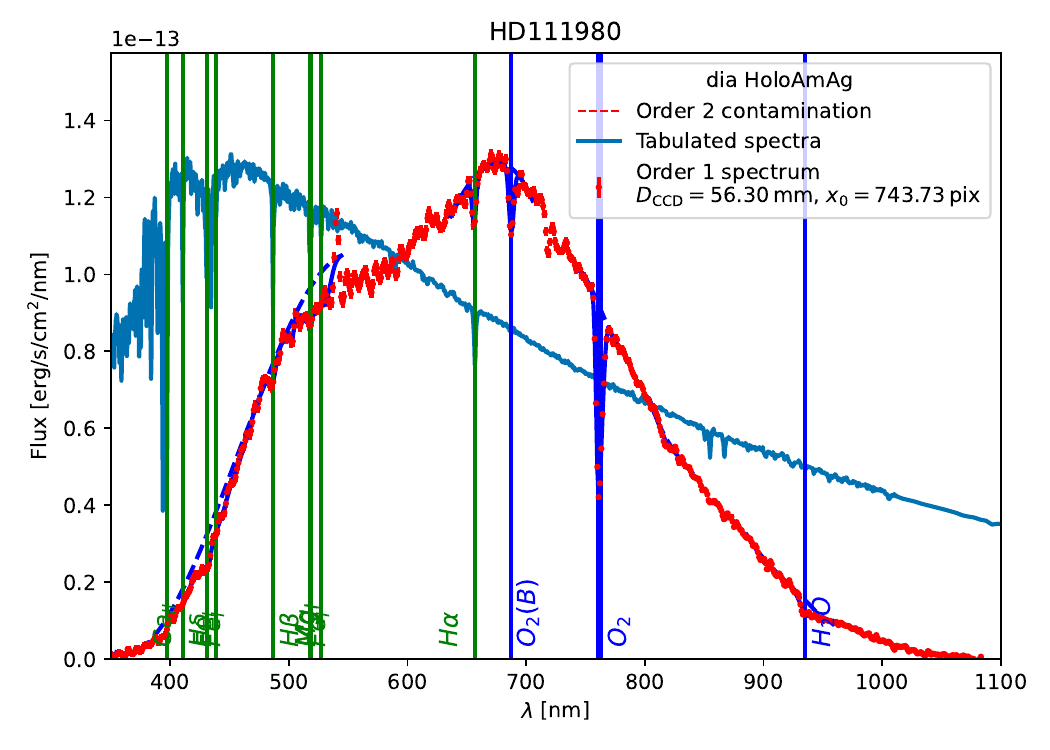}
\end{center}
\caption[] 
{Calibrated spectra of CALSPEC star HD111980 observed at CTIO on 2017 May 30 with a blazed Thorlabs grating 300 lines/mm (left) and an amplitude hologram of 350 lines/mm (right). The CALSPEC SEDs are given for comparison (scaled for convenience). The two dispersers do not have the same transmission curves, which explains the different shapes of the spectra. The vertical lines indicate emission or absorption lines that are detected, positioned at their tabulated values. Locally, the dashed blue lines show the fitted continuum, and the plain blue lines are the Gaussian profiles fitted on absorption lines.}
\label{fig:HD111980_spectra}
\end{figure*}

\section{Spectrum extraction on data}\label{sec:data}

The success of the spectrum extraction on data mostly depends on the model of
the wavelength-dependent PSF of the telescope. If the PSF model correctly
represents the reality, the residuals after the full forward-model fit of the spectrogram
converges towards the expected Gaussian distribution. Otherwise, the extracted
spectrum is distorted when the PSF is too far from reality.

This is illustrated in Figure~\ref{fig:ffm_data}. In the left panels,
a blazed Thorlabs grating with 300 lines/mm was chosen to observe CALSPEC star
HD111980 at CTIO on 2017 May 30. Directly placed in the filter
wheel at $\DCCD\approx 55$\,mm from the CCD, this grating presents a rather strong
defocusing that is poorly modelled by our default circular Moffat PSF, even
with a fourth-order polynomial evolution with wavelength.

The building of an appropriate model for this highly defocused PSF is left for
future work and is presented here as an illustration that the prior
understanding of the telescope can be greatly worthwhile in a forward-fitting approach. These  extractions used a sigma-clipping procedure with a $20\sigma$ threshold in order to reject only the field stars and not the poorly modelled spectrogram pixels.

However, the same PSF kernel as was used to treat the same star observed 5 minutes
later, but with an amplitude hologram optimised to correctly focus the
spectrogram at all wavelength \citep{holo}, leads to residuals between $\pm
5\sigma$ (right figures), mostly dominated by a field star that contaminates the
spectrogram at about $\SI{530}{\nano\meter}$.

The parameters of interest for these two extractions are summarised in
Table~\ref{tab:ffm_reduc}. We realised {a posteriori} that the {S/N} was not sufficient to fit $A_2$, and decided to keep it fixed at 1. The way in which
the ratio $r_{2/1}(\lambda)$ was obtained for these exposures is explained in
Section~\ref{sec:atm}.

\begin{table}[!h]
\caption{Parameters of the CTIO exposures and \Spectractor estimates.}\label{tab:ffm_reduc}
\resizebox{\hsize}{!}{\begin{tabular}{ccc}
\hline\hline
Parameter & CTIO Exposure 1 with & CTIO Exposure 2 with  \\
 & blazed regular grating & amplitude hologram  \\ \hline
Disperser & 300 lines/mm & 350 lines/mm \\
Seeing [$\arcsec$] &  0.68 & 0.65 \\
Airmass & 1.12 & 1.13 \\
$x_0$ [pix] & 767.2 & 779.9  \\
$y_0$ [pix] & 696.7 & 702.6  \\
$\delta y^{(\mathrm{fit})}$ [pix] & $0.2402 \pm 0.0008$ & $-0.217 \pm 0.004 $ \\
$\alpha$ [$^\circ$] & $-0.7980 \pm 0.0001$ & $-1.5650 \pm 0.0004 $ \\
$\left\langle B(\vec r) \right\rangle$ [ADU/s] & 2.4 & 0.2 \\
$\DCCD$ [mm] & 55.53 & 56.33 \\
$\vec A$ size & 562 & 669 \\
$r$ & 0.97 & 0.16 \\
$N_{\mathrm{dof}}$ & 125 & 197 \\
Reduced $\chi^2$ & 9.67 & 1.05 \\
\hline
\end{tabular}}
\end{table}

The calibrated spectra given by \Spectractor at the end of the extraction
process are given in Figure~\ref{fig:HD111980_spectra}. Because of the
strong defocusing, the spectrum from the Thorlabs grating presents broadened
absorption lines, while the amplitude hologram has sharper absorption lines. The
second displays a better spectral resolution that is limited by the atmospheric seeing,
which argues in favour of either controlling the spectrograph PSF at the hardware
level, or of being able to accurately model a defocused PSF kernel. At this point, it seems easier to adjust the spectrograph to obtain a simple PSF model than to guess the complexity of the chromatic PSF with on-sky data. Based on these
spectra or spectrograms, we show in Section~\ref{sec:atm} how to measure the
atmospheric transmission or the instrumental transmission via forward
modelling.

\section{A path toward measuring atmospheric transmission}\label{sec:atm}

One of the main objectives for building a spectrophotometric instrument and its
analysis pipeline is to be able to accurately measure on-site atmospheric
transmission so as to improve the photometric calibration of other telescope on
the same site. For instance, the aim of the Auxiliary Telescope at Cerro Pach\'on
is to measure the atmospheric transmission to correct the photometry of the LSST
survey.

In order to discuss the capabilities of \Spectractor in measuring atmospheric
quantities, we first recall that its main output is a first diffraction order
spectrum,
\begin{equation}
S_1(\lambda)  = T_{\text{inst}, 1}(\lambda) \, T_{\text{atm}}(\lambda | \vec{P}_a) \, S_{*}(\lambda).
\end{equation}
To be able to obtain the atmospheric transmission $T_{\text{atm}}(\lambda |
\vec{P}_a)$, we need to know the stellar SED and the instrumental
transmission, and to have an accurate full forward model for the spectrograph.

Because the most accurate PSF model is achieved with the amplitude hologram at
CTIO because of its focusing properties, we expect better results from its
analysis than from the data acquired with the Thorlabs grating.

In order to inform our forward model, we need to know the disperser on-sky
first-order transmission and their $r_{2/1}(\lambda)$ ratio. While we show below
how to use on-sky data to this end, we also secured the Thorlabs
grating to bring it back on an optical bench at the Laboratoire de Physique Nucléaire et de Hautes \'Energies (LPNHE) and measure its transmission. This
was not possible for the prototype hologram used at CTIO, and we had to recover its transmission with
on-sky data.

While in theory, the forward modelling of the atmospheric transmission could be
based on a perfect a prior knowledge of the instrument and simply needs to
fit each star exposure, in practice, all this is slightly more
complex. We therefore decided to show that our pipeline can be used with intermediary steps in
order to gain increasingly better understanding of the data, up to the point where the
full forward modelling of the atmospheric parameters becomes possible. Because of the limited telescope time and
observations, this first paper relies on a limited amount of data and aims at
presenting the algorithms and procedures. Clearly for
accurate results, many more data are required.

The different steps that we undertook and detail below are as follows:
\begin{itemize}
\item \ref{sec:thorlabs} Laboratory measurement of the blazed grating transmission as a function of $\lambda$;
\item \ref{sec:ctio_transmission} inference of the CTIO $\SI{0.9}{\meter}$ telescope transmission using data taken with the blazed grating during a stable night;
\item  \ref{sec:holoamag} inference of the amplitude hologram transmission during the same stable night with the CTIO $\SI{0.9}{\meter}$ telescope transmission;
\item  \ref{sec:atm_20170530}, \ref{sec:afm}  measurement of  $T_{\text{atm}}(\lambda | \vec{P}_a)$ with data using the amplitude hologram with  $T_{\text{inst}, 1}(\lambda)$, $S_{*}(\lambda)$, and a Moffat PSF kernel.
\end{itemize}

\subsection{Disperser transmission measurement}\label{sec:thorlabs}

\begin{figure}[!th]
\begin{center}
\includegraphics[width=\columnwidth]{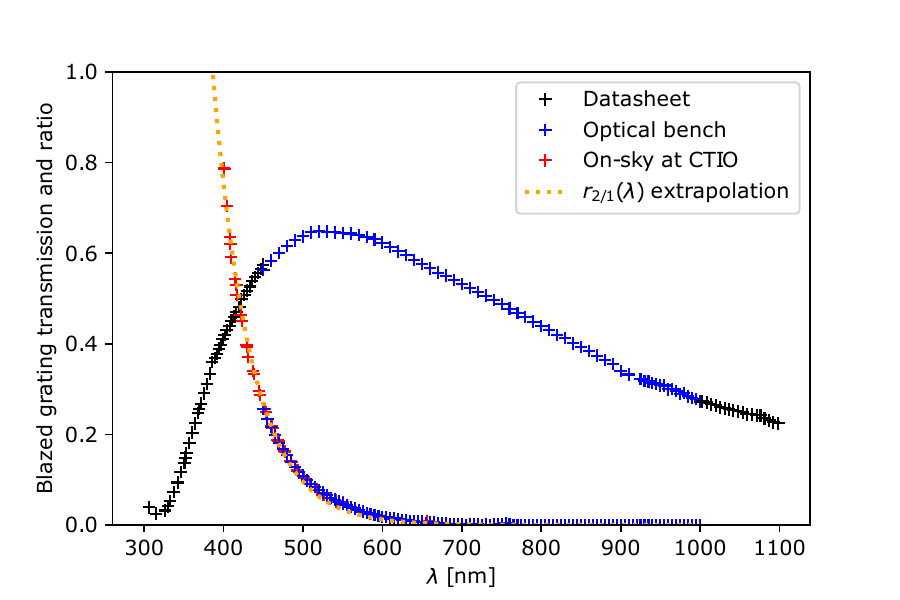}
\end{center}
\caption[] 
{First diffraction-order transmission of the blazed Thorlabs grating with 300 lines/mm (blue and black) and the ratio of transmissions of the order 2 over the order 1 (red, blue, and orange).}
\label{fig:thorlabs}
\end{figure}

Our blazed Thorlabs 300 lines/mm grating was studied two years after the CTIO
campaign on the LPNHE optical test bench. {The optical bench is made of parabolic off-axis mirrors to simulate a monochromatic $f/D=18$ telescope beam.} A monochromator is used to select an
accurately known narrow wavelength interval, and the light, after passing through the
grating mounted on an $xyz$ support, is collected on a CCD device. 

The transmissions for orders 0, 1, and 2 were measured using aperture photometry
at many wavelengths. Because the laboratory bench had been designed
to measure filter transmissions, it could unfortunately not be adjusted to allow a {S/N} that was high enough in the regions below \SI{450}{\nano\meter} and above
\SI{1000}{\nano\meter}. For these wavelength ranges, we therefore used the spreadsheet of the
grating manufacturer.

To measure the $r_{2/1}(\lambda)$ ratio, we used the
optical bench measurement above \SI{450}{\nano\meter} and the CTIO on-sky
measurement from~\cite{holo} below.

In order to measure wavelengths bluer than $<\SI{400}{\nano\meter}$, we extrapolated
the $r_{2/1}(\lambda)$ function with an exponential model
$C\exp{\left[-(\lambda-\lambda_0)/\tau\right]}$ with three free parameters $C$, $\lambda_0$,
and $\tau$ fitted on the laboratory data.

The first-order efficiency curve and the $r_{2/1}(\lambda)$ curve for the blazed Thorlabs 300 lines/mm grating are
represented in Figure~\ref{fig:thorlabs} and were used in \Spectractor when spectra taken with this grating were measured (e.g. in
Figure~\ref{fig:ffm_data} left).

\subsection{Analysis of a photometric night}


To measure the CTIO \SI{0.9}{\meter} telescope transmission, we made use of a set of spectra
acquired at different airmasses during a night with stable photometric
condition. The multiplicity of the airmass conditions and the hypothesis that
the atmospheric transmission spectrum only varies with the quantity of
atmosphere between the source and the observer allowed us to factorise the
atmospheric transmission as an airmass-dependent term and the
instrumental transmission term.

In order to distinguish between the average spectrum of the atmospheric
transmission and the instrumental transmission spectrum, we performed the fit
over the available $N_s$ spectra by simulating the atmospheric transmission with
Libradtran\footnote{\url{http://www.libradtran.org}}
\citep{libradtran2005,libradtran2016} jointly with $N_i$ arbitrary coefficients
to sample the $T_{\mathrm{inst}, 1}(\lambda)$ curve.

\begin{figure}[!h]
\begin{center}
\includegraphics[width=\columnwidth]{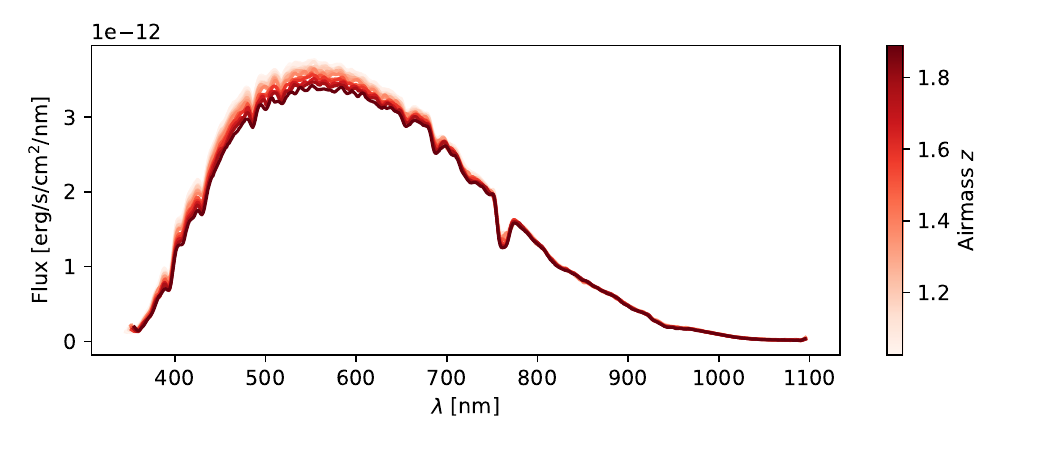}\hfill
\includegraphics[width=\columnwidth]{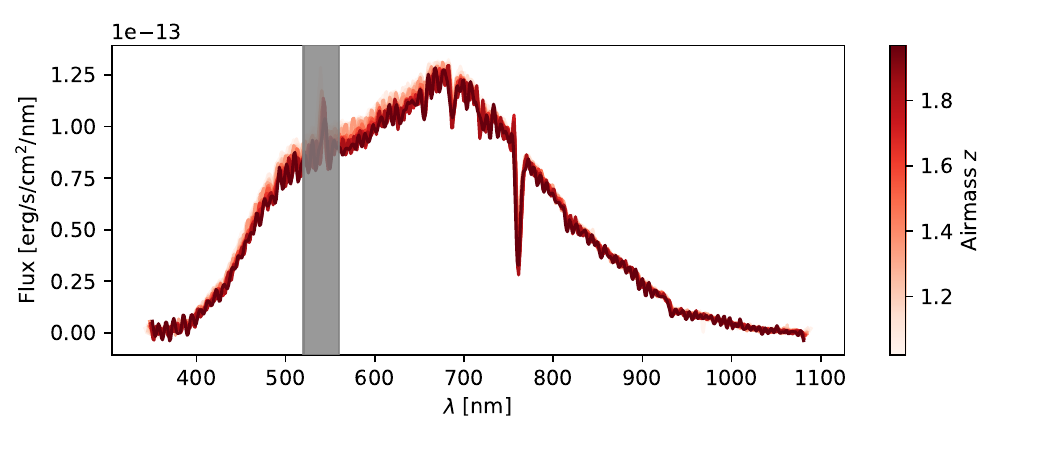}
\end{center}
\caption[] 
{Spectra from the CALSPEC star HD111980 acquired during the night of 2017 May 30 at CTIO, which was  assumed to be photometric, with the blazed Thorlabs grating with 300 lines/mm (top) and an amplitude hologram with 350 lines/mm (bottom). The curves are coloured according to the acquisition airmass $z$. In the amplitude hologram, a field star creates a spike around \SI{530}{\nano\meter}.}
\label{fig:photometric_night_spectra}
\end{figure}

At CTIO on 2017 May 30, the night presented very stable conditions according
to the {in situ} meteorological measurements of temperature, pressure,
humidity, and also a stable seeing around 0.8".
We analysed the spectra of CALPSEC star HD111980 acquired with the Thorlabs and
holographic dispersers under the hypothesis that this night was photometric. The observations cover an airmass range from 1 to 2 (see
Figure~\ref{fig:photometric_night_spectra}).

For each of the dispersers, $N_s$ spectra were acquired and extracted. They
were averaged in \SI{3}{\nano\meter} bins, for which the instrumental transmission is assumed to be very
smooth at this scale.  The main atmospheric and hydrogen absorption lines were masked in this process.

\subsubsection{Analysis of a photometric night in which the instrumental transmission was extracted }\label{sec:ctio_transmission}

{We present here how we estimated the telescope transmission from a photometric night data set and the knowledge of the laboratory measurement of the Thorlabs grating. Basically, we fitted} a telescope transmission model and one atmospheric model on the
collection of spectra extracted from the data collected with this disperser.

From  \SI{300}{\nano\meter} to  \SI{1100}{\nano\meter}, we simultaneously fitted the $N_s=20$ good blazed grating
spectra observed with the model from equation~\eqref{eq:Sp} for $p=1$,
\begin{equation}
S_1(\lambda)  = T_{\text{inst}, 1}(\lambda) \, T_{\text{atm}}(\lambda | \vec{P}_a) \, S_{*}(\lambda),
\end{equation}
where $S_*(\lambda)$ is the binned SED of the CALSPEC star, and $T_{\mathrm{inst},
  1}(\lambda)$ is a vector of $N_i =250$ free linear amplitude parameters.

The Libradtran atmospheric transmission simulation $T_{\mathrm{atm}}(\lambda)$
uses the {in situ} pressure, temperature, and airmass given in each
exposure metadata. In addition, three common parameters $P_a$ were fitted:
\begin{itemize}
\item the precipitable water vapour (PWV; in mm);
\item the ozone quantity (in dobson db);
\item the vertical aerosol optical depth (VAOD).
\end{itemize}

In order to account for a possible small grey variation in the atmospheric
transmission, each spectrum was weighted by a grey factor $A_1^{(n)}$, with
their average constrained to one,
\begin{equation}
\left\langle A_1^{(n)}\right\rangle_n=1.
\end{equation}

The $\chi^2$ we need to minimise thus reads
\begin{equation}
\chi^2 = \sum_{n=1}^{N_s} \left[\vec D_n -A_1^{(n)}S_1(\lambda)\right]^T \mathbf{C}_n^{-1} \left[\vec D_n -A_1^{(n)}S_1(\lambda)\right],
\end{equation}
where $\vec D_n$ is the data vector for spectrum number $n$, and $C_n$ is its
covariance matrix estimated by the \Spectractor extraction pipeline. The
$A_1^{(n)}$ and $P_a$ parameters were fitted jointly via a Gauss-Newton
descent and come with their covariance matrix, while the $T_{\mathrm{inst},
  1}(\lambda)$ linear parameters were computed analytically via the usual algebra
at each descent step. Because the spectrum of the star is assumed to be known, no
regularisation is needed. Because the instrumental transmission is assumed to be smooth,
the descent was repeated with a $5\sigma$ clipping to remove outliers.

This procedure has been tested on simulations, and we verified that it recovered
the injected parameters for instrumental transmission, grey factors, and
atmospheric quantities within the uncertainty ranges. 

\begin{figure}[!h]
\begin{center}
\includegraphics[width=\columnwidth]{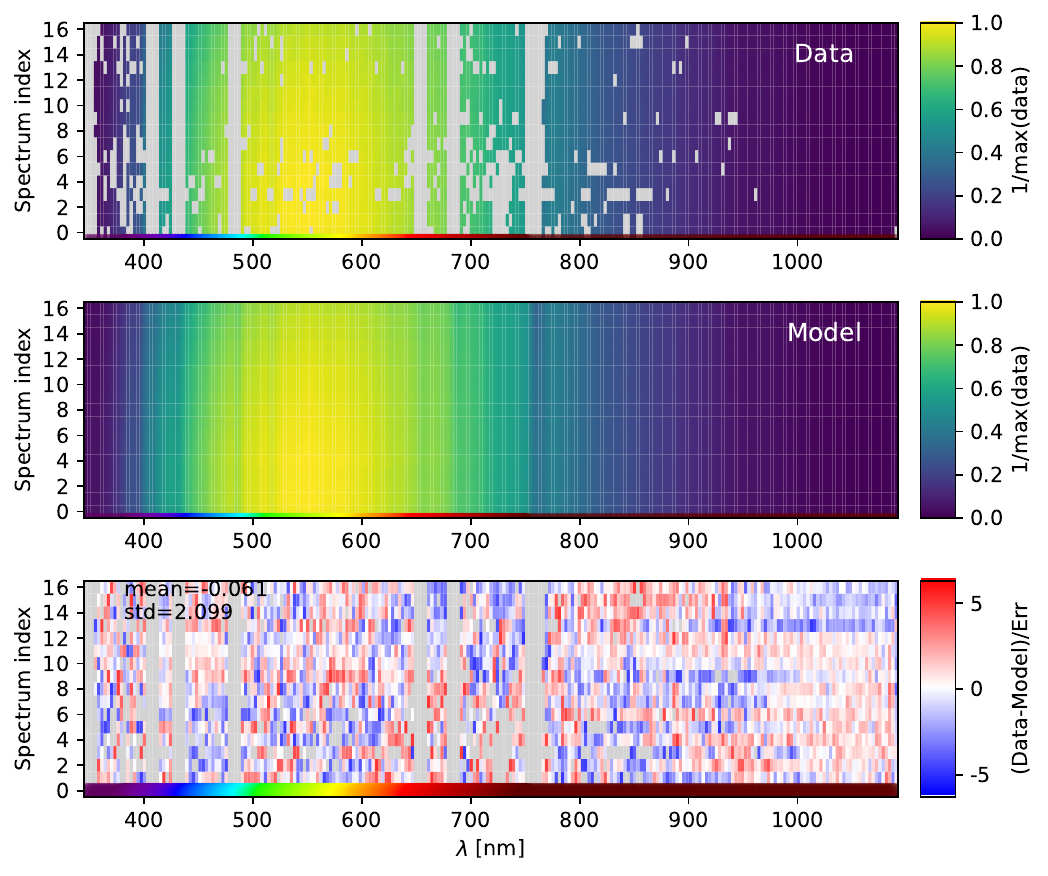}
\end{center}
\caption[] 
{Multispectra fit of a CTIO photometric night using the blazed grating with 300 lines/mm. {Top:} $D_n$ data spectra binned in \SI{3}{\nano\meter} intervals, indexed vertically by their index $n$ and with a coloured amplitude. Masked regions are shown in grey. {Middle:} Best-fitting spectrum models $A_1^{(n)}S_1(\lambda)$  indexed vertically by their index $n$. {Bottom:} Residual map.}
\label{fig:multispectra_thor300}
\end{figure}

The results obtained on the CTIO data are presented in
Figure~\ref{fig:multispectra_thor300}. Approximately 15\% of the 5000 spectral data
points are masked, either because they are close to a spectral line, or
because they are $5\sigma$ outliers. The residuals are structured below the
$2\sigma$ level in the red part of the spectra, either because of an incorrect
PSF model for the redder wavelengths due to defocusing, or because of PWV
variations in the atmosphere, as hinted by the spectra, vertically ordered in
time.

\begin{figure}[!h]
\begin{center}
\includegraphics[width=\columnwidth]{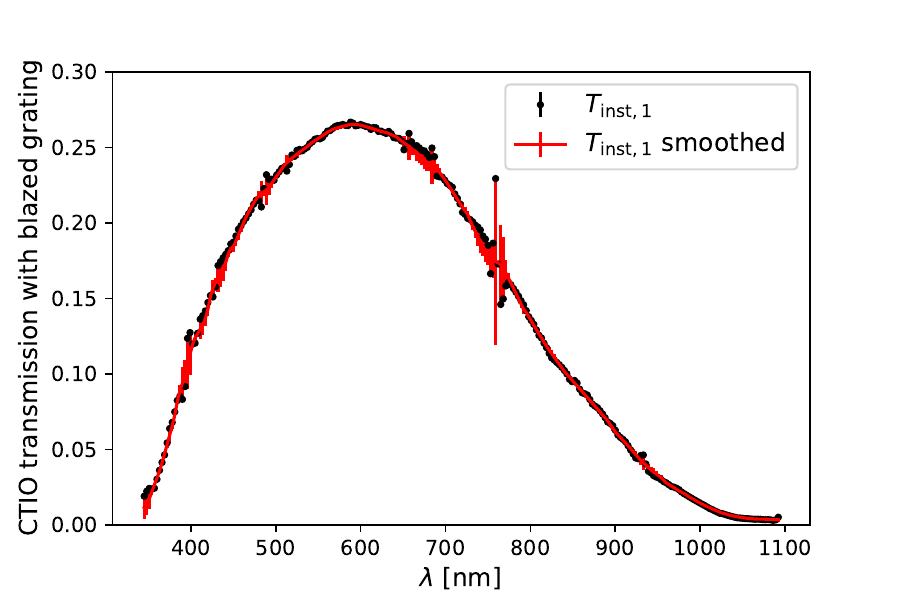}
\end{center}
\caption[] 
{Measured $T_{\mathrm{inst}, 1}(\lambda)$ curve for the CTIO \SI{0.9}{\meter} telescope equipped with a Thorlabs 300 lines/mm blazed grating (black points) from a photometric night. The red curve is a smoothing using a Savitzky–Golay filter of order 1 and a window size 17. }
\label{fig:CTIO_transmission}
\end{figure}

The best $T_{\mathrm{inst}, 1}(\lambda)$ solution that we extracted is
presented in Figure~\ref{fig:CTIO_transmission}. The black points represent the
raw fitted $T_{\mathrm{inst}, 1}(\lambda)$ vector,
and the red curve is smoothed with a Savitzky–Golay filter of order 1 and
window size 17. The error bars result from the combination of raw uncertainties from the fit plus the difference between the smoothed curve and the scattered raw black points. This leads to larger uncertainties for the instrumental throughput where the spectra were masked, around the main absorption lines. 

The transmission curve presents the expected decreases due to the loss of efficiency of
the CCD. Given the measurement of the blazed Thorlabs grating at the laboratory
(Figure~\ref{fig:thorlabs}), we extracted the CTIO \SI{0.9}{\meter} instrumental throughput
from the $T_{\mathrm{inst}, 1}(\lambda)$ smoothed curve.

This fills the lack in {a priori} knowledge of the telescope throughput to
inform our forward model, and it is used in the  following
analysis. 
We obtained the telescope throughput by dividing the fitted instrumental
transmission by the first-order efficiency of the blazed Thorlabs grating. The atmospheric transmission results are detailed in
Section~\ref{sec:atm_20170530}.

A more accurate estimate of the instrumental transmission would need more data,
both to inform a better PSF model, and to constrain the
atmospheric transmission variations better. Because the available data were limited,
our goal was to illustrate that a forward-model approach can be adjusted to
gain more information about the different components of the model. 

We find it also noteworthy that the procedure is symmetric with respect to
atmospheric transmission and telescope transmission: The need for the Libradtran
model as {a priori} to constrain the atmospheric transmission shape could be
replaced by the {a priori} measurement of the telescope transmission.

\subsubsection{Analysis of a photometric night in which amplitude hologram transmissions were obtained}\label{sec:holoamag}

The next step needed to further inform our forward model in order to use the
best-quality data to constrain the atmospheric transmission was to estimate the
holographic disperser transmission. If this transmission is known,  the data gathered with this disperser can be used, and we can take advantage of the
fact that its PSF can be fairly well modelled with a Moffat.

In order to obtain the hologram transmission, we used the same procedure as described
for the Thorlabs disperser above on data that were collected during the same photometric
night, but with the holographic disperser. 
However, the ratio $r_{2/1}(\lambda)$ is still a prior information needed for the full forward model. For the holographic disperser, we built $r_{2/1}(\lambda)$ using the interpolated on-sky data presented in \cite{holo} Figure 21 alone.

\begin{figure}[!h]
\begin{center}
\includegraphics[width=\columnwidth]{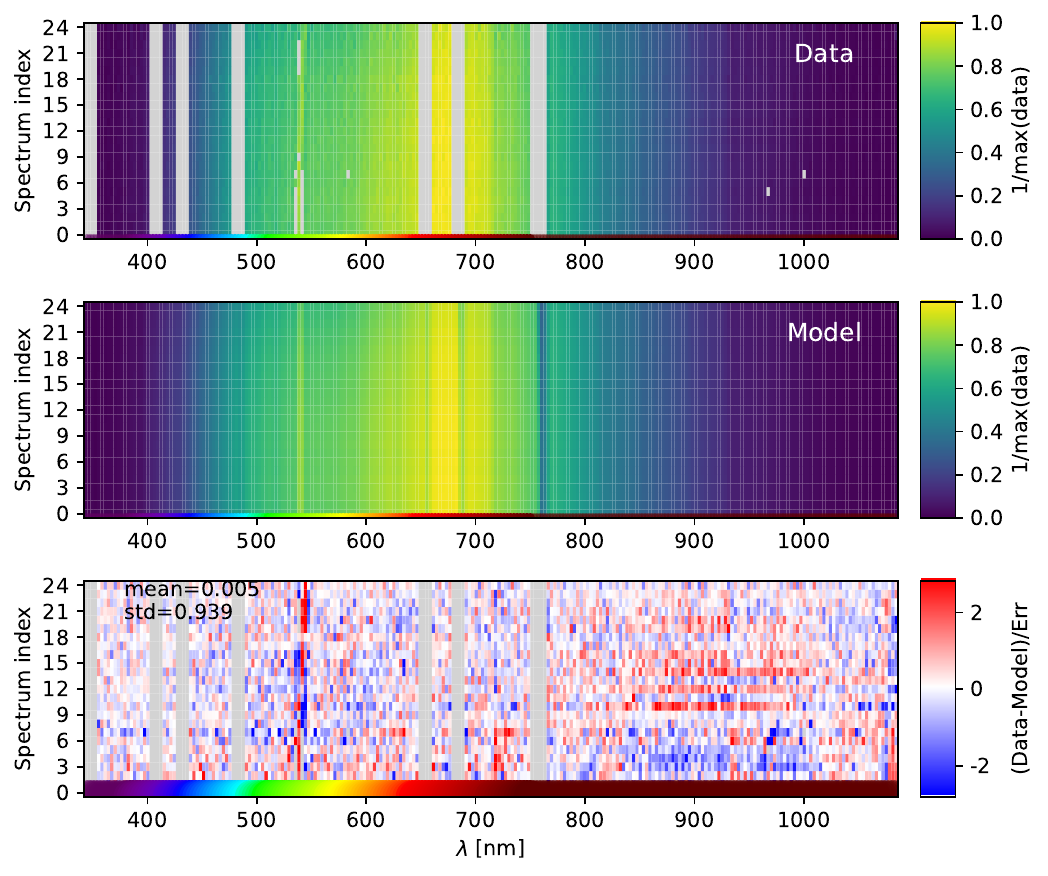}
\end{center}
\caption[] 
{Multispectra fit of a CTIO photometric night using the amplitude hologram with 350 lines/mm. {Top:} $D_n$ data spectra binned in \SI{3}{\nano\meter} intervals, indexed vertically by their index $n$ and with a coloured amplitude. Masked regions are shown in grey. {Middle:} Best-fitting spectrum models $A_1^{(n)}S_1(\lambda)$  indexed vertically by their index $n$. {Bottom:} Residual map.}
\label{fig:multispectra_holoamag}
\end{figure}

The $N_s=27$ spectra are presented in Figure~\ref{fig:photometric_night_spectra},
and the results are presented in
Figure~\ref{fig:multispectra_holoamag}. Approximately 0.5\% of the 6723 data
points are masked. The residuals are below $3\sigma$. A deep absorption feature
is visible around the water-absorption band at \SI{950}{\nano\meter}.

\begin{figure}[!h]
\begin{center}
\includegraphics[width=\columnwidth]{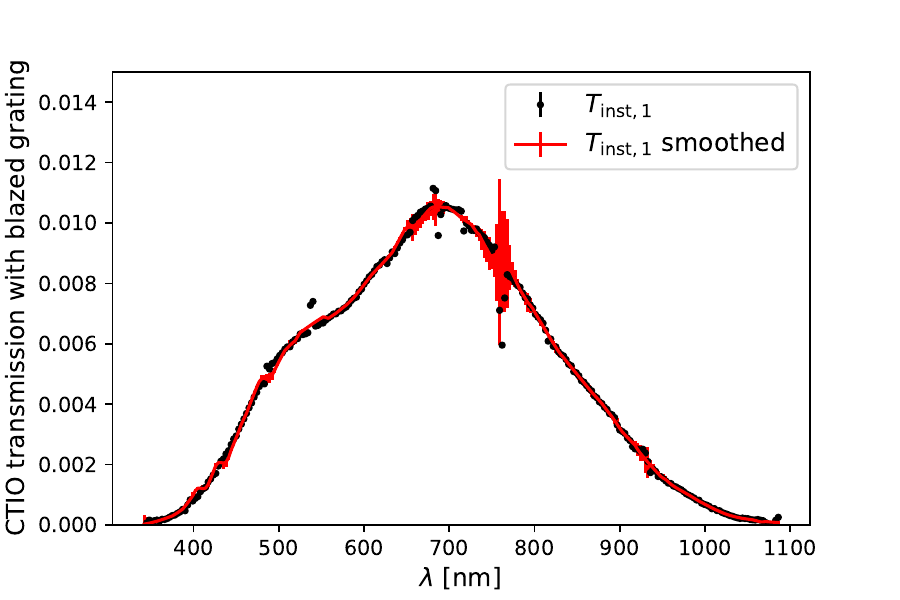}
\end{center}
\caption[] 
{Measured $T_{\mathrm{inst}, 1}(\lambda)$ curve for the CTIO \SI{0.9}{\meter} telescope equipped with an amplitude holographic grating of about 350 lines/mm (black points) from a photometric night. The red curve shows a smoothing using a Savitzky–Golay filter of order 1 and a window size 11.}
\label{fig:holoamg_transmission}
\end{figure}

As in the previous section, from the $T_{\mathrm{inst}, 1}(\lambda)$ best
fit (see Figure~\ref{fig:holoamg_transmission}), we deduced the transmission of
the first diffraction order for the holographic disperser, using the CTIO \SI{0.9}{\meter}
telescope transmission curve determined previously. 
We are well aware of the systematic errors present in these results, and we stress
that they are presented here to illustrate that the forward model approach we
implemented can be used when {a priori} information about crucial components
of the model is lacking.

\subsubsection{Analysis of a photometric night in which atmospheric parameters were extracted}\label{sec:atm_20170530}



In addition to the instrumental transmissions of both dispersers, the procedures
above also yield the parameters describing the mean atmospheric transmission of
the night. Under the assumption that the night was photometric,
these results are presented in Table~\ref{tab:photometric_night_fit}.

\begin{table}[!ht]
    \centering
    \caption[]{Atmospheric parameters $P_a$ fitted for the same photometric night of 2017 May 30, observing CALSPEC star HD111980 with two different dispersers.}
\label{tab:photometric_night_fit}
\begin{center}
\resizebox{\hsize}{!}{\begin{tabular}{ccccc} \hline \hline \\ [-1ex]
 Disperser  & aerosols & ozone & PWV & $\chi^2_{\mathrm{red}}$  \\  [1ex] 
 & VAOD & db & mm &   \\  [1ex]\hline \\ [-1ex]
Thorlabs  & $0.020\pm 0.001$  & $314\pm 2$ & $2.00\pm 0.04$ & 8.3 \\ [1ex]  \hline \\ [-1ex]
Ampl. holo. &  $0.016\pm 0.003$ & $284\pm 9$ &  $2.2\pm 0.1$  & 2.4\\ [1ex]  \hline\hline \\ [-1ex]
MERRA-2 & 0.017 & 265 & 4 -- 5\\ [1ex]  \hline \\ [-1ex]
\end{tabular}}
\tablefoot{Last line gives the MERRA-2 value measured in a \SI{60}{\kilo\meter} wide cell around CTIO.}
\end{center}
\end{table}

The rather low value of the reduced $\chi^2$ for the amplitude hologram
illustrates the focusing properties of this disperser, which allow us to describe
its PSF quite accurately with a simple 2D Moffat. Quantities obtained from the blazed Thorlabs grating data show lower statistical uncertainties than amplitude hologram data because their {S/N} is much higher (because its transmission is much higher). However, they certainly show higher unevaluated PSF systematics than the hologram measurements. The difference between the two estimates of the  atmospheric transmission in Table~\ref{tab:photometric_night_fit} leads to variations in synthesised broad-band magnitudes for the LSST filters of about 8\,mmag in the u, g, and r filters, 3\,mmag in i, 1.5\,mmag in z, and 4\,mmag in y filter for various standard CALSPEC SEDs and supernovae at redshift 0. The millimagnitude accuracy on atmospheric transmission can thus be reached provided that the accuracy of the atmospheric parameters reaches below the difference shown in Table~\ref{tab:photometric_night_fit}: We found that PWV must be fitted with an accuracy better than $\approx\SI{0.05}{\mm}$ to obtain a milli-magnitude accuracy in y band. For VAOD, uncertainties of about 0.001 are required for the u, g, and r bands. For ozone, a 10\,db precision is enough to obtain milli-magnitude precision in r band.

Furthermore, the ozone and VAOD 
parameters we fitted are similar to the estimates of the global meteorological network
MERRA-2\footnote{\url{https://gmao.gsfc.nasa.gov/reanalysis/MERRA-2/}}
\citep{MERRA2} for the CTIO site during that night.
The MERRA-2 PWV value ranges from 4 to 5 mm
during the night of 2017 May 30. As MERRA-2 averages atmospheric quantities in
\SI{60}{\kilo\meter} wide cells, it can be expected that quantities with large
local variations such as water vapour could differ from on-site
measurements. This is even more true for CTIO, which is located at the top of a Chilean
mountain. On the other hand, a high-atmosphere quantity would be expected to
depart less between on-site and satellite measurements. 
We report the MERRA-2 values and compare them to what we extract with our forward model
to illustrate that with a detailed knowledge of the telescope,
the challenging problem of on-site atmospheric transmission measurement can be solved. While
the quoted error bars only propagate the statistical uncertainties and are probably
dominated by systematics, the tentative concordance between the
parameters measured by MERRA-2 and our forward-model results supports the
algorithm we developed. 

\begin{figure}[!h]
\begin{center}
\includegraphics[width=\columnwidth]{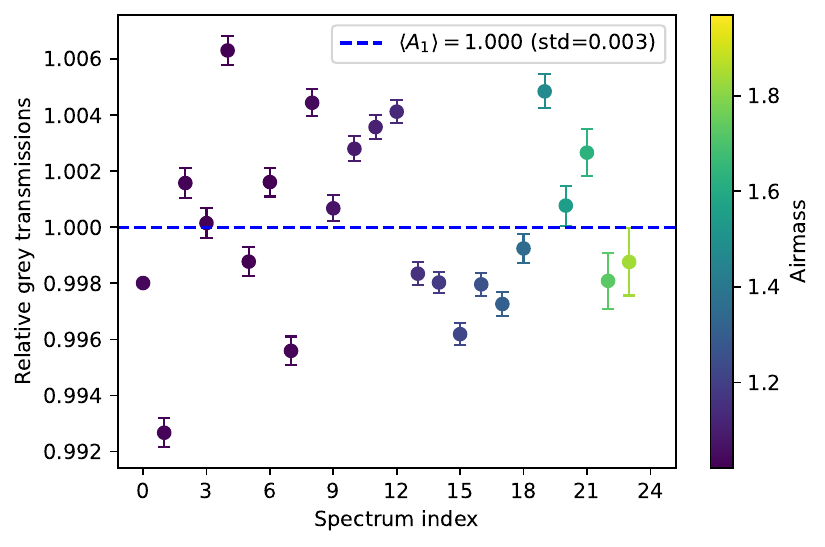}
\end{center}
\caption[] 
{Measured $A_1^{(n)}$ grey absorption factors for the CTIO night of 2017 May 30, in which CALSPEC star HD111980 was observed, for each spectrum (ordered in time).}
\label{fig:A1s}
\end{figure}

\begin{figure}[!h]
\begin{center}
\includegraphics[width=\columnwidth]{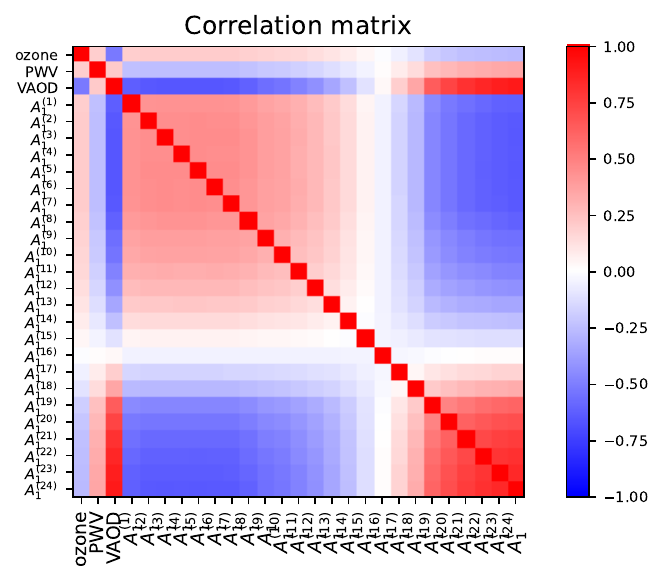}
\end{center}
\caption[] 
{Correlation matrix for the multi-spectrum parameters fitted  for the CTIO night of 2017 May 30, in which CALSPEC star HD111980 was observed.}
\label{fig:multispectra_cov}
\end{figure}

For completeness, we also present the
evolution of the grey parameters $A_1^{(n)}$ through the night in
Figure~\ref{fig:A1s} for the holographic disperser data, together with the final correlation matrix of the fitted
parameters in Figure~\ref{fig:multispectra_cov}. The variation in the 27 $A_1^{(n)}$ factors is lower than 1\%. This supports the first-order approximation of the night as
being photometric and again shows that the procedure is able to improve our
understanding of the data. It also offers venues to improve the model.

Finally, we note that the correlation matrix shows that the VAOD aerosol
parameter is particularly strongly correlated to the spectrum amplitudes. This is expected because this quantity is mostly determined by the spectrum
slope for $\lambda \approx \SI{400}{\nano\meter}$. Any systematic on the amplitude of the spectrum therefore directly affects the estimate of aerosols.

\subsection{Atmospheric forward-model approach}\label{sec:afm}

After illustrating that the forward-model approach can be used to measure the
telescope and disperser transmissions, which yields a set of stellar and
atmospheric transmission spectra, we can proceed one step further. Assuming that
our measurement of these crucial components of the forward model had been done
with enough data to be accurate, we might skip the part in which the spectrum
is extracted and directly fit the atmospheric parameters on the raw
spectrogram.

At the cost of having access to the transmissions described above, when we model the spectrum $S_1(\lambda)$
as the product of a known instrumental transmission $T_{\text{inst}, 1}(\lambda)
$, a Libradtran atmospheric model $ T_{\text{atm}}(\lambda |
\vec{P}_a)$, and the SED $S_{*}(\lambda)$ of a known CALSPEC star, we can describe
any observed spectrogram as
\begin{align}
\vec I(\vec{Z} | \vec A,\vec r_{c}, \vec{P}) & = \tilde{\mathbf{M}}(\vec{Z} | \vec r_{c}, \vec{P} )\, \vec{A} \\
\vec A (\lambda) & = A_1 T_{\text{inst}, 1}(\lambda) \, T_{\text{atm}}(\lambda | \vec{P}_a) \, S_{*}(\lambda),
\end{align}
with $A_1$ a grey factor.

As before, the parameters and their covariance matrix were estimated via a
Gauss-Newton descent by minimising a $\chi^2$ calculated over a single
spectrogram. The fitted parameters were $A_1, A_2, \delta y^{(\mathrm{fit})},
P_a, \DCCD, \alpha$, and all the polynomial coefficients that model the
wavelength dependence of the PSF kernel. Each spectrogram was fitted
independently. We call this a {spectrogram fit}. As a comparison, and as a way to assess the quality of the stellar spectrum
forward model, we also directly fit $S_1(\lambda)$ and the atmospheric transmission on the stellar spectra extracted at this step. We call this a
{spectrum fit}.

\subsubsection{Qualification on simulations}

The direct extraction of atmospheric parameters from a spectrogram was tested
on simulations. The parameters we chose to simulate were the extracted parameters found from fitting all data spectrograms of the photometric night of 2017 May 30, involving the amplitude hologram.
With these parameters, we simulated spectrograms of a CALPSEC star spanning
an airmass $\approx 1$ to $\approx 2$ with the same seeing and atmospheric
conditions as that of our data. 
We arbitrarily fixed the unknown $P_a$ parameters to 300\;db
for ozone, 0.03 for VAOD, and \SI{5}{\milli\meter} for PWV.

\begin{figure*}[!h]
\begin{center}
\textbf{Simulations}\\
\includegraphics[width=0.75\textwidth]{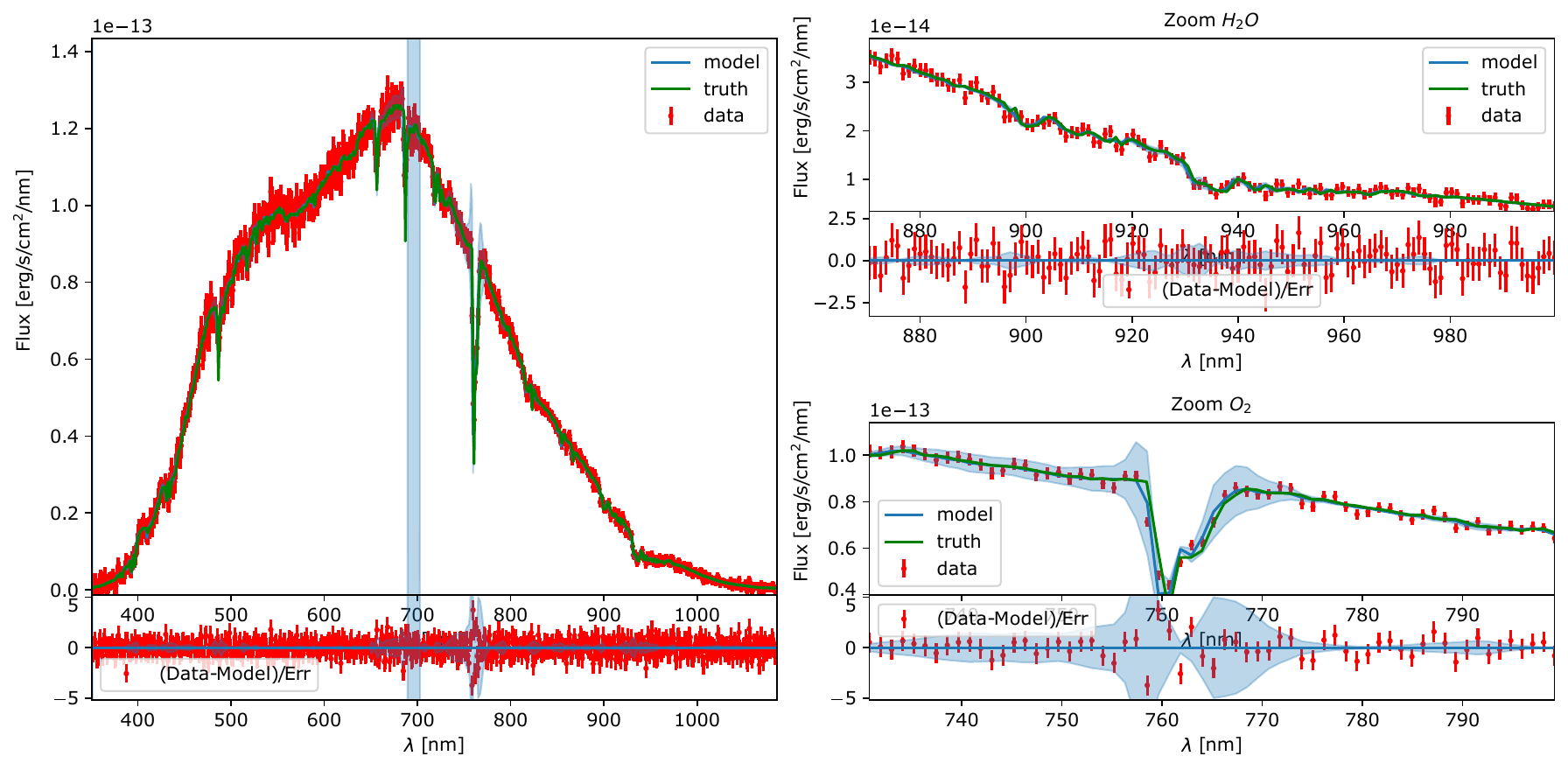}\hfill\includegraphics[width=0.25\textwidth]{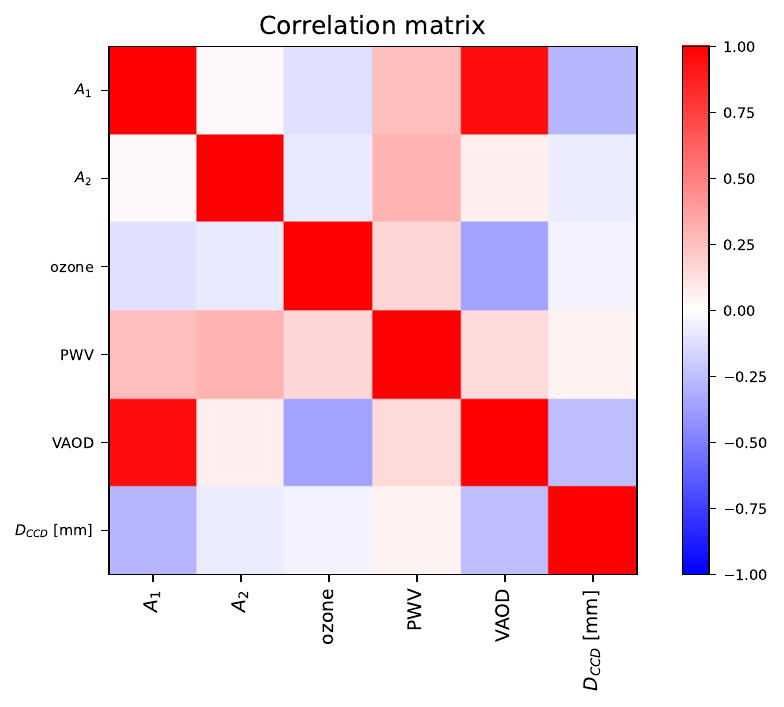}
\end{center}
\caption[] 
{Results from the $S_1(\lambda)$ model fit on the spectrum of CALSPEC star HD111980, extracted from a simulated spectrogram. {Left:} Spectrum data (red) compared with the best-fitting model (blue) and the model uncertainties (light blue band) due to the CTIO telescope transmission uncertainties, the true injected spectrum in the simulation (green), and the residuals (bottom).  {Middle:} Zoom around the dioxygen line at $\SI{762}{\nano\meter}$ and zoom on the $H_2O$ absorption band around $\SI{950}{\nano\meter}$. {Right:} Correlation matrix of the fitted parameters.}
\label{fig:sim_20170530_134_spectrum_fit}
\end{figure*}

\begin{figure*}[!h]
\begin{center}
\textbf{Simulations}\\
\includegraphics[width=0.9\textwidth]{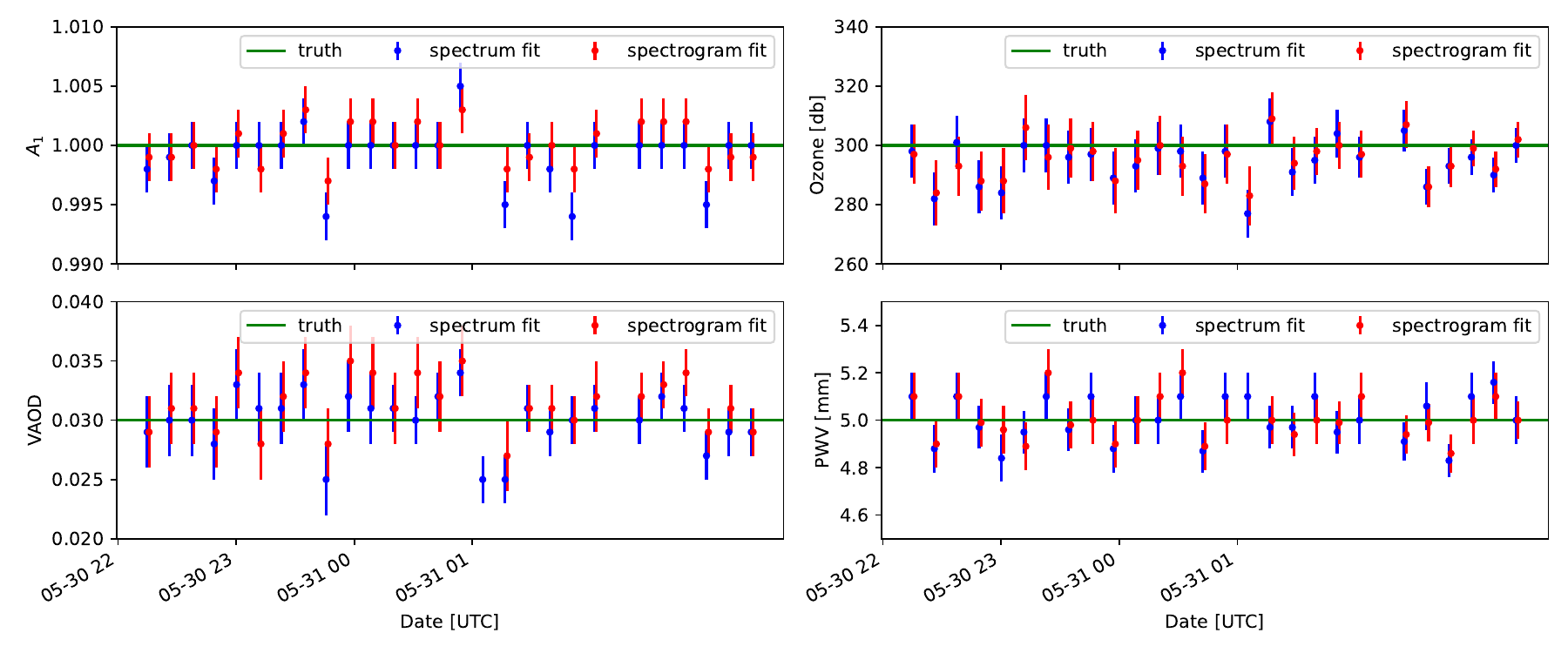}
\end{center}
\caption[] 
{Evolution of the atmospheric parameters in the simulation of the night of 2017 May 30, in which CALSPEC star HD111980 was observed with an amplitude hologram. The blue points show the spectrum fits, while the orange points show the spectrogram fits. The green line gives the true values injected in the simulations.}
\label{fig:night_20170530_sim}
\end{figure*}

The result of the {spectrogram fit} is very similar to what was presented in
Figure~\ref{fig:deconvolution_residuals}. For comparison, we present the result
of the {spectrum fit} of the extracted spectrum from one of the simulated
spectrograms in Figure~\ref{fig:sim_20170530_134_spectrum_fit}. The extracted spectrum (red points), the best-fitting spectrum model
(blue), and the true spectrum (green) all agree within the quoted
uncertainties. 

In addition, the recovered atmospheric values are compatible with the true
injected values within the uncertainties, with a strong correlation between the
grey parameter $A_1$ and the aerosols, as seen before on real data.
The nightly behaviour is presented in Figure~\ref{fig:night_20170530_sim}. All
values agree with the true values for both methods and correctly account for the
variable simulated conditions.

We again note the strong correlation between the VAOD and the $A_1$
parameter. As mentioned before, this is an expected behaviour because aerosols
specifically affect the spectrum slope in the blue and the global spectrum amplitude.

In addition to validating the spectrogram fit, theses results also show that the
forward-model process and all the pipeline steps presented above do not bias the
measurement of the atmospheric parameters.

\subsubsection{Data analysis}

\begin{figure*}[!h]
\begin{center}
\textbf{Data}\\
\includegraphics[width=0.65\textwidth]{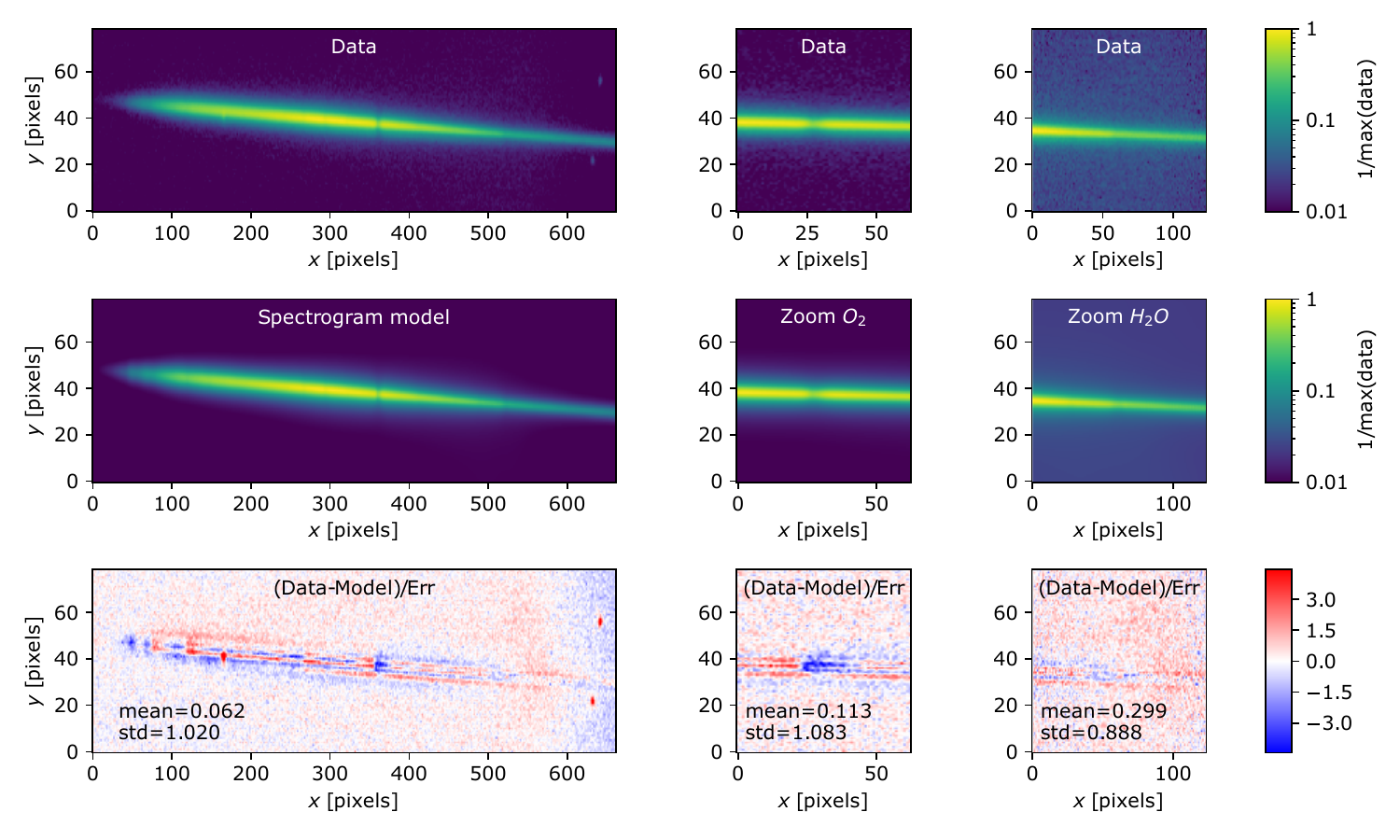}\hfill\includegraphics[width=0.35\textwidth]{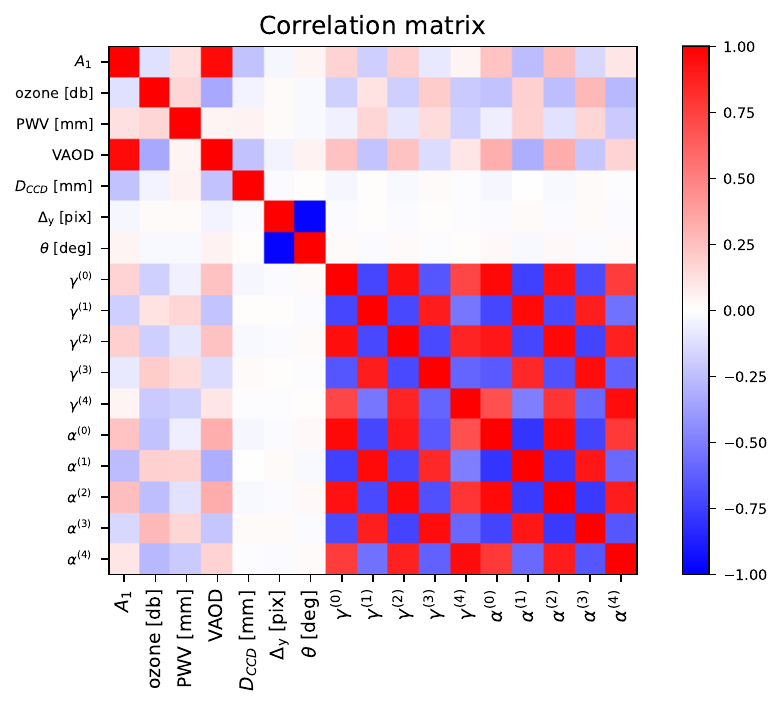}
\end{center}
\caption[] 
{Results from the atmospheric forward model of CALSPEC star HD111980 with a Moffat PSF kernel (the shape parameters evolve as a fourth-order polynomial function), observed with an amplitude hologram with 350 lines/mm. {Left:} Spectrogram data (top), best-fitting spectrogram model (middle), and residuals in units of $\sigma$ (bottom) {Middle:} Zoom around the dioxygen line at $\SI{762}{\nano\meter}$ and zoom on the $H_2O$ absorption band around $\SI{950}{\nano\meter}$. All colour maps are normalised by the maximum of the simulated spectrogram. {Right:} Correlation matrix of the fitted parameters.}
\label{fig:reduc_20170530_134_spectrogram_fit}
\end{figure*}

\begin{figure*}[!h]
\begin{center}
\textbf{Data}\\
\includegraphics[width=0.75\textwidth]{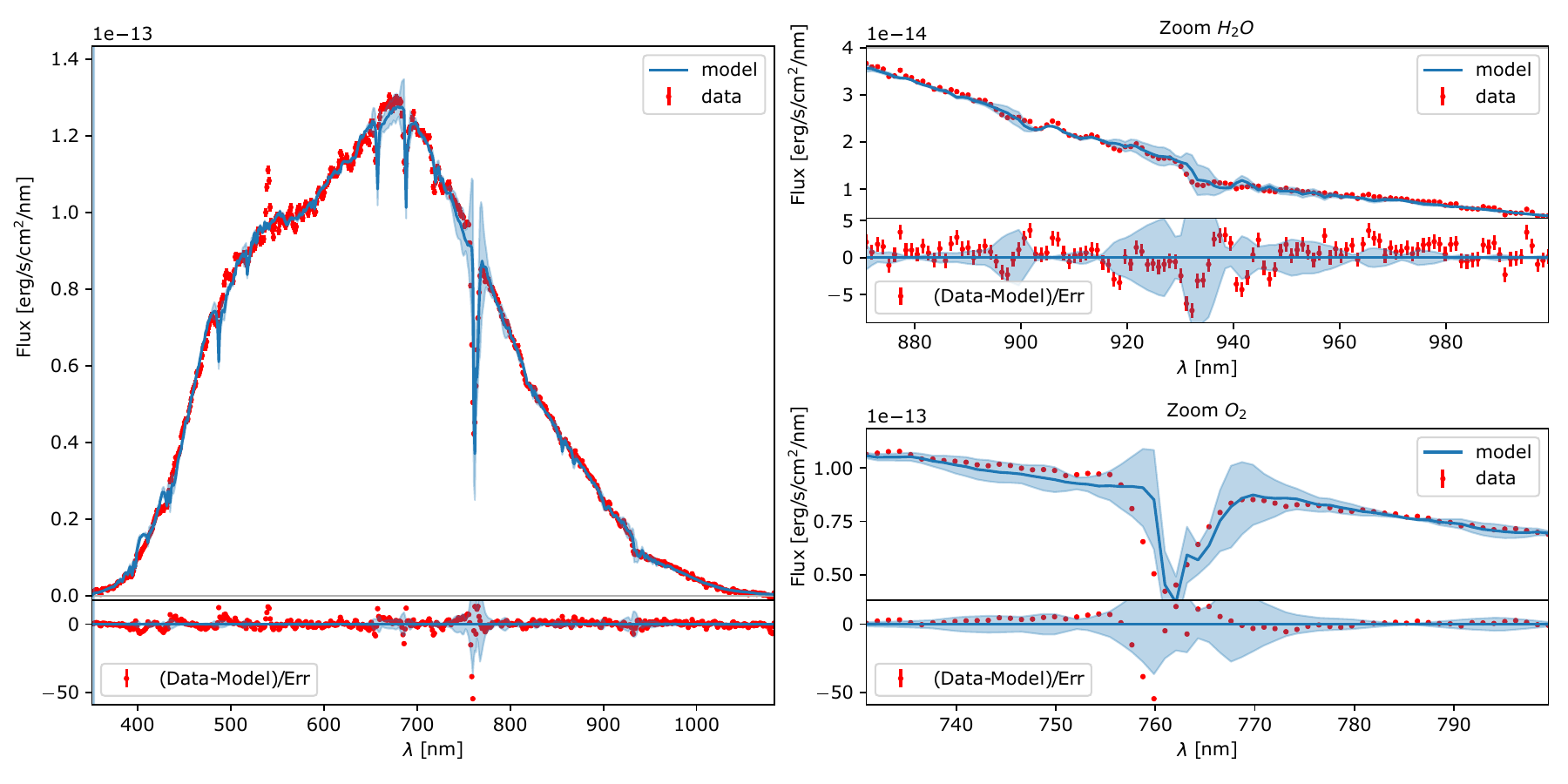}\hfill\includegraphics[width=0.25\textwidth]{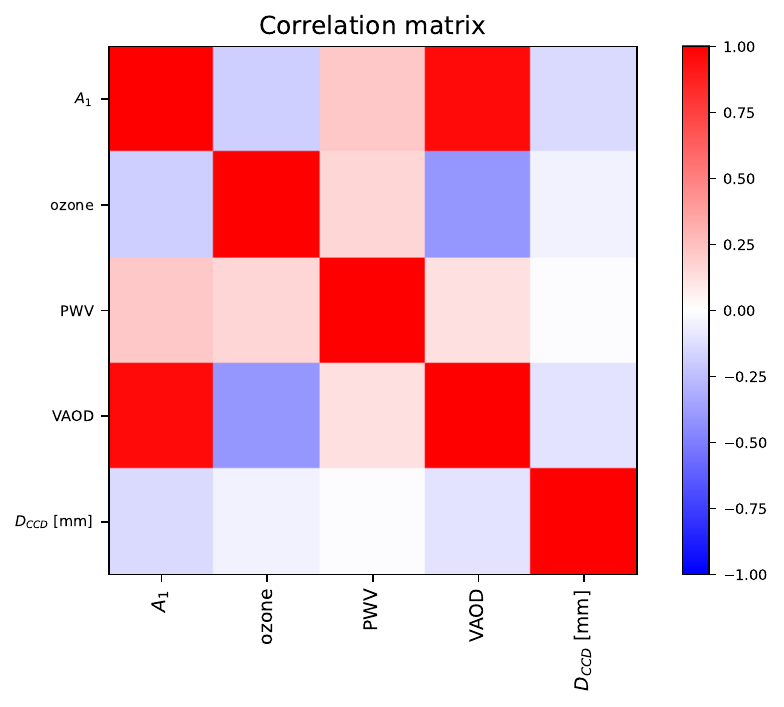}
\end{center}
\caption[] 
{Results from the $S_1(\lambda)$ model fit on the spectrum of CALSPEC star HD111980. {Left:} Spectrum data (red) compared with the best-fitting model (blue, mostly behind the green curve) and the model uncertainties (light blue band) due to the CTIO telescope transmission uncertainties, and the residuals (bottom).  {Middle:} Zoom on the dioxygen line at $\SI{762}{\nano\meter}$ and zoom on the $H_2O$ absorption band around $\SI{950}{\nano\meter}$. {Right:} Correlation matrix of the fitted parameters.}
\label{fig:reduc_20170530_134_spectrum_fit}
\end{figure*}

\begin{figure*}[!h]
\begin{center}
\textbf{Data}\\
\includegraphics[width=0.9\textwidth]{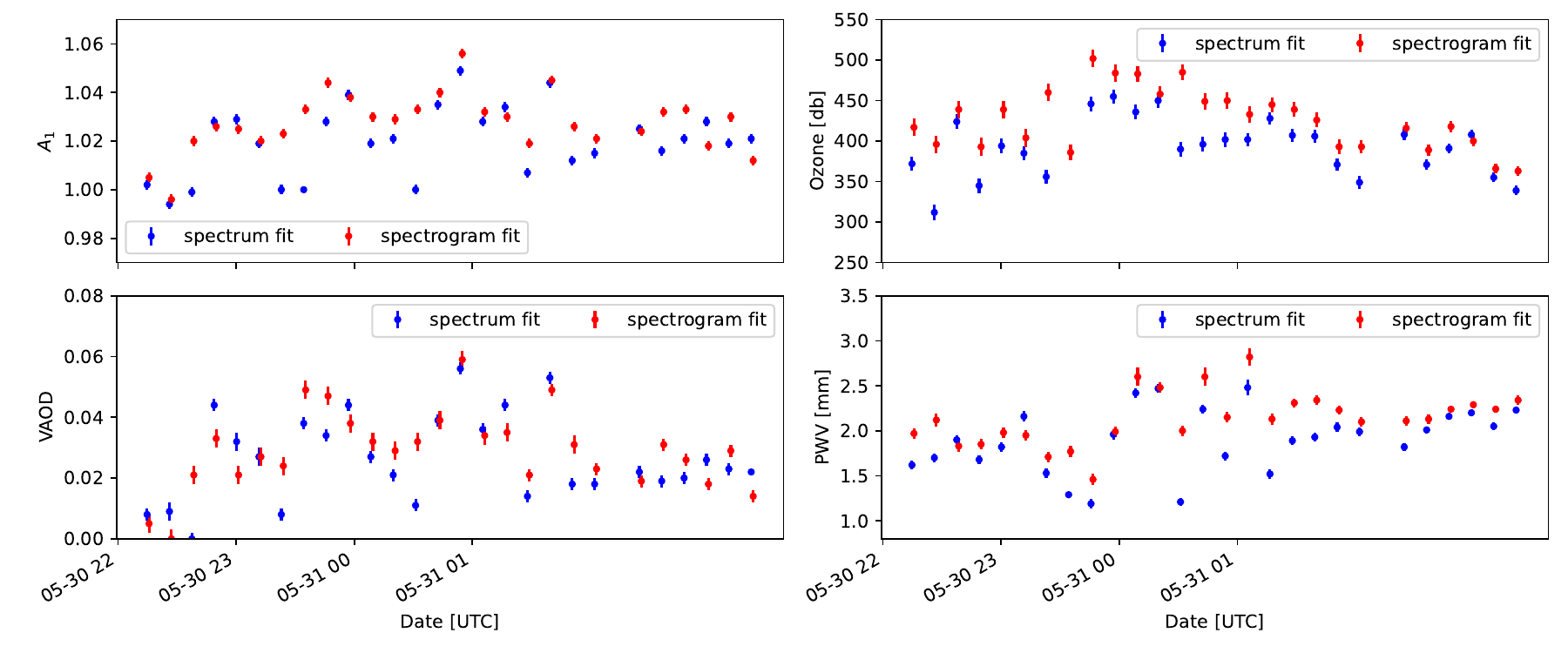}
\end{center}
\caption[] 
{Evolution of the atmospheric parameters as a function of the exposure time during the night of 2017 May 30, in which CALSPEC star HD111980 was observed with an amplitude hologram. The blue points show the spectrum fits, and the orange points show the spectrogram fits.}
\label{fig:night_20170530_reduc}
\end{figure*}

The individual fits of the spectrograms and spectra extracted from CTIO data are
presented in Figures~\ref{fig:reduc_20170530_134_spectrogram_fit},
\ref{fig:reduc_20170530_134_spectrum_fit}, and~\ref{fig:night_20170530_reduc}.

The atmospheric forward modelling fits the data at the $5\sigma$ uncertainty level, but the PSF model imprints structured residuals similarly to what happens in the full forward-model case (Figure~\ref{fig:ffm_data}). This effect is visible throughout the spectrogram and inside the dioxygen absorption line.

The spectrum presented in Figure~\ref{fig:reduc_20170530_134_spectrum_fit}
is globally well fitted by the
$S_1(\lambda)$ model. The fit residuals around the main dioxygen line for data and for the simulation are compatible with the instrumental throughput uncertainties. All spectrograms and spectra of the night show the same
residual patterns.
The two methods yield very similar values for the atmospheric parameters. The values of the {spectrogram fit} are smooth in time, with a visible correlation between VAOD and $A_1$ parameters, while the {spectrum fit} values are shifted and more scattered, probably due to the higher sensitivity of this simpler procedure to outliers such as the field-star contamination of the spectra around \SI{530}{\nano\meter}.

We recall that in the {spectrogram fit}, the raw spectrogram data are directly
fitted with a model that contains the instrumental transmission for diffraction
orders 1 and 2, a Libradtran atmospheric model, and models for the dispersion
relation and PSF kernel.

At this point, the smoothness of the atmospheric parameter curves and the
reasonable values that we obtained (low ozone and a few millimetres of precipitable
water vapour) are as close to reality as can be expected with the
quality and size of the data set at hand.

We again acknowledge that these atmospheric results are affected by systematic
uncertainties and choices (e.g. the circularity of the PSF model, the PSF size of the second
diffraction order, or the blazed grating transmission model) that affect
the absolute value of the quoted parameters. Unfortunately, we do not have enough
data to estimate systematics and proceed. We leave an analysis of the
atmospheric transmission for a future paper, for example, based on the high-quality
data set promised by AuxTel, which is dedicated to measuring atmospheric
transmission at the Rubin Observatory site. With its mirror with a diameter of $\SI{1.2}{\meter}$, its high-quality CCD sensor, and its fast readout electronics, the spectra of CALSPEC standards or stars with similar magnitudes can be acquired at a rate of approximately one or two per minute. Its overall observation strategy is not yet defined, but it has to infer the atmospheric transmission for each LSST pointing, presumably observing a grid of stars (CALSPEC stars or others) to infer directional atmospheric transmissions and interpolate them for the LSST pointing.

\section{Summary and conclusions}\label{sec:conclusion}

Slitless spectrophotometry with forward modelling opens a path towards the
acquisition of spectra with imaging telescopes that can be simply transformed into
spectrographs by inserting a disperser on the light path.

We demonstrated on simulations that building a forward model of a spectrogram
allows for accurate spectrophotometry, with a spectral resolution that only
depends on the width of the PSF along the dispersion axis. The key of the process is a
regularisation algorithm, which is fed with as much as prior information as
possible (regularity of the searched spectrum, PSF parametrisation, ADR, and grating
efficiency). The two key functions of the model are the dispersion function
$\Delta_p(\lambda)$ and the PSF model $\phi(\vec r, \lambda)$, together with the knowledge of the $r_{p/1}(\lambda)$ ratio of diffraction-order transmissions.

We exemplified that this procedure functions on real data, with tentatively very
promising results. We are aware of the limits of the data set in our possession,
and we exemplified that the forward-model procedure can be used to
improve our knowledge of the data, and by doing so, to inform the forward
model.

We can also summarise some of the important lessons learned while implementing
the \Spectractor pipeline as follows.
\begin{itemize}
\item Forward modelling provides a modular approach in which each brick is a
  physical or empirical model that can be changed or improved, depending on the
  data particularities and {S/N}. The residuals indicate how the model can be improved (data rules).
\item   When it is implemented, forward modelling can easily simulate data
  sets to test new algorithms.
\item The second diffraction order is not a contamination, but a signal that helps
  recover the blue part of the first-order spectrum. It should be taken advantage
  of whenever possible. This in particular means that we need to rethink the common wisdom
  of spectroscopy by increasing the efficiency of the grating in the second
  order, and use a field rotator (if available) in order to more easily
  separate the different diffraction orders on the sensor through ADR.
\item The accurate knowledge of the PSF is
  thus crucial and requires dedicated data and an analysis that need to be
  carefully budgeted for. The PSF width sets the spectral resolution and the number of degrees of
  freedom that can be extracted from the data, and thus, decreasing the width is crucial.
 \end{itemize}

Because the main scientific driver for the development of \Spectractor is the
measurement of on-site atmospheric transmission, we pushed our analysis to that point. We showed that our procedure allows us to measure the on-sky
telescope transmission to the direct extraction of atmospheric
parameters from spectrograms.

While the atmospheric parameters are dominated by systematic uncertainties, in
particular, by our partial knowledge of the instrumental transmission and of
the PSF, the comparison with satellite data shows a promising tentative
agreement. Work that goes beyond this paper, which was devoted to presenting the
spectrophotometry method, is a more intensive study of on-site atmosphere
transmission. This requires in particular access to many more data and
a specific detailed analysis to obtain accurate instrumental transmissions and PSF
model. 

We would finally like to acknowledge that many elements of the forward model can
and will be improved as new data become available. In particular, the background might be estimated directly in the forward model, including other diffraction orders, by modelling the contamination of the field-star spectrogram and integrating the chromatic flat-fielding in the model to account for pixel efficiencies, and so. If the flat screen in the observatory can be illuminated with a system of several LEDs or lasers, a cube of wavelength-dependent relative transmission might be obtained that could be directly included in the forward model. Instead of dividing the exposures by one of them, each layer of the PSF cube $\phi_p(\vec r, \lambda)$ could then be multiplied by the flat corresponding to the correct wavelength before the spectrogram model is integrated over $\lambda$. Using the forward model, we can thus solve the exact problem of deflation in spectrophotometry.

These ideas are worth implementing in the algorithm if required by data. On
the other hand, many hardware solutions can also be implemented
to increase the {a priori} knowledge of the instrument and to greatly improve the
forward-model analysis. For instance we showed that holographic dispersers such as
those presented in \cite{holo} improve the focusing of the spectrogram on the
sensor on the whole visible and near-infrared range, which facilitates the PSF modelling. Its narrow width also allows a better spectral resolution. Another improvement would be using a collimated beam projector {(CBP)} \citep{Coughlin2016,Souverin2022} to measure the telescope transmission at the per mill level,
and monitor its evolution with time.

In conclusion, we have presented the theoretical tools, together
with a detailed implementation example to add spectro-photometric ability to an
imager by the insertion of a disperser on the light path. This comes at some
computational cost, which is readily available today, but it also requires either {a
priori} knowledge of the instrument or dedicated data and an analysis to bring the
model to the required level of accuracy.



\begin{acknowledgements}
We are grateful to the CTIO technical staff members
Hernan Tirado and Manuel Hernandez for their help during our tests with the CTIO 0.9\,m telescope. We also thank M\'elanie Chevance for her participation to the observations and Augustin Guyonnet for fruitful advice for the CTIO image reduction. The cost of the observations have been shared by the IJCLab (IN2P3-CNRS) and the Department of Physics and Harvard-Smithsonian Center for Astrophysics, Harvard University. F.B. is part of the FP2M federation (CNRS FR 2036) and of the project Labex MME-DII (ANR11-LBX-0023-01).



This paper has undergone internal review in the LSST Dark Energy Science Collaboration. The internal reviewers were Marc Betoule, Andres Plazas-Malagon and David Rubin.\\

J.Neveu is the primary author of the paper and of the Spectractor software, leading the analysis toward measurement of atmosphere transmission measurement. V.Brémaud implemented the atmospheric differential refraction in the forward model, and wrote the pipeline to determine the CTIO telescope throughput. F.Barret brought the mathematical frame for the regularisation procedure. S.Bongard contributed with general discussions on the spectrum extraction and atmospheric physics. Y.Copin developed the theoretical framework of slitless spectrophotometry and of the forward modelling. S.Dagoret-Campagne and M.Moniez contributed with general discussions on spectrum extraction, the data taking and analysis of CTIO data. L.Le Guillou built the specific system to measure the disperser transmission on the LPNHE optical bench and measured the dispersers. P.Antilogus, C.Juramy and E.Sepulveda built and maintained the LPNHE optical bench.  



The DESC acknowledges ongoing support from the Institut National de 
Physique Nucl\'eaire et de Physique des Particules in France; the 
Science \& Technology Facilities Council in the United Kingdom; and the
Department of Energy, the National Science Foundation, and the LSST 
Corporation in the United States.  DESC uses resources of the IN2P3 
Computing Center (CC-IN2P3--Lyon/Villeurbanne - France) funded by the 
Centre National de la Recherche Scientifique; the National Energy 
Research Scientific Computing Center, a DOE Office of Science User 
Facility supported by the Office of Science of the U.S.\ Department of
Energy under Contract No.\ DE-AC02-05CH11231; STFC DiRAC HPC Facilities, 
funded by UK BEIS National E-infrastructure capital grants; and the UK 
particle physics grid, supported by the GridPP Collaboration.  This 
work was performed in part under DOE Contract DE-AC02-76SF00515.


\end{acknowledgements}

\bibliographystyle{aa}
\bibliography{spectractor}

\begin{thebibliography}{42}
\expandafter\ifx\csname natexlab\endcsname\relax\def\natexlab#1{#1}\fi

\bibitem[{Ade {et~al.}(2016)Ade, Aghanim, Arnaud, Ashdown, Aumont, Baccigalupi,
  Banday, Barreiro, Bartolo, Battaner, Benabed, Benoit-L{\'{e}}vy, Bernard,
  Bersanelli, Bielewicz, Bonaldi, Bonavera, Bond, Borrill, Bouchet, Boulanger,
  Bracco, Burigana, Calabrese, Cardoso, Catalano, Chamballu, Chary, Chiang,
  Christensen, Colombo, Combet, Crill, Curto, Cuttaia, Danese, Davies, Davis,
  de~Bernardis, de~Rosa, de~Zotti, Delabrouille, Delouis, Dickinson, Diego,
  Dole, Donzelli, Dor{\'{e}}, Douspis, Dunkley, Dupac, Efstathiou, Elsner,
  En{\ss}lin, Eriksen, Falgarone, Ferri{\`{e}}re, Finelli, Forni, Frailis,
  Fraisse, Franceschi, Frolov, Galeotta, Galli, Ganga, Ghosh, Giard,
  Gjerl{\o}w, Gonz{\'{a}}lez-Nuevo, G{\'{o}}rski, Gruppuso, Guillet, Hansen,
  Harrison, Helou, Hern{\'{a}}ndez-Monteagudo, Herranz, Hildebrandt, Hivon,
  Hornstrup, Hovest, Huang, Huffenberger, Hurier, Jaffe, Jones, Juvela,
  Keih{\"{a}}nen, Keskitalo, Kisner, Kneissl, Knoche, Kunz, Kurki-Suonio,
  Lamarre, Lasenby, Lattanzi, Lawrence, Leonardi, Le{\'{o}}n-Tavares, Levrier,
  Liguori, Lilje, Linden-V{\o}rnle, L{\'{o}}pez-Caniego, Lubin,
  Mac{\'{i}}as-P{\'{e}}rez, Maffei, Maino, Mandolesi, Maris, Martin,
  Mart{\'{i}}nez-Gonz{\'{a}}lez, Masi, Matarrese, McGehee, Melchiorri,
  Mennella, Migliaccio, Miville-Desch{\^{e}}nes, Moneti, Montier, Morgante,
  Mortlock, Munshi, Murphy, Naselsky, Nati, Natoli, Novikov, Novikov,
  Oppermann, Oxborrow, Pagano, Pajot, Paoletti, Pasian, Perdereau, Pettorino,
  Piacentini, Piat, Pierpaoli, Plaszczynski, Pointecouteau, Polenta, Ponthieu,
  Pratt, Prunet, Puget, Rachen, Reach, Rebolo, Reinecke, Remazeilles, Renault,
  Renzi, Ristorcelli, Rocha, Rosset, Rossetti, Roudier,
  Rubi{\~{n}}o-Mart{\'{i}}n, Rusholme, Sandri, Santos, Savelainen, Savini,
  Scott, Serra, Soler, Stolyarov, Sudiwala, Sunyaev, Suur-Uski, Sygnet, Tauber,
  Terenzi, Toffolatti, Tomasi, Tristram, Tucci, Umana, Valenziano, Valiviita,
  {Van Tent}, Vielva, Villa, Wade, Wandelt, Wehus, Yvon, Zacchei, \&
  Zonca}]{PlanckFilament}
Ade, P. A.~R., Aghanim, N., Arnaud, M., {et~al.} 2016, Astronomy {\&}
  Astrophysics, 586, A141

\bibitem[{{Bertin} \& {Arnouts}(1996)}]{1996A&AS..117..393B}
{Bertin}, E. \& {Arnouts}, S. 1996, \aaps, 117, 393

\bibitem[{Betoule {et~al.}(2013)Betoule, Marriner, Regnault, Cuillandre,
  Astier, Guy, Balland, {El Hage}, Hardin, Kessler, {Le Guillou}, Mosher, Pain,
  Rocci, Sako, \& Schahmaneche}]{Betoule2013}
Betoule, M., Marriner, J., Regnault, N., {et~al.} 2013, Astronomy \&
  Astrophysics, 552, A124

\bibitem[{{Betoule, M.} {et~al.}(2014){Betoule, M.}, {Kessler, R.}, {Guy, J.},
  {Mosher, J.}, {Hardin, D.}, {Biswas, R.}, {Astier, P.}, {El-Hage, P.},
  {Konig, M.}, {Kuhlmann, S.}, {Marriner, J.}, {Pain, R.}, {Regnault, N.},
  {Balland, C.}, {Bassett, B. A.}, {Brown, P. J.}, {Campbell, H.}, {Carlberg,
  R. G.}, {Cellier-Holzem, F.}, {Cinabro, D.}, {Conley, A.}, {D\'{}Andrea, C.
  B.}, {DePoy, D. L.}, {Doi, M.}, {Ellis, R. S.}, {Fabbro, S.}, {Filippenko, A.
  V.}, {Foley, R. J.}, {Frieman, J. A.}, {Fouchez, D.}, {Galbany, L.}, {Goobar,
  A.}, {Gupta, R. R.}, {Hill, G. J.}, {Hlozek, R.}, {Hogan, C. J.}, {Hook, I.
  M.}, {Howell, D. A.}, {Jha, S. W.}, {Le Guillou, L.}, {Leloudas, G.},
  {Lidman, C.}, {Marshall, J. L.}, {M\"oller, A.}, {Mour\~ao, A. M.}, {Neveu,
  J.}, {Nichol, R.}, {Olmstead, M. D.}, {Palanque-Delabrouille, N.},
  {Perlmutter, S.}, {Prieto, J. L.}, {Pritchet, C. J.}, {Richmond, M.}, {Riess,
  A. G.}, {Ruhlmann-Kleider, V.}, {Sako, M.}, {Schahmaneche, K.}, {Schneider,
  D. P.}, {Smith, M.}, {Sollerman, J.}, {Sullivan, M.}, {Walton, N. A.}, \&
  {Wheeler, C. J.}}]{Betoule2014}
{Betoule, M.}, {Kessler, R.}, {Guy, J.}, {et~al.} 2014, A\&A, 568, A22

\bibitem[{{Betoule, Marc} {et~al.}(2023){Betoule, Marc}, {Antier, Sarah},
  {Bertin, Emmanuel}, {Blanc, Pierre \'Eric}, {Bongard, S\'ebastien}, {Cohen
  Tanugi, Johann}, {Dagoret-Campagne, Sylvie}, {Feinstein, Fabrice}, {Hardin,
  Delphine}, {Juramy, Claire}, {Le Guillou, Laurent}, {Le Van Suu, Auguste},
  {Moniez, Marc}, {Neveu, J\'er\'emy}, {Nuss, \'Eric}, {Plez, Bertrand},
  {Regnault, Nicolas}, {Sepulveda, Eduardo}, {Sommer, K\'elian}, {Souverin,
  Thierry}, \& {Wang, Xiao Feng}}]{StarDICE_bench}
{Betoule, Marc}, {Antier, Sarah}, {Bertin, Emmanuel}, {et~al.} 2023, A\&A, 670,
  A119

\bibitem[{{Birch} \& {Downs}(1993)}]{1993Metro..30..155B}
{Birch}, K.~P. \& {Downs}, M.~J. 1993, Metrologia, 30, 155

\bibitem[{Birch \& Downs(1994)}]{Birch_1994}
Birch, K.~P. \& Downs, M.~J. 1994, Metrologia, 31, 315

\bibitem[{Bohlin {et~al.}(2014)Bohlin, Gordon, \& Tremblay}]{Bohlin_2014}
Bohlin, R.~C., Gordon, K.~D., \& Tremblay, P.-E. 2014, Publications of the
  Astronomical Society of the Pacific, 126, 711

\bibitem[{{Bohlin} {et~al.}(2020){Bohlin}, {Hubeny}, \& {Rauch}}]{Bohlin_2020}
{Bohlin}, R.~C., {Hubeny}, I., \& {Rauch}, T. 2020, \aj, 160, 21

\bibitem[{{Bolton} \& {Schlegel}(2010)}]{2010PASP..122..248B}
{Bolton}, A.~S. \& {Schlegel}, D.~J. 2010, \pasp, 122, 248

\bibitem[{Bradley {et~al.}(2020)Bradley, Sip{\H o}cz, Robitaille, Tollerud,
  Vin{\'{\i}}cius, Deil, Barbary, Wilson, Busko, G{\"u}nther, Cara, Conseil,
  Bostroem, Droettboom, Bray, Bratholm, Lim, Barentsen, Craig, Pascual, Perren,
  Greco, Donath, de~Val-Borro, Kerzendorf, Bach, Weaver, D'Eugenio, Souchereau,
  \& Ferreira}]{photutils}
Bradley, L., Sip{\H o}cz, B., Robitaille, T., {et~al.} 2020, astropy/photutils:
  1.0.0

\bibitem[{Burke {et~al.}(2010)Burke, Axelrod, Blondin, Claver, Ivezi{\'{c}},
  Jones, Saha, Smith, Smith, \& Stubbs}]{Burke_2010}
Burke, D.~L., Axelrod, T., Blondin, S., {et~al.} 2010, The Astrophysical
  Journal, 720, 811

\bibitem[{Burke {et~al.}(2017)Burke, Rykoff, Allam, Annis, Bechtol, Bernstein,
  Drlica-Wagner, Finley, Gruendl, James, Kent, Kessler, Kuhlmann, Lasker, Li,
  Scolnic, Smith, Tucker, Wester, Yanny, Abbott, Abdalla, Benoit-L{\'{e}}vy,
  Bertin, Rosell, Kind, Carretero, Cunha, D'Andrea, da~Costa, Desai, Diehl,
  Doel, Estrada, Garc{\'{\i}}a-Bellido, Gruen, Gutierrez, Honscheid, Kuehn,
  Kuropatkin, Maia, March, Marshall, Melchior, Menanteau, Miquel, Plazas, Sako,
  Sanchez, Scarpine, Schindler, Sevilla-Noarbe, Smith, Smith, Soares-Santos,
  Sobreira, Suchyta, Tarle, \& and}]{Burke_2017}
Burke, D.~L., Rykoff, E.~S., Allam, S., {et~al.} 2017, The Astronomical
  Journal, 155, 41

\bibitem[{Coughlin {et~al.}(2016)Coughlin, Abbott, Brannon, Claver, Doherty,
  Fisher-Levine, Ingraham, Lupton, Mondrik, \& Stubbs}]{Coughlin2016}
Coughlin, M., Abbott, T. M.~C., Brannon, K., {et~al.} 2016, in Observatory
  Operations: Strategies, Processes, and Systems {VI}, ed. A.~B. Peck, R.~L.
  Seaman, \& C.~R. Benn ({SPIE})

\bibitem[{{Edl{\'e}n}(1966)}]{1966Metro...2...71E}
{Edl{\'e}n}, B. 1966, Metrologia, 2, 71

\bibitem[{Emde {et~al.}(2016)Emde, Buras-Schnell, Kylling, Mayer, Gasteiger,
  Hamann, Kylling, Richter, Pause, Dowling, \& Bugliaro}]{libradtran2016}
Emde, C., Buras-Schnell, R., Kylling, A., {et~al.} 2016, Geoscientific Model
  Development, 9, 1647

\bibitem[{Gelaro {et~al.}(2017)Gelaro, McCarty, Suárez, Todling, Molod,
  Takacs, Randles, Darmenov, Bosilovich, Reichle, Wargan, Coy, Cullather,
  Draper, Akella, Buchard, Conaty, da~Silva, Gu, Kim, Koster, Lucchesi,
  Merkova, Nielsen, Partyka, Pawson, Putman, Rienecker, Schubert, Sienkiewicz,
  \& Zhao}]{MERRA2}
Gelaro, R., McCarty, W., Suárez, M.~J., {et~al.} 2017, Journal of Climate, 30,
  5419

\bibitem[{Golub {et~al.}(1979)Golub, Heath, \& Wahba}]{Golub1979}
Golub, G.~H., Heath, M., \& Wahba, G. 1979, Technometrics, 21, 215

\bibitem[{Hall(1966)}]{Hall:66}
Hall, J.~T. 1966, Appl. Opt., 5, 1051

\bibitem[{Hansen(1992)}]{Hansen1992}
Hansen, P. 1992, SIAM Review, 34, 561

\bibitem[{Hansen(2010)}]{hansen2010discrete}
Hansen, P. 2010, Discrete Inverse Problems: Insight and Algorithms,
  Fundamentals of Algorithms (Society for Industrial and Applied Mathematics)

\bibitem[{Hazenberg(2019)}]{hazenberg:tel-02950846}
Hazenberg, F. 2019, Theses, {Sorbonne Universit{\'e}}

\bibitem[{{Horne}(1986)}]{Horne1986}
{Horne}, K. 1986, \pasp, 98, 609

\bibitem[{Houston \& Rice(2006)}]{Houston_2006}
Houston, J.~M. \& Rice, J.~P. 2006, Metrologia, 43, S31

\bibitem[{{Ingraham} {et~al.}(2020){Ingraham}, {Clements}, {Ribeiro}, {Reuter},
  {Fisher-Levine}, {Hoblitt}, {Lupton}, {Thomas}, {Stubbs}, {Arndt},
  {Callahan}, {Claver}, {Colleoni}, {Corral}, {Doherty}, {Economou}, {Fausti
  Neto}, {Jenness}, {Khine}, {Krughoff}, {Mondrik}, {Menanteau}, {Mills},
  {O'Mullane}, {Reil}, {Rivera}, {Stalder}, {Sebag}, {Shipsey}, {Tighe},
  {Thornton}, {Villalobos}, \& {Wiecha}}]{2020SPIE11452E..0UI}
{Ingraham}, P., {Clements}, A.~W., {Ribeiro}, T., {et~al.} 2020, in Society of
  Photo-Optical Instrumentation Engineers (SPIE) Conference Series, Vol. 11452,
  Society of Photo-Optical Instrumentation Engineers (SPIE) Conference Series,
  114520U

\bibitem[{{Lang} {et~al.}(2010){Lang}, {Hogg}, {Mierle}, {Blanton}, \&
  {Roweis}}]{astrometrynet}
{Lang}, D., {Hogg}, D.~W., {Mierle}, K., {Blanton}, M., \& {Roweis}, S. 2010,
  \aj, 139, 1782

\bibitem[{{Li} {et~al.}(2019){Li}, {Li}, {Lv}, {Duan}, {Haerken}, \&
  {Zhao}}]{2019MNRAS.484.2403L}
{Li}, M., {Li}, G., {Lv}, K., {et~al.} 2019, \mnras, 484, 2403

\bibitem[{Mayer \& Kylling(2005)}]{libradtran2005}
Mayer, B. \& Kylling, A. 2005, Atmospheric Chemistry and Physics, 5, 1855

\bibitem[{Moniez {et~al.}(2021)Moniez, Neveu, Dagoret-Campagne, Gentet,
  Le~Guillou, \& Collaboration}]{holo}
Moniez, M., Neveu, J., Dagoret-Campagne, S., {et~al.} 2021, Monthly Notices of
  the Royal Astronomical Society, 506, 5589

\bibitem[{Murty(1962)}]{Murty:62}
Murty, M. V. R.~K. 1962, J. Opt. Soc. Am., 52, 768

\bibitem[{{Neveu} {et~al.}(2021){Neveu}, {Br{\'e}maud}, {Dagoret-Campagne}, \&
  {Fisher-Levine Merlin}}]{spectractorascl}
{Neveu}, J., {Br{\'e}maud}, V., {Dagoret-Campagne}, S., \& {Fisher-Levine
  Merlin}. 2021, {Spectractor: Spectrum extraction tool for slitless
  spectrophotometry}

\bibitem[{{Outini} \& {Copin}(2020)}]{Outini2019}
{Outini}, M. \& {Copin}, Y. 2020, \aap, 633, A43

\bibitem[{Perlmutter {et~al.}(1999)Perlmutter, Aldering, Goldhaber, Knop,
  Nugent, Castro, Deustua, Fabbro, Goobar, Groom, Hook, Kim, Kim, Lee, Nunes,
  Pain, Pennypacker, Quimby, Lidman, Ellis, Irwin, McMahon, Ruiz‐Lapuente,
  Walton, Schaefer, Boyle, Filippenko, Matheson, Fruchter, Panagia, Newberg,
  Couch, \& Project}]{Perlmutter1999}
Perlmutter, S., Aldering, G., Goldhaber, G., {et~al.} 1999, The Astrophysical
  Journal, 517, 565

\bibitem[{Riess {et~al.}(1998)Riess, Filippenko, Challis, Clocchiatti, Diercks,
  Garnavich, Gilliland, Hogan, Jha, Kirshner, Leibundgut, Phillips, Reiss,
  Schmidt, Schommer, Smith, Spyromilio, Stubbs, Suntzeff, \& Tonry}]{Riess1998}
Riess, A.~G., Filippenko, A.~V., Challis, P., {et~al.} 1998, The Astronomical
  Journal, 116, 1009

\bibitem[{{Robertson}(1986)}]{Robertson1986}
{Robertson}, J.~G. 1986, \pasp, 98, 1220

\bibitem[{Rubin {et~al.}(2022)Rubin, Aldering, Antilogus, Aragon, Bailey,
  Baltay, Bongard, Boone, Buton, Copin, Dixon, Fouchez, Gangler, Gupta, Hayden,
  Hillebrandt, Kim, Kowalski, Kuesters, Leget, Mondon, Nordin, Pain, Pecontal,
  Pereira, Perlmutter, Ponder, Rabinowitz, Rigault, Runge, Saunders, Smadja,
  Suzuki, Tao, Taubenberger, Thomas, Vincenzi, \& Factory}]{Rubin2022}
Rubin, D., Aldering, G., Antilogus, P., {et~al.} 2022, Uniform Recalibration of
  Common Spectrophotometry Standard Stars onto the CALSPEC System using the
  SuperNova Integral Field Spectrograph

\bibitem[{Ryan {et~al.}(2018)Ryan, Casertano, \& Pirzkal}]{Ryan2018}
Ryan, R.~E., Casertano, S., \& Pirzkal, N. 2018, Publications of the
  Astronomical Society of the Pacific, 130, 034501

\bibitem[{Schroeder \& Inc(2000)}]{schroeder2000astronomical}
Schroeder, D. \& Inc, E.~I. 2000, Astronomical Optics, Electronics \&
  Electrical (Elsevier Science)

\bibitem[{Scolnic {et~al.}(2018)Scolnic, Jones, Rest, Pan, Chornock, Foley,
  Huber, Kessler, Narayan, Riess, Rodney, Berger, Brout, Challis, Drout,
  Finkbeiner, Lunnan, Kirshner, Sanders, Schlafly, Smartt, Stubbs, Tonry,
  Wood-Vasey, Foley, Hand, Johnson, Burgett, Chambers, Draper, Hodapp, Kaiser,
  Kudritzki, Magnier, Metcalfe, Bresolin, Gall, Kotak, McCrum, \&
  Smith}]{Scolnic2018}
Scolnic, D.~M., Jones, D.~O., Rest, A., {et~al.} 2018, The Astrophysical
  Journal, 859, 101

\bibitem[{Souverin {et~al.}(2022)Souverin, Neveu, Betoule, Bongard,
  Brownsberger, Cohen-Tanugi, Dagoret-Campagne, Feinstein, Juramy, Guillou,
  Van~Suu, Blanc, Hazenberg, Nuss, Plez, Sepulveda, Sommer, Stubbs, Regnault,
  \& Urbach}]{Souverin2022}
Souverin, T., Neveu, J., Betoule, M., {et~al.} 2022, Measurement of telescope
  transmission using a Collimated Beam Projector

\bibitem[{{The Nearby Supernova Factory} {et~al.}(2013){The Nearby Supernova
  Factory}, {Buton, C.}, {Copin, Y.}, {Aldering, G.}, {Antilogus, P.}, {Aragon,
  C.}, {Bailey, S.}, {Baltay, C.}, {Bongard, S.}, {Canto, A.}, {Cellier-Holzem,
  F.}, {Childress, M.}, {Chotard, N.}, {Fakhouri, H. K.}, {Gangler, E.}, {Guy,
  J.}, {Hsiao, E. Y.}, {Kerschhaggl, M.}, {Kowalski, M.}, {Loken, S.}, {Nugent,
  P.}, {Paech, K.}, {Pain, R.}, {P\'econtal, E.}, {Pereira, R.}, {Perlmutter,
  S.}, {Rabinowitz, D.}, {Rigault, M.}, {Runge, K.}, {Scalzo, R.}, {Smadja,
  G.}, {Tao, C.}, {Thomas, R. C.}, {Weaver, B. A.}, \& {Wu, C.}}]{Buton_2013}
{The Nearby Supernova Factory}, {Buton, C.}, {Copin, Y.}, {et~al.} 2013, A\&A,
  549, A8

\bibitem[{Wahba(1990)}]{wahba1990spline}
Wahba, G. 1990, Spline Models for Observational Data, CBMS-NSF Regional
  Conference Series in Applied Mathematics (Society for Industrial and Applied
  Mathematics (SIAM, 3600 Market Street, Floor 6, Philadelphia, PA 19104))

\end{thebibliography}

\appendix

\section{Astrometry}\label{sec:astrometry}

It is crucial in many regards to accurately anchor the wavelength calibration. For
atmospheric transmission measurement, a small shift of about $\SI{1}{\nm}$ can significantly bias the estimate of
the aerosol parameters. Unfortunately, a shift like this can occur because of the poor determination of the position of the
{zeroth order} of the spectrum, which is usually saturated with long bleeding
spikes. It is difficult to localise it accurately and might not be robust enough to
achieve a centroid determination precision that is better than the pixel scale for every
image in every circumstances.

We thus found it useful to explore how the field stars might be used to set a precise
astrometry using the \texttt{astrometry.net}
library\footnote{\url{http://astrometry.net/}}.

The field star centroids were first extracted from the image using the
\texttt{IRAFStarFinder} method from the \texttt{photutils} library
\citep{photutils}, using a $5\sigma$ clipping above the threshold.

The \texttt{astrometry.net} \texttt{solve-field} function was then called:
Patches of stars were compared to known asterisms to obtain
the precise location of the image on the sky as well as the transformation between
image coordinates and sky coordinates in the form of a World Coordinate System
(WCS) description.

This procedure may not yield subpixel precision for the centroid of the target star,
in particular, if it has a high proper motion. To improve the precision, we
compared the first source catalogue, whose positions were converted into sky
coordinates, with the star positions from the {Gaia} DR2 catalogue, corrected
for their proper motion.

The difference between the 50 brightest field-star coordinates in the two
catalogues was subtracted from the WCS solution in order to lock it on the
{Gaia} catalogue.

We then removed the star with the largest distance to the {Gaia}
catalogue, reapplied \texttt{astrometry.net} followed by the {Gaia}
catalogue centring and repeated this operation ten times. The
astrometric solution in which the distance between the image stars and the
{Gaia} stars was smallest was kept. This procedure minimises the effect of stars with poorly reconstructed centroids (due to saturation
effects) or high proper motions. It ends with a scatter that is evaluated at $\approx 0.15\arcsec$ RMS (Figure~\ref{fig:astrometry_residuals}).

\section{Gauss-Newton minimisation algorithm}\label{sec:newton}

The gradient descent to minimise a $\chi^2$ using the Gauss-Newton algorithm
works as follows. We consider a data set with $N$ data points gathered into a
vector $\vec{D}$ with their uncertainties (correlated or uncorrelated), represented by a
matrix $W$. We wish to model these data with a model
$m(\vec{P})$ depending on a set of parameters $\vec{P}$. For a parameter vector $\vec P$, the
$\chi^2$ is defined as
\begin{equation}
 \chi^2(\vec P) = \left(\vec{M}(\vec{P}) - \vec{D}\right)^T W \left(\vec{M}(\vec{P}) - \vec{D}\right) = \vec{R}^T(\vec{P}) W  \vec{R}(\vec{P}), 
\end{equation}
where $\vec{M}(\vec{P})$ is the vector of the model-predicted values for the
$N$ data points. The vector $\vec R(\vec{P})$ is the residual vector.

In order to find the set of parameters $\hat{\vec{P}}$ that minimises the
$\chi^2$ function, we searched for the zero of the $\chi^2$ gradient that verifies
$\vec{\nabla}_{\vec P} \chi^2(\hat{\vec{P}}) = 0$. The algorithm we used is the
iterative multi-dimensional Gauss-Newton method, which we describe hereafter.

We started the minimisation with a first-guess value for the parameters
$\vec{P}_0$. A Taylor expansion at first order of the $\vec{\nabla}_{\vec P}
\chi^2$ function can be performed around the starting point
$\vec{P}_0$ and gives
\begin{equation}
\vec{\nabla}_{\vec P} \chi^2(\vec{P}_1)\stackrel{\vec P \approx\vec P_0}{\approx}  2 \vec J_0^T \vec W \vec{R}_0 +2\vec J_0^T \vec W \vec J_0 \vec{\delta P_1} + \cdots,
\end{equation}
with $\vec{\delta P_1} = \vec{P}_1- \vec{P}_0$ and $\vec
J_0=\vec\nabla_{\vec P} \vec M(\vec P_0)$ the Jacobian matrix of the model
evaluated at $\vec{P}_0$. For a linear model, that is, a model that can
be written as $m(\vec{P} | \vec{Z}) = \sum_{i=1}^N P_i f(\vec{Z})$, the
$\vec{\nabla}_{\vec P} \chi^2(\vec{P})$ is exactly equal to its first-order
Taylor expansion.

The zero of the function is then approached by solving the equation $
\vec{\nabla}_{\vec P} \chi^2 (\vec{P}_1) = \vec 0$,
\begin{equation}
\vec{\nabla}_{\vec P} \chi^2(\vec{P}_1)  = \vec 0 \Rightarrow \vec{P}_1 = \vec{P}_0 - \left( \vec J_0^T \vec W \vec J_0\right)^{-1}\vec J_0^T \vec W \vec{R}_0.
\end{equation}
Because the approximation comes from the Taylor expansion and because the
numerical accuracy of the Jacobian matrix computation is finite, it is unlikely that the
$\vec{P}_1$ that is found exactly cancels the $\chi^2$ gradient.

We then searched for the $\alpha$ value that minimises the $\chi^2$ function along
the line parametrised by the vector $\alpha_1 \vec{\delta P_1}$ where
$\alpha_1$ is a real number.  The $\vec P_1$ value solution then reads
\begin{equation}
\vec{P}_{1} = \vec{P}_0 - \hat\alpha_1 \left( \vec J_0^T \vec W \vec J_0\right)^{-1}\vec J_0^T \vec W \vec{R}_0.
\end{equation}
The process was iterated $K$ times,
\begin{equation}
 \vec{P}_{k+1} = \vec{P}_k - \hat\alpha_{k+1} \left( \vec J_k^T \vec W \vec J_k\right)^{-1}\vec J_k^T \vec W \vec{R}_k, 
\end{equation}
until a convergence criterion is reached. For example when the value of
$\chi^2(\vec{P}_k)$ or $\vec{P}_k$ evolution with $k$ drops below a certain
threshold.

The best-fitting model is then considered to be parametrised by the
$k$th vector: $\hat{\vec{P}} \approx \vec{P}_k$. The covariance matrix of
the $\vec{P}_k$ parameters is obtained as the Hessian matrix at the
minimum $\chi^2$,
\begin{equation}
 \vec C(\hat{\vec{P}}) =  \left( \vec J_k^T \vec W \vec J_k\right)^{-1}.
\end{equation}

In \Spectractor, we also implemented the possibility to limit the $\vec{P}$
search within given bounds (e.g. we can impose that the amplitudes are all
positive). We found that these bounds help the algorithm to converge.

\section{Second-order penalisation}\label{sec:laplacian_reg}

Regularisation techniques involving the total variation are often used in image analysis for denoising or deconvolution while recovering sharp edges. One way to justify the use of a penalisation with the discretised Laplacian is to understand that it entails an automatic bound on the total variation. We show this below. 

We recall some notations. The complete cost to minimise is  
\begin{align}
\mathcal{E}(\vec{A} | \vec r_c, \vec P)& = \left(\vec{D} - \mathbf{M}\, \vec{A}\right)^T \mathbf{W}
    \left(\vec{D} - \mathbf{M}\,\vec{A} \right) \notag \\ &\ \ \  +  r (\vec A - \vec A_{0} )^T \mathbf{Q} (\vec A - {\vec A}_{0}) \notag \\ 
    & = \chi^2(\vec{A} | \vec r_c, \vec P) + r \chi^2_{\rm pen}(\vec{A} | \vec A_{0}),\\
     \quad \vec A_0 = {\vec{A}}^{(1D)}.
\end{align}
$\vec{D}$ is the data vector, $\mathbf{M}$ is the design matrix, and $\vec A$ is the amplitude vector that gives the spectrogram and that we wish to obtain.
$\mathbf{W}$ is the inverse of the covariance matrix, so that when we have all the true parameters for $\vec{A}$, $\vec r_c$, and $\vec P$, then $\chi^2(\vec{A} | \vec r_c, \vec P)$ is the sum of the squares of the residuals, and thus is the realisation of a random variable following a $\chi^2$ law with $N_xN_y$ degrees of freedom, hence the name of the cost,  $\chi^2$.

The penalisation term $\chi^2_{\rm pen}(\vec{A} | \vec A_{0})=(\vec A - \vec A_{0} )^T \mathbf{Q} (\vec A - {\vec A}_{0})$ is also a quadratic term, and for $\mathbf{Q}=\mathbf{L}^T \mathbf{U}^T \mathbf{U} \mathbf{L}$ with the Laplacian operator $\mathbf{L}=-\mathbf{D}^T\mathbf{D}$,
\begin{align}
\mathbf{L} & = \begin{pmatrix}
-1 & 1 & 0 & 0 & \cdots & 0 & 0\\
1 & -2 & 1 & 0 &\cdots & 0 & 0\\
0 & 1 & -2 & 1 &\cdots & 0 & 0\\
\vdots & \ddots & \ddots & \ddots & \vdots& \vdots \\
0 & 0 & 0 &0 & \cdots & -2 & 1\\ 
0 & 0 & 0 & 0 &\cdots & 1 & -1\\ 
\end{pmatrix}\\
\mathbf{D} & = \begin{pmatrix}
1 & -1 & 0 & 0 & \cdots & 0 & 0\\
 0& 1 & -1 & 0 &\cdots & 0 & 0\\
0 & 0 & 1 & -1 &\cdots & 0 & 0\\
\vdots & \ddots & \ddots & \ddots & \vdots& \vdots \\
0 & 0 & 0 &0 & \cdots & 1 & -1\\ 
0 & 0 & 0 & 0 &\cdots & 0 & 1\\ 
\end{pmatrix},
\end{align}
and $ \mathbf{U}$ is such that 
\begin{equation}\label{eq:Qsimple}
 \mathbf{U} = \begin{pmatrix}
1/\sigma_{A_{1D}^{(1)}} & 0 & \cdots & 0\\
0 & 1/\sigma_{A_{1D}^{(2)}} & \cdots & 0\\
\vdots & \ddots & \ddots & \vdots \\
0 & 0 & \cdots & 1/\sigma_{A_{1D}^{(N_x)}}\\ 
\end{pmatrix}
\end{equation}
we obtain $\chi^2_{\rm pen}(\vec{A} | \vec A_{0})=(\mathbf{UL}(\vec A - \vec A_{0}) )^T  (\mathbf{UL}(\vec A - {\vec A}_{0}))$, and it is the quadratic norm (or the squared Euclidean norm) of the vector $\mathbf{UL}(\vec A - \vec A_{0})$, denoted $\|\mathbf{UL}(\vec A - \vec A_{0})\|_2^2$.

When we interpret $\vec A$ as the discretisation of a continuous spectrogram $a$ by setting $A^{(i)}=a\left(\frac{i}{N_x}\right)$, and when $\sigma$ is a function such that $\sigma\left(\frac{i}{N_x}\right)=\sigma_{A_{1D}^{(i)}}$, a continuous analogue of this term would be a term of the form 
$$\|(a-a_0)''\|_{2,\sigma}^2=\int_0^1\left[(a-a_0)''(x)\right]^2\frac\dx{\sigma^2(x)}$$,
where $a, a_0$ would be functions whose $\vec A, \vec A_0$ are discretisations. As a simple consequence, 
\begin{equation}
\lim_{N_x\to\infty}{N^3_x}\chi^2_{\rm pen}(\vec{A} | \vec A_{0}) =\|(a-a_0)''\|_{2,\sigma}^2
\end{equation}
because {$N_x\mathbf{D}\approx-\frac{d}{dx}$}.

The total variation distance is defined as the (weighted) norm-1 of the gradient operator as a functional term, $$\|(a-a_0)'\|_{1,\sigma}=\int_0^1|(a-a_0)'(x)|\frac\dx{\sigma(x)}.$$
We also note that
\begin{equation}
\lim_{N_x\to\infty}\sum_{i=1}^{N_x}|\mathbf{UD}(\vec{A} -\vec A_{0})| =\|(a-a_0)'\|_{1,\sigma}.
\end{equation}

However, by a simple argument, we can show that the 2-norm of the second derivative controls the 1-norm of the first derivative, so that minimising the former means that we also minimise the latter. In order to prove it, let $f(x)=a(x)-a_0(x)$, we obtain
\begin{align}
f'(x)&=f'(0)+\int_0^xf''(s)\ds=f'(0)+\int_0^xf''(s)\sigma(s)\frac{\ds}{\sigma(s)}\\
&\leqslant f'(0)+ \left(\int_0^x(f''(s))^2\frac{\ds}{\sigma(s)^2}\right)^{1/2}\left(\int_0^x\sigma^2(s)\ds\right)^{1/2}.
\end{align}
Thus, under the supplementary constraint that $a'(0)=a_0'(0)$, we have
\begin{align}
\|(a-a_0)'\|_{1,\sigma}
&\leqslant\left(\int_0^1\sigma^2(s)\ds\right)^{1/2}\|(a-a_0)''\|_{2,\sigma}.
\end{align}
The discrete analogue asymptotically in $N_x\to +\infty$ is
\begin{align}
\sum_{i=1}^{N_x}|\mathbf{U D}(\vec A-\vec A_0)|_{1,\sigma}
&\leqslant N_x\sqrt{\mathrm{Tr}(\mathbf{U} ^{-2})\chi^2_{\rm pen}(\vec{A} | \vec A_{0})}.
\end{align}
This shows that regularisation using the weighted quadratic second-order derivative automatically ensures an upper bound of the weighted total variation norm. This means that while it is computationally much faster, regularising by the weighted quadratic norm of the second-order derivative ensures an upper bound on the regularisation via the weighted total variation norm. Because the usual advantages of the weighted total variation norm (no assumption of a second-order derivative or research of a sparse minimiser) are not important here, the choice of the weighted quadratic norm of the second-order derivative as a loss function is completely pertinent.

\section{Atmospheric differential refraction}\label{sec:adr}

The ADR mostly depends on the pressure,
the temperature, and the airmass and only loosely on the atmospheric humidity.

In our wavelength-calibration process for CALSPEC stars, the absorption line
that weights most in the fit is the main dioxygen line at 762.1\,nm. When the ADR
is not correctly modelled and taken into account in the wavelength calibration,
shifts of the absorption line minima towards the blue part of the spectrum can
be observed throughout the night while the airmass of the star changes.

\begin{figure}[!ht]
\begin{center}
\includegraphics[width=0.48\columnwidth]{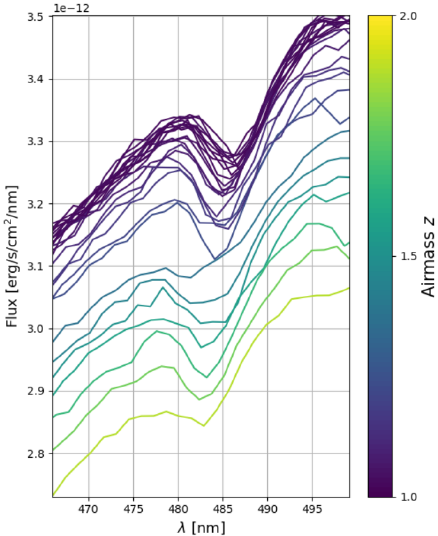}
\includegraphics[width=0.48\columnwidth]{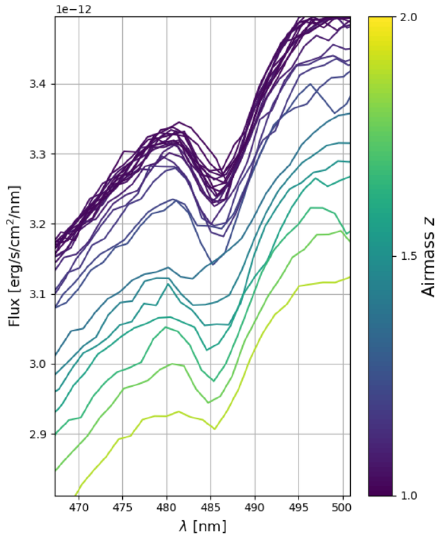}
\end{center}
\caption[] 
{Absorption line $H\beta$ at 486.3\,nm of CALSPEC star HD111980 observed during the night of 2017 May 30 while it goes from an airmass of 1 to an airmass of approximately 2. {Left:} Without modelling the ADR effect in the wavelength-calibration process. {Right:} With the ADR effect in the wavelength-calibration process. The profiles are better aligned.}
\label{fig:adr1}
\end{figure}

\begin{figure}[!ht]
\begin{center}
\includegraphics[width=0.48\columnwidth]{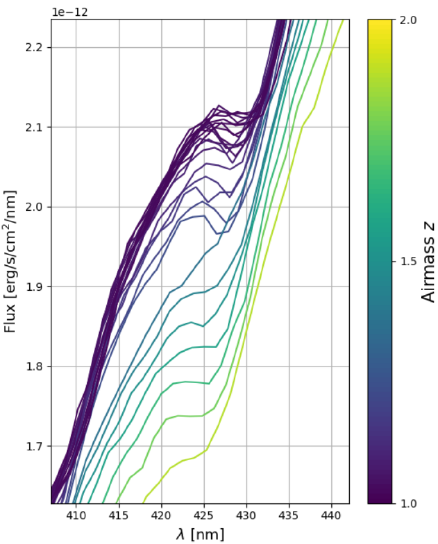}
\includegraphics[width=0.48\columnwidth]{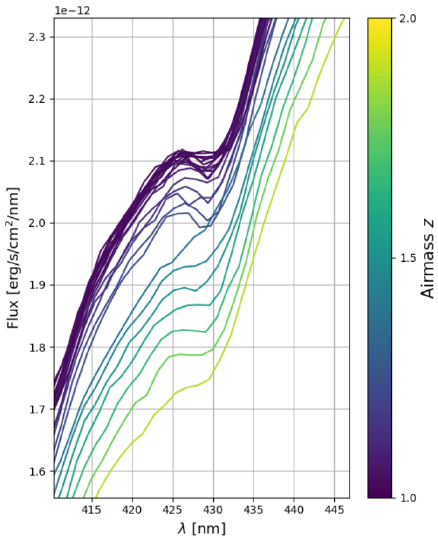}
\end{center}
\caption[] 
{Same as Figure~\ref{fig:adr1}, but for the $Fe$ absorption line at 430.8\,nm.}
\label{fig:adr2}
\end{figure}

This is illustrated in the left panels of Figures~\ref{fig:adr1}
and~\ref{fig:adr2}. In the right panels, the ADR effect is included in the
wavelength-calibration process through a wavelength-dependent shift of the zeroth-order
centroid $\delta u_0^{(\mathrm{ADR})}(\lambda)$. This
procedure absorbs most of the line shifts when the dispersion axis is not orthogonal to the zenith direction.


\begin{figure}[!th]
\centering
\includegraphics[width=0.4\textwidth]{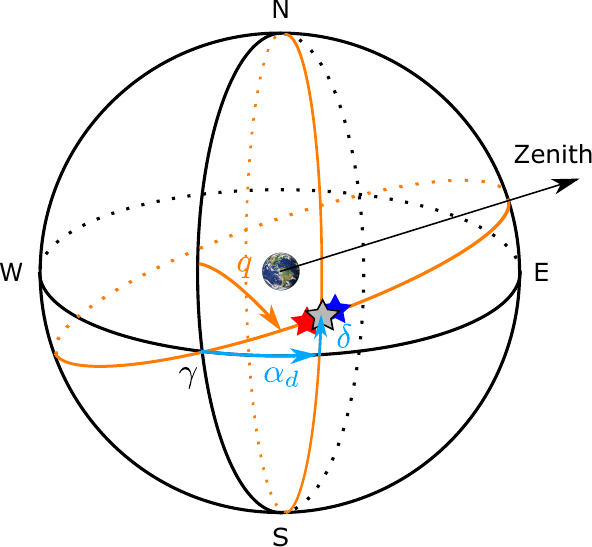}
\includegraphics[width=0.5\textwidth]{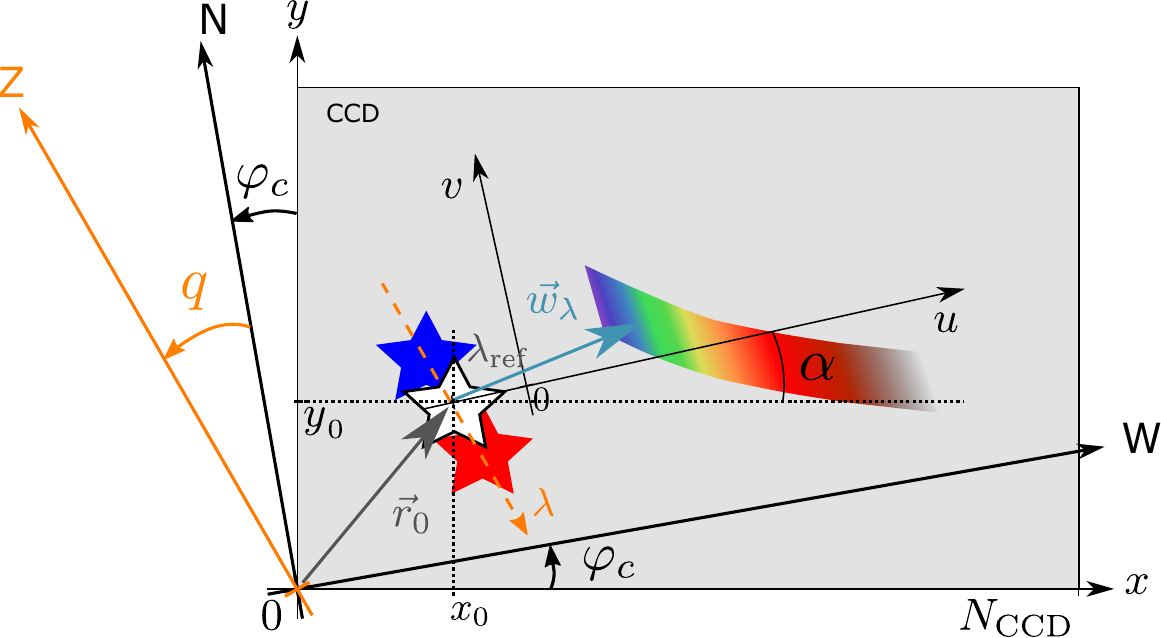}
\caption{Angle conventions for the prediction of the ADR effect in the spectrogram. {Top:} Celestial coordinates in right ascension $\alpha_d$ (RA) and declination $\delta$ (DEC) system, with $q$ the parallactic angle that sets the direction of the local zenith positively eastward. {Bottom:} Angle conventions in the \Spectractor image, with $\varphi_c$ the north-west axis angle with respect to the horizontal axis of the camera.}\label{fig:celestial_coordinates}
\end{figure}

For completeness, we show in Figure~\ref{fig:celestial_coordinates} the
angle conventions that are used in \Spectractor to correctly compute the zenith direction
in the image.

\end{document}